\documentclass[twocolumn,showpacs,pra,longbibliography]{revtex4-1}

\usepackage{graphicx}
\usepackage{colordvi}
\usepackage{color}

\usepackage{url}
\usepackage[colorlinks]{hyperref}
\hypersetup{
    plainpages=true,
    breaklinks=true,
    hypertexnames=false,
    pageanchor=true,
    colorlinks=true,
    linkcolor={blue},
    citecolor={red},
    urlcolor={blue},
    anchorcolor={black}
}

\begin{document}

\title{Nonclassical light at exceptional points of a quantum $\mathcal{PT}$-symmetric two-mode system}

\author{Jan Pe\v{r}ina Jr.}
\email{jan.perina.jr@upol.cz} \affiliation{Joint Laboratory of
Optics, Faculty of Science, Palack\'{y} University, Czech
Republic, 17. listopadu 12, 771~46 Olomouc, Czech Republic}

\author{{Anton\' \i n} Luk\v{s}}
\affiliation{Joint Laboratory of Optics, Faculty of Science,
Palack\'{y} University, Czech Republic, 17. listopadu 12, 771~46
Olomouc, Czech Republic}

\author{Joanna K. Kalaga}
\affiliation{Joint Laboratory of Optics, Faculty of Science,
Palack\'{y} University, Czech Republic, 17. listopadu 12, 771~46
Olomouc, Czech Republic; Quantum Optics and Engineering Division,
Faculty of Physics and Astronomy, University of Zielona G\' ora,
Prof. Z. Szafrana 4a, 65-516 Zielona G\' ora, Poland}

\author{Wies\l aw Leo\' nski}
\affiliation{Joint Laboratory of Optics, Faculty of Science,
Palack\'{y} University, Czech Republic, 17. listopadu 12, 771~46
Olomouc, Czech Republic; Quantum Optics and Engineering Division,
Faculty of Physics and Astronomy, University of Zielona G\' ora,
Prof. Z. Szafrana 4a, 65-516 Zielona G\' ora, Poland}

\author{Adam Miranowicz}
\affiliation{Faculty of Physics, Adam Mickiewicz University,
61-614 Pozna\' n, Poland}

\begin{abstract}
A two-mode optical parity-time ($\mathcal{PT}$) symmetric system, with gain and damping, described by a quantum
quadratic Hamiltonian with additional small Kerr-like nonlinear terms, is analyzed from the point of view of
nonclassical-light generation. Two kinds of stationary states with different types of (in)stability are revealed.
Properties of one of these are related to the presence of semiclassical exceptional points, i.e., exotic degeneracies
of the non-Hermitian Hamiltonian describing the studied system without quantum jumps. The evolution of the logarithmic
negativity, principal squeezing variances, and sub-shot-noise photon-number correlations, considered as entanglement
and non-classicality quantifiers, is analyzed in the approximation of linear-operator corrections to the classical
solution. Suitable conditions for nonclassical-light generation are identified in the oscillatory regime, especially at
and around exceptional points that considerably enhance the nonlinear interaction and, thus, the non-classicality of
the generated light. The role of quantum fluctuations, inevitably accompanying attenuation and amplification in the
evolution of quantum states, is elucidated. The evolution of the system is analyzed for different initial conditions.
\end{abstract}

\maketitle

\section{Introduction}

Parity-time ($\mathcal{PT}$) symmetric Hamiltonians have attracted
a great deal of attention after the occurrence of the paper by
Bender and Boettcher~\cite{Bender1998, Bender1999,Bender2003} in
which they showed how such Hamiltonians can be applied for
describing physical systems. We note that Hamiltonian $ \hat{H} $
is considered to be $\mathcal{PT}$-symmetric if it is invariant
under the combined action of the parity $ \hat{P} $ and the
time-reversal $ \hat{T} $ operators (i.e., $ \hat{H} $ commutes
with the $ \hat{P}\hat{T} $ operator.) Such Hamiltonians, although
involving damping and amplification, are endowed with real
eigenvalues in certain ranges of their
parameters~\cite{Bender2005}. This is so due to certain balance
between the damping and amplification and other parameters
describing an open quantum system. Such Hamiltonians have been
successfully used for describing numerous classical physical
systems involving optical coupled structures~\cite{El-Ganainy2007,
Ramezani2010,Zyablovsky2014,Ogren2017}, optical
waveguides~\cite{Turitsyna2017,Xu2018}, optical
lattices~\cite{Graefe2011, Miri2012,Ornigotti2014,Shui2019},
coupled optical
microresonators~\cite{Peng2014,Peng2014a,Liu2016,Zhou2016,Arkhipov2019},
quantum-electrodynamics circuits (QED)~\cite{Quijandria2018},
optomechanical systems~\cite{Tchodimou2017,Wang2019}, systems with
complex potentials~\cite{Guo2009}, photonics
molecules~\cite{El-Ganainy2014}, etc. Enhanced sensing in such
systems has been demonstrated~\cite{Chen2017,Liu2016}. Also
$\mathcal{PT}$-symmetric chaotic
systems~\cite{Rubinstein2007,Bittner2012,Lu2015} and systems
exhibiting bidirectional invisibility \cite{Koutserimpas2018} have
been addressed.

Balance between damping and amplification leads, among others, to considerable effective enhancement of both linear and
nonlinear interactions in physical systems for suitable parameters, especially at and around exceptional points
(EPs)~\cite{Ozdemir2019,Miri2019}. This is appealing for quantum physicists as it may result in the enhancement of
nonclassical behavior of such systems. Here, the question arises what is the counterpart of classical
$\mathcal{PT}$-symmetric Hamiltonians in quantum
physics~\cite{Agarwal2012,He2015,Vashahri2017,Scheel2018,Perinova2019,Minganti2019} and how they behave. The crucial
problem here is the conservation of commutation relations among the field operators that is disturbed by the processes
of both damping and amplification. The fluctuation-dissipation theorem tells us that both damping and amplification are
accompanied by fluctuating forces that affect an analyzed system~\cite{Scheel2018}. Whereas damping and amplification
can compensate (equilibrate) each other in the classical evolution, the fluctuating forces related to both of them
cannot due to their randomness and quantumness~\cite{Scheel2018}. Especially the noise occurring in the amplification
represents a problem~\cite{Scheel2018} that can be circumvented in passive quasi-$\mathcal{PT}$-symmetric
systems~\cite{Zhong2016}. Although the presence of fluctuating forces disturbs the balance between damping and
amplification to a certain extent, the enhancement of interactions remains. Strong effective nonlinear interactions can
then be used to support the generation of nonclassical light~\cite{He2015,Vashahri2017}. Moreover, other purely quantum
effects have been predicted in quantum $\mathcal{PT}$-symmetric systems including the generation of
entanglement~\cite{Antonosyan2018} and the quantum Zeno effect~\cite{Naikoo2019}. Also their application in
quantum-information processing has been discussed in Ref.~\cite{Croke2015}.

The main goal of this work is the analysis of nonclassical properties of light generated in a typical two-mode
$\mathcal{PT}$-symmetric system with the emphasis on their improvement due to the enhancement of nonlinear interactions
in the vicinity of EPs. Here we study the standard (i.e., semiclassical) EPs, corresponding to $\mathcal{PT}$ phase
transitions, which are degeneracies of non-Hermitian Hamiltonians describing effectively the system in its
semiclassical regime (i.e., without quantum jumps). The motivation for this goal is a rapidly growing research interest
and progress, both theoretical and experimental, on EPs in the last decade, as summarized in two recent
reviews~\cite{Ozdemir2019,Miri2019}. This interest has been stimulated by the observation of nontrivial and often
counter-intuitive optical and condensed-matter phenomena in the vicinity of EPs, together with the enhancement of
interactions. These phenomena can be used to enhance light-matter interactions, as well as to control the generation,
transfer, and detection of light. These EP-induced effects include: enhancement of
sensing~\cite{Liu2016,Chen2017,Hodaei2017}, loss-induced photon~\cite{Brands2014a,Peng2014a,Chang2014} and
phonon~\cite{Jing2014,Lu2017} lasing, nonreciprocal light transmission~\cite{Peng2014,Chang2014}, unidirectional
invisibility~\cite{Lin2011,Regen2012}, chiral modes and directional lasing~\cite{Peng2016}, lasing with enhanced-mode
selectivity~\cite{Feng2014,Hodaei2014}, asymmetric mode switching~\cite{Doppler2016}, group velocity control via
optomechanically-induced transparency~\cite{Jing2015}, and enhanced optomechanical cooling~\cite{Jing2017}, among many
other effects. Applications of EPs are not limited to standard photonics, but also have been proposed for, e.g.,
microwave photonics using superconducting quantum circuits~\cite{Quijandria2018}, quantum plasmonics~\cite{Benisty2011}
(for a review see~\cite{Tame2013}), electronics~\cite{Schindler2011,Chen2019}, metamaterials~\cite{Kang2013}, cavity
optomechanics~\cite{Jing2014,Harris2016,Jing2017}, and acoustics~\cite{Zhu2014,Alu2015}. The EPs, which correspond to
$\mathcal{PT}$ phase transitions, are useful to reveal and describe dynamical phase transitions in condensed-matter
open quantum systems and to classify their topological
phases~\cite{LeykamPRL17,GonzalesPRB17,HuPRB17,GaoPRL18,LiuPRL19,BliokhNat19,MoosSciPost19,Ge2019} or topological
energy transfers~\cite{Xu2016}. Inspired by these various applications of EPs in the vast majority of semiclassical
systems, we now study the behavior of a typical purely quantum $\mathcal{PT}$-symmetric system at and around EPs.

Although we consider a typical quantum-optical $\mathcal{PT}$-symmetric system composed of two optical modes, the
underlying dynamical operator equations and the resulting behavior are relevant for effective description of analogous
systems including matter-field interactions and discussed in some of the above references in the semiclassical
approach. Two considered optical modes, one damped and the other amplified, interact via the linear coupling (for
scrutinizing the results, see Ref.~\cite{Morales2016}). A Kerr-type nonlinearity, typical for physical
$\mathcal{PT}$-symmetric systems and originating, e.g., in the gain saturation, is assumed in both
modes~\cite{He2015,Vashahri2017}. Nonclassical behavior of the analyzed system that stems from the Kerr nonlinearity is
accompanied by that originating in parametric down-conversion~\cite{Boyd2003,Mandel1995}. Whereas the role of
parametric down-conversion in nonclassical-light generation dominates for lower field intensities, the Kerr
nonlinearity gives considerable contribution to nonclassical properties at higher field intensities.

In optics, such models comprising two mutually linearly interacting Kerr nonlinear oscillators are referred to as the
Kerr couplers~\cite{PerinaJr2000,ElOrany2004}. In their usual form of application, i.e. without considering mode
amplification and even its balance with damping, they already allow for the observation of squeezing and antibunching
as well as the generation of quantum entanglement~\cite{Thapliyal2014}. Such quantum Kerr oscillators, as nonlinear
systems, also exhibit quantum chaotic behavior under suitable
conditions~\cite{Milburn1986,Milburn1991,Kowalewska2008,Kowalewska2009,Shahinyan2013}. When analyzed at the
single-photon level, the so-called photon (phonon) blockade effects may occur in the Kerr coupled systems. They result
in truncation of the Hilbert space (nonlinear quantum
scissors)~\cite{Leonski1994,Miranowicz2006,Hovsepyan2014,Miranowicz2014} and subsequent generation of highly entangled
states~\cite{Miranowicz2006,Kowalewska2012,Thapliyal2014}. Various forms of quantum correlations such as the
entanglement and Einstein-Podolsky-Rosen (EPR) steering were also predicted in short chains of quantum Kerr
oscillators~\cite{Olsen2015a,Olsen2015b,Kalaga2016,Kalaga2017}. On the other hand, the process of parametric
down-conversion analyzed alone without any damping or amplification gives rise to both single-mode squeezing and
entanglement between both modes expressed through sub-shot-noise photon-number correlations~\cite{PerinaJr2016b}.

In our quantum $\mathcal{PT}$-symmetric model, parametric down-conversion is treated exactly, contrary to the Kerr
nonlinearity that is incorporated by using linear-operator corrections to the classical solution obtained for mean
values. Properties of steady states including their stability are related to the presence of EPs. In our analysis, we
focus on the region with oscillations in the evolution, where real eigenvalues of the Hamiltonian are found and which
is the most promising for nonclassical-light generation. EPs lie at the border of this region and provide the greatest
effective enhancement of nonlinear interactions. The analysis of similar systems in the regime with amplification, in
which complex eigenvalues occur, can be found in Refs.~\cite{He2015,Vashahri2017} including squeezed-light generation.
The oscillatory regime with $\mathcal{PT}$-symmetry may also be broken due to larger mode intensities, as shown in
Ref.~\cite{Sukhorukov2010} by numerical simulations of a classical nonlinear evolution.

The paper is organized as follows. In Sec. II, we introduce the
model and find its solution in the approximation of
linear-operator corrections to the classical solution. Also
entanglement and non-classicality quantifiers used in the analysis
are introduced and determined in  Sec. II. Steady states of the
corresponding classical nonlinear equations and their stability
are discussed in Sec.~III. Section~IV brings the analytical
solution of the simplified model that includes just the linear
coupling between modes and parametric down-conversion as the
source of non-classicality and entanglement. The evolution of
quantum states, including the effect of entanglement sudden depth
and rebirth, is discussed in general in Sec.~V. Section~VI is
devoted to the generation of quantum states at EPs. Finally,
Sec.~VII brings conclusions. In Appendix~A, stability analysis of
steady states is given.

\section{Two-mode $\mathcal{PT}$-symmetric system with parametric
down-conversion and its quantum evolution}

\subsection{Hamiltonian}

We consider a quantum system composed of two optical oscillator
modes, labelled as 1 and 2, with identical frequencies $ \omega $
described by the following Hamiltonian $ \hat{H} $ written in the
interaction representation~\cite{Perina1991}:
\begin{eqnarray}  
 \hat{H} &=& -i\gamma_1 \hat{a}_1^\dagger\hat{a}_1 -i\gamma_2
  \hat{a}_2^\dagger\hat{a}_2 + \left[ \epsilon \hat{a}_1^\dagger\hat{a}_2 +
   \kappa \hat{a}_1\hat{a}_2 + {\rm h.c.} \right] \nonumber  \\
 & & + \beta_1 \hat{a}_1^{\dagger 2}\hat{a}_1^2 +
   \beta_2 \hat{a}_2^{\dagger 2}\hat{a}_2^2 +
   \beta_c \hat{a}_1^{\dagger}\hat{a}_2^{\dagger}\hat{a}_1\hat{a}_2.
\label{1}
\end{eqnarray}
In Eq.~(\ref{1}), symbol $ \hat{a}_j $ ($ \hat{a}_j^\dagger $)
stands for the annihilation (creation) operator of a photon in
mode $ j $, $ j=1,2 $, and h.c. replaces the Hermitian conjugated
terms. Whereas mode 1 dissipates with the rate $ \gamma_1 \ge 0 $,
mode 2 is amplified at the rate $ -\gamma_2 \ge 0 $. Both modes
are linearly connected via the coupling constants $ \epsilon $ and
$ \kappa $. The coupling constant $ \epsilon $ describes the
exchange of photons between both modes, that conserves the energy.
It originates, e.g., in the overlap of evanescent waves of
wave-guided fields. On the other hand, the coupling constant $
\kappa $ characterizes parametric down-conversion with the
simultaneous creation (or annihilation) of two photons, one in
mode 1, the other in mode 2. This constant, that quantifies adding
or removing the energy in/from the system, is responsible for
nonclassical behavior. We note that this term also usually occurs
in QED Hamiltonians when non-rotating-wave-approximation (non-RWA)
solutions are considered~\cite{Quijandria2018}. Moreover, also
small Kerr nonlinear terms with nonlinear coupling constants $
\beta_1 $ and $ \beta_2 $ appropriate for modes 1 and 2,
respectively, as well as a small cross-Kerr nonlinear term ($
\beta_c $) are written in Hamiltonian $ \hat{H} $ in
Eq.~(\ref{1}). These Kerr terms occur naturally in nonlinear
photonic structures~\cite{Boyd2003} and also when an optical field
interacts with two-level atoms to experience damping or
amplification~\cite{Lamb1974}.

A $\mathcal{PT}$-symmetric form of Hamiltonian $ \hat{H} $ written
in Eq.~(\ref{1}) requires that the exchange of operators of modes
1 and 2 in Eq.~(\ref{1}) transforms the Hamiltonian $ \hat{H} $
into its Hermitian conjugated operator $ \hat{H}^\dagger $. This
occurs provided that all constants appearing in Eq.~(\ref{1}) are
real and
\begin{equation}  
 \gamma_2 = -\gamma_1,  \hspace{1cm} \beta_2 = \beta_1 .
\label{2}
\end{equation}
Moreover, to be useful in physics, the $\mathcal{PT}$-symmetric
form of Hamiltonian $ \hat{H} $ should have real eigenvalues [see
the condition in Eq.~(\ref{6}) below]. We note that the role of
quantum noise in nonlinear evolution of the system under the
conditions in Eq.~(\ref{2}) and $ \kappa = 0 $ was addressed
in~\cite{Perinova2019}.

\subsection{Heisenberg approach without Langevin forces}

To analyze the dynamics of the considered system, we invoke the
usual canonical commutation relations for field
operators~\cite{Meystre2007} and derive the Heisenberg equations
from Hamiltonian $ \hat{H} $ in Eq.~(\ref{1}):
\begin{eqnarray}  
 \frac{d\hat{a}_1}{dt} &=& -\gamma_1\hat{a}_1 -i\epsilon\hat{a}_2
  -i\kappa\hat{a}_2^\dagger -i\beta_c \hat{a}_2^\dagger\hat{a}_2\hat{a}_1
  -2i\beta_1 \hat{a}_1^\dagger\hat{a}_1^2 , \nonumber \\
 \frac{d\hat{a}_2}{dt} &=& -\gamma_2\hat{a}_2 -i\epsilon\hat{a}_1
  -i\kappa\hat{a}_1^\dagger -i\beta_c\hat{a}_1^\dagger\hat{a}_1\hat{a}_2
  -2i\beta_2 \hat{a}_2^\dagger\hat{a}_2^2 , \nonumber \\
 & &
\label{3}
\end{eqnarray}
together with the Hermitian-conjugated ones. However, the solution
of the operator equations in the form of Eq.~(\ref{3}) does not
conserve the commutation relations because of the damping and
amplification terms. To overcome this problem, we have to include
the additional fluctuating Langevin operator forces with suitable
properties [see Eqs.~(\ref{9}) and (\ref{10}) and the text
below]~\cite{Perina1991,Agarwal2012,Perinova2019}.

Neglecting small nonlinear terms in Eqs.~(\ref{3}) ($
\beta_1=\beta_2=\beta_c =0 $), we arrive at the following system
of linear differential operator equations with constant
coefficients:
\begin{equation} 
 \frac{d}{dt} \left[ \begin{array}{c} \hat{a}_1(t) \\
  \hat{a}_1^\dagger(t) \\ \hat{a}_2(t) \\
  \hat{a}_2^\dagger(t) \end{array} \right] = -i
  \left[ \begin{array}{cccc}
   -i\gamma_1 & 0 & \epsilon & \kappa \\
   0 & -i\gamma_1 & -\kappa & -\epsilon \\
   \epsilon & \kappa & -i\gamma_2 & 0 \\
   -\kappa & -\epsilon & 0 & -i\gamma_2 \end{array} \right]
  \left[ \begin{array}{c} \hat{a}_1(t) \\
   \hat{a}_1^\dagger(t) \\ \hat{a}_2(t) \\
   \hat{a}_2^\dagger(t) \end{array} \right] .
\label{4}
\end{equation}
Algebraic analysis of the dynamical matrix of Eq.~(\ref{4})
results in two doubly degenerate eigenfrequencies $ \bar\nu_{1,2}
$:
\begin{equation}  
 \bar\nu_{1,2} = -\frac{i}{2}(\gamma_1+\gamma_2) \pm \sqrt{
 \epsilon^2 - \kappa^2 - \frac{(\gamma_1-\gamma_2)^2}{4} }.
\label{5}
\end{equation}
According to Eq.~(\ref{5}), real frequencies occur provided that $
\gamma_2 = -\gamma_1 \equiv -\gamma $ and
\begin{equation}  
 \mu^2 \equiv \epsilon^2 - \kappa^2 - \gamma^2 \ge 0.
\label{6}
\end{equation}
In this case, the eigenvectors $ \bar Y^\pm_{\nu_j} $ belonging to
the doubly degenerate eigenfrequencies $ \bar\nu_{j} $, $ j=1,2 $,
can be derived in the following form:
\begin{eqnarray}  
 \bar Y^\pm_{\nu_1} &=& \frac{1}{2\sqrt{\epsilon}} \left(
  \zeta^\pm, -\zeta^\mp, \pm \frac{\zeta^\pm (\mu + i\gamma)}{\xi}
  , \mp \frac{\zeta^\mp (\mu + i\gamma)}{\xi}
   \right), \nonumber \\
 \bar Y^\pm_{\nu_2} &=& \frac{1}{2\sqrt{\epsilon}} \left(
  \zeta^\pm, -\zeta^\mp, \mp \frac{\zeta^\pm (\mu - i\gamma)}{\xi}
  , \pm \frac{\zeta^\mp (\mu - i\gamma)}{\xi}
   \right),
\label{7}
\end{eqnarray}
where $ \xi = \sqrt{\epsilon^2-\kappa^2} $, $ \zeta^\pm =
\sqrt{\epsilon \pm \xi} $, and $ \mu = \sqrt{\xi^2-\gamma^2} $.
Equality in Eq.~(\ref{6}) identifies semiclassical
EPs~\cite{Bender1998} that form the boundary in the space of
parameters. For the EPs, the eigenvectors belonging in
Eq.~({\ref{7}) to different eigenfrequencies coincide, i.e. $ \bar
Y^\pm_{\nu_1} = \bar Y^\pm_{\nu_2} $. We note that Eq.~(\ref{3})
does not include quantum Langevin forces. Thus, the calculated EPs
are semiclassical, which might be completely different from
quantum EPs related to the Liouvillians, except for the special
case of the Hamiltonian in Eq.~(\ref{1}) with $ \epsilon > 0 $ and
$ \kappa = \beta_1 = \beta_2 = \beta_c = 0 $. Indeed, as shown in
Ref.~\cite{Minganti2019}, in this case the quantum and
semiclassical EPs are essentially equivalent for larger
excitations.

The corresponding classical equations are written for complex
amplitudes $ \alpha_1 $ and $ \alpha_2 $ of fields (in the
coherent states $ |\alpha_1\rangle $ and $ |\alpha_2\rangle $) in
modes 1 and 2, respectively:
\begin{eqnarray}  
 \frac{d\alpha_1}{dt} &=& -\gamma_1\alpha_1 -i\epsilon\alpha_2
  -i\kappa\alpha_2^* -i \left[\beta_c |\alpha_2|^2
  +2\beta_1 |\alpha_1|^2\right] \alpha_1 , \nonumber \\
 \frac{d\alpha_2}{dt} &=& -\gamma_2\alpha_2 -i\epsilon\alpha_1
  -i\kappa\alpha_1^* -i\left[\beta_c|\alpha_1|^2
  +2\beta_2 |\alpha_2|^2\right]\alpha_2 . \nonumber \\
 & &
\label{8}
\end{eqnarray}
These equations allow in general only for numerical solution.

\subsection{Langevin-Heisenberg approach for quantum amplitude
corrections}

To solve the nonlinear operator equations in Eq.~(\ref{3}), we
adopt the method of linear-operator corrections to the classical
solution. Some of the quantum features of the evolving states are
then described by the equations for the operator-amplitude
corrections $ \delta\hat{a}_j $, $ j=1,2 $, to the classical
solution ($ \hat{a}_j = \alpha_j + \delta\hat{a}_j $). For the
Hamiltonian in Eq.~(\ref{1}), they attain the form:
\begin{eqnarray}   
 \frac{d\delta\hat{a}_1}{dt} &=& -\left(\gamma_1
  +4i\beta_1|\alpha_1|^2  +i\beta_c|\alpha_2|^2 \right)\delta\hat{a}_1
  -2i\beta_1|\alpha_1|^2\delta\hat{a}_1^\dagger  \nonumber \\
  & & \hspace{-5mm} \mbox{}
  - i\left(\epsilon+\beta_c\alpha_1\alpha_2^*\right)\delta\hat{a}_2
  - i\left(\kappa+\beta_c\alpha_1\alpha_2\right)\delta\hat{a}_2^\dagger
  + \hat{l}_1, \nonumber \\
 \frac{d\delta\hat{a}_2}{dt} &=&
  - i\left(\epsilon+\beta_c\alpha_1^*\alpha_2\right)\delta\hat{a}_1
  - i\left(\kappa+\beta_c\alpha_1\alpha_2\right)\delta\hat{a}_1^\dagger
   \nonumber \\
  & & \hspace{-5mm} \mbox{} -\left(\gamma_2+4i\beta_2|\alpha_2|^2+i\beta_c|\alpha_1|^2\right)\delta\hat{a}_2
  -2i\beta_2|\alpha_2|^2\delta\hat{a}_2^\dagger  \nonumber \\
  & & \hspace{-5mm} \mbox{}  + \hat{l}_2.
\label{9}
\end{eqnarray}
The newly introduced fluctuating Langevin operator forces $ l_1 $
and $ l_2 $ independently compensate for the damping in mode 1 and
the amplification in mode 2. We assume their statistical
properties in the Markov and Gaussian
form~\cite{Meystre2007,Agarwal2012,Perinova2019}:
\begin{eqnarray} 
 \langle\hat{l}_1^\dagger(t)\hat{l}_1(t')\rangle = 0,
  \hspace{3mm} \langle\hat{l}_1(t)\hat{l}_1^\dagger(t')\rangle =
  2\gamma_1 \delta(t-t'), \nonumber \\
 \langle\hat{l}_2^\dagger(t)\hat{l}_2(t')\rangle = -2\gamma_2 \delta(t-t'),
  \hspace{3mm} \langle\hat{l}_2(t)\hat{l}_2^\dagger(t')\rangle =0,
\label{10}
\end{eqnarray}
where $ \delta $ stands for the Dirac function.

Equations (\ref{9}) together with those for the
Hermitian-conjugated operator amplitude corrections $
\delta\hat{a}_j^\dagger $, $ j=1,2 $, represent a closed set of
four linear operator equations having the following matrix form:
\begin{equation}  
 \frac{d \delta\hat{\bf A}(t) }{dt} = {\bf M}(t) \delta\hat{\bf A}(t)
  + \hat{\bf L}(t),
\label{11}
\end{equation}
where $ \delta\hat{\bf A}^{\rm T}
\equiv(\delta\hat{a}_1,\delta\hat{a}_1^\dagger,\delta\hat{a}_2,\delta\hat{a}_2^\dagger)$,
$ \hat{\bf L}^{\rm T}\equiv
(\hat{l}_1,\hat{l}_1^\dagger,\hat{l}_2,\hat{l}_2^\dagger)$, and $
{\bf M} $ is the appropriate matrix. The solution of
Eq.~(\ref{11}) is written as~\cite{PerinaJr2000}:
\begin{eqnarray}  
 \delta\hat{\bf A}(t) &=& {\bf P}(t,0)\delta\hat{\bf A}(0)
    + \hat{\bf F}(t),
\label{12}   \\
  \hat{\bf F}(t) &=& \int_{0}^t d\tilde{t} {\bf P}(t,\tilde{t})
  \hat{\bf L}(\tilde{t}).
\label{13}
\end{eqnarray}
The evolution matrix $ {\bf P}(t,t') $ is obtained as a solution
of the equation
\begin{equation} 
 d{\bf P}(t,t')/dt = M(t){\bf P}(t,t')
\label{14}
\end{equation}
assuming $ {\bf P}(t',t') $ is the unity matrix. The correlation
functions of the fluctuating operator forces $ \hat{\bf F}(t) $
defined in Eq.~(\ref{13}) are derived from those in Eq.~(\ref{10})
using the formula~\cite{PerinaJr2000}:
\begin{eqnarray} 
 \langle \hat{\bf F}(t)\hat{\bf F}^{\rm T}(t)\rangle &=& \int_{0}^t d\tilde{t}
  \int_{0}^t d\tilde{t}' {\bf P}(t,\tilde{t}) \langle \hat{\bf L}(\tilde{t})
  \hat{\bf L}^{\rm T}(\tilde{t}')\rangle {\bf P}^{\rm
  T}(t,\tilde{t}'). \nonumber \\
 & &
\label{15}
\end{eqnarray}

Using Eq.~(\ref{12}), we express the solution of Eq.~(\ref{9}) for
the original operator amplitude corrections in the simplified
form:
\begin{equation}  
 \delta\hat{\bf a}(t) = {\bf U}(t)\delta\hat{\bf a}(0)
  + {\bf V}(t)\delta\hat{\bf a}^\dagger(0) + \hat{\bf f}(t),
\label{16}
\end{equation}
where $ \delta\hat{\bf a}^{\rm T} \equiv
(\delta\hat{a}_1,\delta\hat{a}_2) $, $ U_{j,k}(t) =
P_{2j-1,2k-1}(t,0) $, $ V_{jk}(t) = P_{2j-1,2k}(t,0) $, and $
\hat{f}_j(t) = \hat{F}_{2j-1}(t) $, $ j,k=1,2 $.

In our analysis, in which we assume incident vacuum or coherent
states, the generated states remain Gaussian and so we express
their normal characteristic function $ C_{\cal N} $
as~\cite{Perina1991}
\begin{eqnarray} 
 C_{\cal N}(\mu_1,\mu_2,t) &=& \exp\Biggl\{ \sum_{j=1,2} \Bigl[
   \left( \alpha_j^*(t)\mu_j -{\rm c.c.}\right)
   \nonumber\\
 & & \hspace{-5mm} \mbox{} -B_j(t)|\mu_j|^2 + \left(C_j(t)\mu_j^{2*} - {\rm c.c.}\right)/2 \Bigr]
  \nonumber \\
 & & \hspace{-5mm} \mbox{} + \left( D(t)\mu_1^*\mu_2^* + \bar{D}(t)\mu_1\mu_2^* + {\rm
   c.c.}\right) \Bigr\},
\label{17}
\end{eqnarray}
where c.c. replaces the complex-conjugated term. The functions $
B_j $, $ C_j $ ($j=1,2 $), $ D $, and $ \bar{D} $ introduced in
Eq.~(\ref{17}) are given for the incident coherent states as:
\begin{eqnarray}  
 B_j(t) &\equiv& \langle\delta\hat{a}_j^\dagger(t)\delta\hat{a}_j(t)\rangle
  = \sum_{l=1,2} \left[ |V_{jl}(t)|^2 +
  \langle\hat{f}_j^\dagger(t)\hat{f}_j(t)\rangle \right], \nonumber \\
 C_j(t) &\equiv& \langle [\delta\hat{a}_j(t)]^2\rangle
  = \sum_{l=1,2} \left[ U_{jl}(t) V_{jl}(t) +
  \langle[\hat{f}_j(t)]^2\rangle \right], \nonumber \\
 D(t) &\equiv& \langle\delta\hat{a}_1(t)\delta\hat{a}_2(t)\rangle
   \nonumber \\
 & & = \sum_{l=1,2} \left[ U_{1l}(t)V_{2l}(t) +
  \langle\hat{f}_1(t)\hat{f}_2(t)\rangle \right], \nonumber \\
 \bar{D}(t) &\equiv& -\langle\delta\hat{a}_1^\dagger(t)\delta\hat{a}_2(t)\rangle
  \nonumber \\
 & &  = -\sum_{l=1,2} \left[V_{1l}^*(t)V_{2l}(t)  +
   \langle\hat{f}_1^\dagger(t)\hat{f}_2(t)\rangle \right].
\label{18}
\end{eqnarray}

As the functions $ D $ and $ \bar{D} $ occur in the characteristic
function $ C_{\cal N} $ in Eq.~(\ref{18}), the described state may
exhibit entanglement. It can be quantified via the logarithmic
negativity $ E_N $ determined from the symmetric covariance matrix
as described in Ref.~\cite{Adesso2007}. This entanglement may lead
to quantum correlations between the photon numbers $ n_1 $ and $
n_2 $ of both modes. These correlations are usually quantified by
the sub-shot-noise parameter $ R < 1 $ where $ R\equiv \langle
\Delta( \hat{n}_1-\hat{n}_2)^2\rangle /(\langle \hat{n}_1\rangle +
\langle\hat{n}_2\rangle) $, $ \hat{n}_j =
\hat{a}_j^\dagger\hat{a}_j $, $ j=1,2 $ and $ \Delta\hat{x} =
\hat{x} - \langle\hat{x}\rangle $ for an arbitrary operator $
\hat{x} $. For the characteristic function $ C_{\cal N} $ in
Eq.~(\ref{17}) with the coefficients in Eq.~(\ref{18}), the
sub-shot-noise parameter $ R $ attains the form:
\begin{eqnarray} 
 R &=& 1 + \Bigl\{ \sum_{j=1,2}\left[ b_j^2 + |c_j|^2\right]
  -2|d|^2 -2|\bar{d}|^2 \nonumber \\
 & & \mbox{} -2\left(
  |\alpha_1|^2-|\alpha_2|^2\right)^2 \Bigr\} /
   \left( b_1+b_2 \right)
\label{19}
\end{eqnarray}
and $ b_j = |\alpha_j|^2 + B_j $, $ c_j = \alpha_j^2 + C_j $ ($
j=1,2 $), $ d = \alpha_1\alpha_2 + D $, and $ \bar{d} =
\alpha_1^*\alpha_2 - \bar{D} $.

The generated states exhibit their non-classicality also as the
squeezing of phase fluctuations of their complex amplitudes. Both
single-mode squeezing quantified by the principal squeezing
variances $ \lambda_1 $ and $ \lambda_2 $ smaller than
1~\cite{Luks1988} and two-mode squeezing described by the
principal squeezing variance $ \lambda < 2 $~\cite{Perina1991} may
be observed:
\begin{eqnarray}  
 \lambda_j &=& 1+ 2\left( B_j - |C_j| \right), \hspace{5mm} j=1,2,
\label{20}  \\
 \lambda &=& 1+ 2\left( B_1 + B_2 - \bar{D} - \bar{D}^* -
  | C_1+C_2+2D| \right). \nonumber \\
 & &
\label{21}
\end{eqnarray}

\section{Steady states and their stability analysis}

For the general form of Hamiltonian $ \hat{H} $ in Eq.~(\ref{1}),
we find two nontrivial steady states and analyze their stability.
Expressing the complex amplitudes $ \alpha_j $ as $
\varrho_j\exp(i\varphi_j) $, $ j=1,2 $, we transform the nonlinear
equations in Eq.~(\ref{3}) into the form:
\begin{eqnarray}   
 \frac{d\varrho_1}{dt} &=& -\gamma_1\varrho_1 + \left[\epsilon\sin(\varphi)
  - \kappa\sin(\psi)\right] \varrho_2, \nonumber \\
 \frac{d\varrho_2}{dt} &=& -\gamma_2\varrho_2 - \left[\epsilon\sin(\varphi)
  + \kappa\sin(\psi)\right] \varrho_1, \nonumber \\
 \frac{d\varphi_1}{dt} &=& -\left[ \epsilon\cos(\varphi)
  +\kappa\cos(\psi)\right] \frac{\varrho_2}{\varrho_1} - \beta_c\varrho_2^2
  -2\beta_1\varrho_1^2, \nonumber \\
 \frac{d\varphi_2}{dt} &=& -\left[\epsilon\cos(\varphi)
  +\kappa\cos(\psi)\right] \frac{\varrho_1}{\varrho_2} - \beta_c\varrho_1^2
  -2\beta_2\varrho_2^2, \nonumber \\
 & &
\label{22}
\end{eqnarray}
where $ \psi = \varphi_2 + \varphi_1 $ and $ \varphi = \varphi_2 -
\varphi_1 $.

To reveal steady states of Eqs.~(\ref{22}), we set
\begin{equation}  
 \frac{d\varrho_1}{dt} = \frac{d\varrho_2}{dt} = \frac{d\varphi_1}{dt}
   = \frac{d\varphi_2}{dt} = 0.
\label{23}
\end{equation}
The subtraction of the fourth equation multiplied by $ \varrho_1^2
$ from the third one multiplied by $ \varrho_2^2 $ leads us to the
relation between the amplitudes $ \varrho_1 $ and $ \varrho_2 $:
\begin{equation}  
 \varrho_2^{\rm st} =  \beta_{12} \varrho_1^{\rm st},
\label{24}
\end{equation}
where $ \beta_{12} \equiv \sqrt[4]{\beta_1/\beta_2} $ (we assume $
\beta_1, \beta_2 > 0 $). For the steady states and their stability
analysis in the special case in which only the cross-Kerr
nonlinearity is present ($ \beta_1 = \beta_2 = 0 $, $ \beta_c \ne
0 $), see Ref.~\cite{PerinaJr2019x}.

Using this relation, the first and second equations in
Eqs.~(\ref{22}) are transformed into the form:
\begin{eqnarray}  
 \gamma_2 \beta_{12} &=& -\epsilon\sin(\varphi^{\rm st})
  -\kappa\sin(\psi^{\rm st}) , \nonumber \\
 \gamma_1 / \beta_{12} &=& \epsilon\sin(\varphi^{\rm st})
  -\kappa\sin(\psi^{\rm st}).
\label{25}
\end{eqnarray}
The solution of the linear algebraic equations in Eq.~(\ref{25})
with respect to $ \sin(\varphi^{\rm st}) $ and $ \sin(\psi^{\rm
st}) $ leaves us with the formulas for angles $ \varphi^{\rm st} $
and $ \psi^{\rm st} $:
\begin{eqnarray} 
 \psi^{\rm st} &=&  \arcsin\left[\frac{ - \gamma_2 \beta_{12}
   - \gamma_1 /\beta_{12} }{ 2\kappa}\right] , \nonumber \\
 \varphi^{\rm st} &=& \arcsin\left[\frac{ - \gamma_2 \beta_{12}
   + \gamma_1 /\beta_{12} }{ 2\epsilon} \right].
\label{26}
\end{eqnarray}

Finally, applying Eq.~(\ref{24}) in the third equation in
Eqs.~(\ref{22}) we arrive at the formula for $ \varrho_1^{\rm st}
$ (we assume $ \beta_c > - 2\sqrt{\beta_1 \beta_2} $)
\begin{eqnarray}  
 \varrho_1^{\rm st} &=& \sqrt{ \frac{-c_\epsilon - c_\kappa }{ \beta_c \beta_{12} + 2
  \beta_1/\beta_{12} }}
\label{247}
\end{eqnarray}
that identifies two distinct steady states. We have for the first
one
\begin{equation} 
 c_\epsilon = \mp\epsilon|\cos(\varphi^{\rm st})| \hspace{2mm} {\rm and}
 \hspace{2mm} c_\kappa = \mp\kappa|\cos(\psi^{\rm st})|,
\label{28}
\end{equation}
where the combination of signs has to guarantee that $
\varrho_1^{\rm st} \ge 0 $. On the other hand, the second steady
state is identified via one combination of signs assuring that $
\varrho_1^{\rm st} \ge 0 $:
\begin{equation} 
 c_\epsilon = \pm\epsilon|\cos(\varphi^{\rm st})| \hspace{2mm} {\rm and}
 \hspace{2mm} c_\kappa = \mp\kappa|\cos(\psi^{\rm st})|.
\label{29}
\end{equation}

A numerical analysis of the first steady state of Eq.~(\ref{28})
reveals the expected physical behavior. The state is stable
[unstable] if the damping in the system dominates [is dominated
by] its amplification ($ \gamma_1 > |\gamma_2| $) [($ \gamma_1 <
|\gamma_2| $)]. For the $\mathcal{PT}$-symmetric case $ \gamma_2 =
-\gamma_1 $, the frequencies in the stability analysis are real
(for the analytical results in this case, see Appendix~A).

On the other hand, the second steady state of Eq.~(\ref{29}) is
unstable independently of the relation between the values of $
\gamma_1 $ and $ |\gamma_2| $. Only in EPs occurring in
$\mathcal{PT}$-symmetric systems, all frequencies in the stability
analysis are zero (see Fig.~\ref{fig1}). We note that the analyzed
EPs are semiclassical, i.e. they belong to open quantum models in
their semiclassical limit, where quantum jumps are ignored. In
open models with quantum jumps, quantum EPs can be analyzed using
the Liouvillian spectra, as recently demonstrated
in~\cite{Minganti2019}. In the analyzed semiclassical EPs, the
amplitudes $ \varrho_1^{\rm st} $ and $ \varrho_2^{\rm st} $ of
the steady state vanish, i.e., the steady state coincides with the
trivial steady state. In our opinion, the presence of such
specific steady states in nonlinear systems reveals the same
values of parameters as those identified by the linear analysis as
EPs [compare Eq.~(\ref{6})].
\begin{figure} 
 (a)\includegraphics[width=0.4\textwidth]{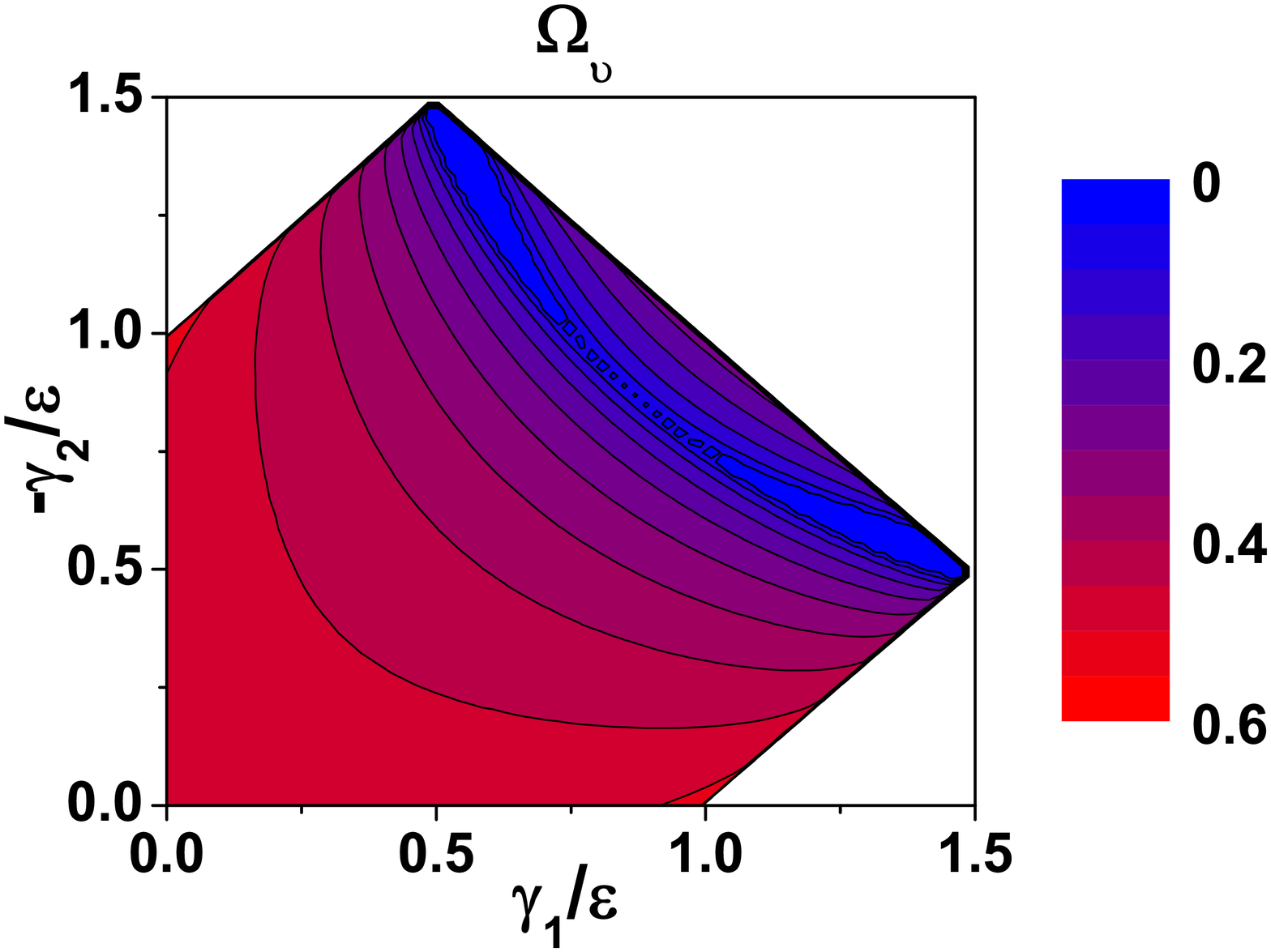}
 \vspace{2mm}
 (b)\includegraphics[width=0.4\textwidth]{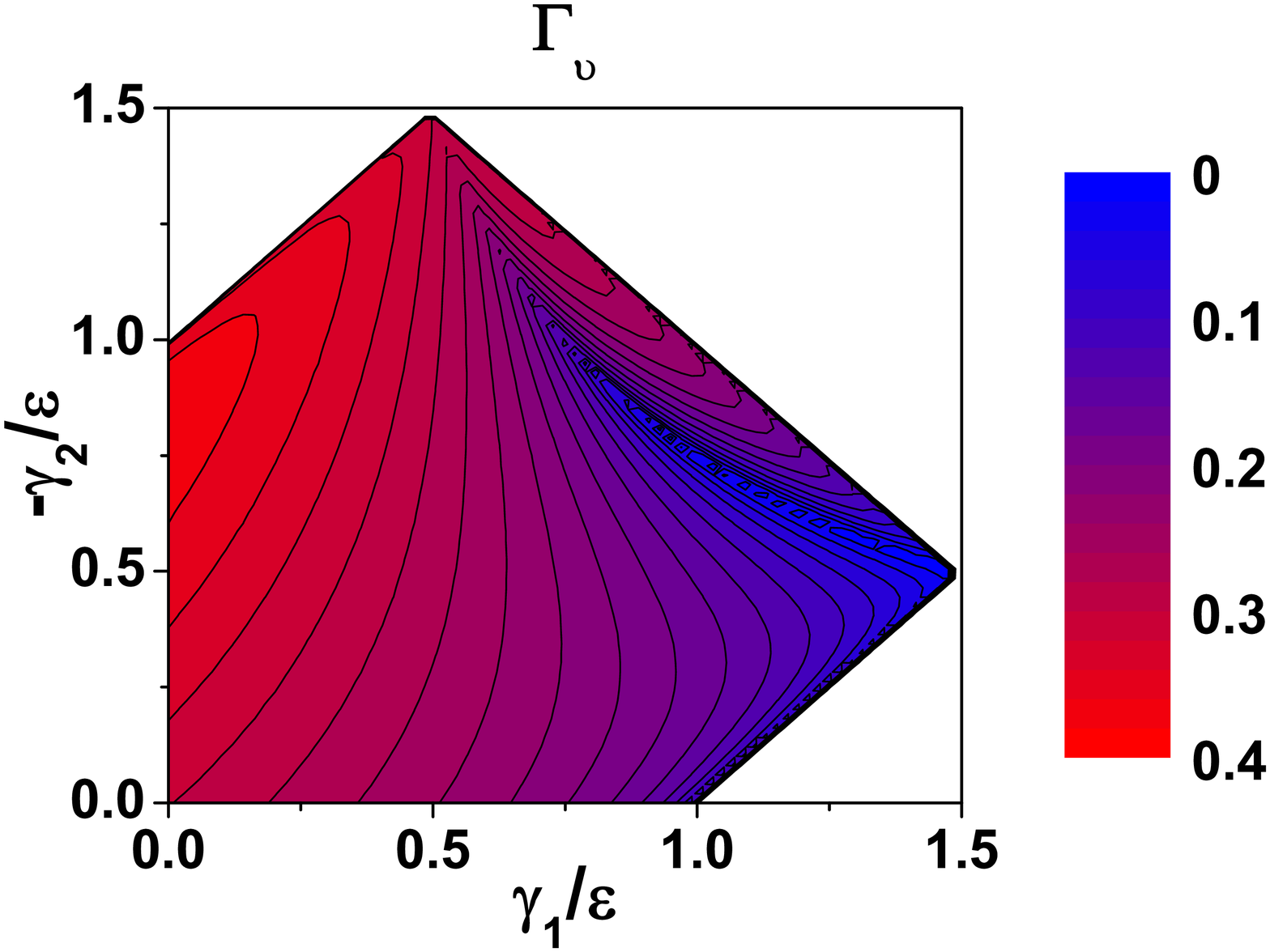}

 \caption{Maximal values of the real and imaginary parts of four complex
 frequencies $ \tilde\nu $ from the stability analysis plotted as
 functions (a) $ \Omega_{\tilde\nu} = \log[1+|{\rm Re}(\tilde\nu)|] $
 and (b) $ \Gamma_{\tilde\nu} = {\rm sign}[{\rm Im}(\tilde\nu)]\log[1+|{\rm Im}(\tilde\nu)|] $
 depending on the varying dimensionless damping and amplification parameters $
 \gamma_1/\epsilon $ and $ -\gamma_2/\epsilon $; $ {\rm sign} $ gives the sign of its
 argument, $ \log $ means the decimal logarithm, $ \kappa = 0.5\epsilon $,
 $ \beta_1=\beta_2 = 0.05\epsilon $, and $ \beta_c =
 0 $.}
\label{fig1}
\end{figure}

If a steady state is considered as the classical solution, the
matrix $ {\bf M} $ of the linear operator equations in
Eq.~(\ref{11}) is time independent and the evolution can be
formulated in terms of its eigenvalues $ (-i\tilde\nu_j) $ and
corresponding eigenvectors $ {\bf Y}_j $, $ j=1,\ldots,4$. Forming
a matrix $ {\bf Y} $ from the eigenvectors $ {\bf Y}_j $ placed as
columns and introducing the matrix $ {\bf y} $ as its inverse, we
can express the elements of the matrices $ {\bf U} $ and $ {\bf V}
$ in Eq.~(\ref{16}) and the matrix $ \langle \hat{\bf F}\hat{\bf
F}^{\rm T}\rangle $ in Eq.~(\ref{15}) as follows:
\begin{eqnarray} 
 U_{jk}(t) &=& \sum_{l=1}^4 \exp(-i\tilde\nu_l t) Y_{2j-1,l}y_{l,2k-1} ,
  \nonumber \\
 V_{jk}(t) &=& \sum_{l=1}^4 \exp(-i\tilde\nu_l t) Y_{2j-1,l}y_{l,2k} ,
  \nonumber \\
 \langle\hat{\bf F}_{k}(t)\hat{\bf F}^{\rm T}_{k'}(t)\rangle &=& 2i
  \sum_{l,l'=1}^4 Y_{kl} Y_{k'l'} \frac{\exp[-i(\tilde\nu_l+\tilde\nu_{l'})t]
  -1}{\tilde\nu_l+\tilde\nu_{l'} } \nonumber \\
 & & \mbox{} \times \left[ \gamma_1y_{l1}y_{l'2} -
   \gamma_2y_{l4}y_{l'3} \right].
\label{30}
\end{eqnarray}

\section{Parametric down-conversion as the source of
non-classicality in $\mathcal{PT}$-symmetric system}

Before we analyze the behavior of the system under general
conditions numerically, we derive analytical formulas for the
considered quantities in the most simplified form of Hamiltonian $
\hat{H} $ in which only $ \epsilon $ and $ \kappa $ are nonzero.
Zero values of $ \gamma_1 $ and $ \gamma_2 $ mean no fluctuating
forces. This analysis provides an insight into the behavior of
non-classicality.

In this case, the field operators obey Eq.~(\ref{4}) with $
\gamma_1 = \gamma_2 = 0 $. Equation~(\ref{5}) gives us two doubly
degenerate real eigenfrequencies $ \bar{\nu}_{1,2} = \pm\xi $, $
\xi = \sqrt{\epsilon^2-\kappa^2} $. The solution of the Heisenberg
equations is then obtained as
\begin{eqnarray}   
 \hat{a}_1(t) &=& \cos(\xi t)\hat{a}_1(0) \nonumber \\
 & & \mbox{} -i\sin(\xi t)\left[
  (\epsilon/\xi)\hat{a}_2(0) + (\kappa/\xi)\hat{a}_2^\dagger(0)
  \right], \nonumber \\
 \hat{a}_2(t) &=& \cos(\xi t)\hat{a}_2(0) \nonumber \\
 & & \mbox{} -i\sin(\xi t)\left[
  (\epsilon/\xi)\hat{a}_1(0) + (\kappa/\xi)\hat{a}_1^\dagger(0)
  \right].
\label{31}
\end{eqnarray}

For the solution in Eq.~(\ref{31}), the parameters of the
characteristic function $ C_{\cal N} $ defined in Eqs.~(\ref{18})
attain the form:
\begin{eqnarray}  
 & B_1(t)=B_2(t) = (\kappa/\xi)^2 \sin^2(\xi t),& \nonumber \\
 & C_1(t)=C_2(t) = -(\epsilon\kappa/\xi^2) \sin^2(\xi t),& \nonumber \\
 & D(t) = -i(\kappa/\xi) \sin(2\xi t)/2, & \nonumber \\
 & \bar{D} = 0. &
\label{32}
\end{eqnarray}

Further, the symmetric covariance matrix $ {\bf C}_{\cal S} $ is
derived in the following form:
\begin{equation} 
 {\bf C}_{\cal S} = \left[ \begin{array}{cccc}
  1 + d_1(t) & 0 & 0 & d_3(t) \\
  0 & 1 + d_2(t) & d_3(t) & 0 \\
  0 & d_3(t) & 1 + d_1(t) & 0 \\
  d_3(t) & 0 & 0 & 1 + d_2(t) \end{array} \right],
\label{33}
\end{equation}
where $ d_1(t) = -[2\kappa/(\epsilon+\kappa)]\sin^2(\xi t) $, $
d_2(t) = [2\kappa/(\epsilon-\kappa)]\sin^2(\xi t) $, and  $ d_3(t)
= -[\kappa/\xi]\sin(2\xi t) $. The symplectic eigenvalues $
\sigma_{\pm} $ of the positive-partial-transposed (PPT) covariance
matrix $ {\bf C}_{\cal S} $ with $ \det ({\bf C}_{\cal S}) = 1 $
fulfill $ \sigma_{\pm}^2 = ( \Delta \pm \sqrt{\Delta^2-4})/2
$~\cite{Hill1997,Adesso2007} where the seralian $ \Delta $ (the
second global invariant of matrices $4\times4$, that accompanies
the determinant, see, e.g.,~\cite{Adesso2007}) is derived as:
\begin{equation} 
 \Delta = 2 + (2\kappa/\xi)^2 \sin^2(2\xi t).
\label{34}
\end{equation}
Using these results, the positive logarithmic negativity $ E_N
$~\cite{Hill1997,Adesso2007} is determined by the formula $ E_N =
-\log(\sigma_-^2)/2 $ applicable if $ |\sigma_-| < 1 $:
\begin{equation} 
 E_N(t) = -\frac{1}{2}\log\left( 1+ 2\frac{ 1 -\sqrt{1+ g^2(t)}}{
 g^2(t)} \right)
\label{35}
\end{equation}
and $ g(t) = \xi/[\kappa\sin(2\xi t)] $.

Assuming the incident vacuum states, Eq.~(\ref{19}) provides the
sub-shot-noise parameter R as
\begin{equation} 
 R(t) = 2 (\epsilon/\xi)^2\sin^2(\xi t).
\label{36}
\end{equation}

The principal squeezing variances, determined along
Eqs.~(\ref{20}) and (\ref{21}), are obtained for arbitrary
incident coherent states as:
\begin{eqnarray}  
 \lambda_{1}(t) = \lambda_{2}(t) &=& 1-2[\kappa/(\epsilon+\kappa)]\sin^2(\xi t),
  \nonumber \\
 \lambda(t) &=& 2\left[1+2(\kappa/\xi)^2\sin^2(\xi t)
   -2(\kappa/\xi^2)|\sin(\xi t)| \right. \nonumber \\
  & & \left. \mbox{} \times \sqrt{\epsilon^2-\kappa^2\cos^2(\xi
   t)} \right].
\label{37}
\end{eqnarray}

For the semiclassical EP occurring for $ \epsilon = \kappa $, the
above formulas simplify to
\begin{eqnarray}  
 E_N(t) &=& -\frac{1}{2}\log\left( 1+8\epsilon^2 t^2 - 4\epsilon
  t\sqrt{1+4\epsilon^2t^2} \right),   \nonumber \\
 R(t) &=& 2\epsilon^2 t^2, \nonumber \\
 \lambda_1 &=& \lambda_2 = 1, \nonumber \\
 \lambda(t) &=& 2\left( 1+2\epsilon^2 t^2 - 2\epsilon t\sqrt{1+\epsilon^2t^2}
  \right).
\label{38}
\end{eqnarray}
According to Eq.~(\ref{38}), the non-classicality is 'encoded'
only in the correlations between the modes. Whereas the
sub-shot-noise parameter $ R $ is nonclassical only for short
times, two-mode squeezing gradually develops with time and it
becomes maximal for $ t\rightarrow \infty $. Except for $ t=0 $,
the states are entangled and the logarithmic negativity $ E_N $
even goes to infinity for $ t\rightarrow \infty $.

For a system with general parameters ($ \epsilon \neq \kappa $),
the maximal value $ E_N^{\rm max} $ of the logarithmic negativity
and the minimal values of the sub-shot-noise parameter $ R^{\rm
min} $ and the principal squeezing variances ($ \lambda_{1,2}^{\rm
min} $, $ \lambda^{\rm min} $) taken over $ t\in\langle
0,\infty\rangle $ reveal the system potential to generate
nonclassical states:
\begin{eqnarray} 
 E_N^{\rm max} &=& -\frac{1}{2}\log\left( \frac{\epsilon-\kappa}{\epsilon+\kappa}\right),   \nonumber \\
 R^{\rm min} &=& 0, \nonumber \\
 \lambda_1^{\rm min} &=& \lambda_2^{\rm min} = \lambda^{\rm min}/2 = \frac{\epsilon-\kappa}{\epsilon+\kappa}.
\label{39}
\end{eqnarray}
The analysis of formulas in Eq.~(\ref{39}) confirms that the best
conditions for the generation of nonclassical states occur at the
EP, i.e. if $ \epsilon = \kappa $. Moreover, the larger the
distance $ \epsilon-\kappa $ from the EP in the space of
parameters is, the worse the non-classicality parameters are. It
is worth noting that the extremal values in Eq.~(\ref{39}) are
reached at different time instants: $ \xi t = \pi/4 + n\pi $ for $
E_N^{\rm max} $, $ \xi t = \pi/2 + n\pi $ for $ \lambda_1^{\rm
min} $, $ \lambda_2^{\rm min} $ and $ \lambda^{\rm min} $, and $
\xi t = (n+1)\pi $ for $ R^{\rm min} $, $ n = 0, 1,\ldots $.

\section{Nonclassical states and their time evolution}

Based on numerical analysis, we extend the discussion of
nonclassical properties of the generated states from Sec.~IV to
the case with damping and amplification. First, we consider only
weak damping and amplification. In this case, the system with
parametric down-conversion as a source of non-classicality has
enough time to evolve an initial classical state into highly
nonclassical states endowed with entanglement, squeezing, and
sub-shot-noise photon-number correlations, as shown in
Fig.~\ref{fig2}. Even, provided that we consider an initial state
close to the second stationary state of Eq.~(\ref{29}) we observe
the effect of entanglement sudden death and rebirth in the
evolution of the logarithmic negativity $ E_N $ [see
Fig.~\ref{fig2}(a)]. A comparison of the curves in
Figs.~\ref{fig2}(a) and \ref{fig2}(b) reveals that the states
exhibit single mode non-classicality evidenced by the single-mode
principal squeezing variances $ \lambda_1 $ and $ \lambda_2 $ even
at time instants in which the entanglement vanishes. Also, the
time evolution of the two-mode principal squeezing variance $
\lambda $ in Fig.~\ref{fig2}(b) documents that its nonclassical
values may originate either in single-mode non-classicalities or
two-mode entanglement. In general, the curves in Fig.~\ref{fig2}
show that even small values of damping and amplification result in
a gradual loss of non-classicality during the time evolution.
\begin{figure} 
 (a)\includegraphics[width=0.35\textwidth]{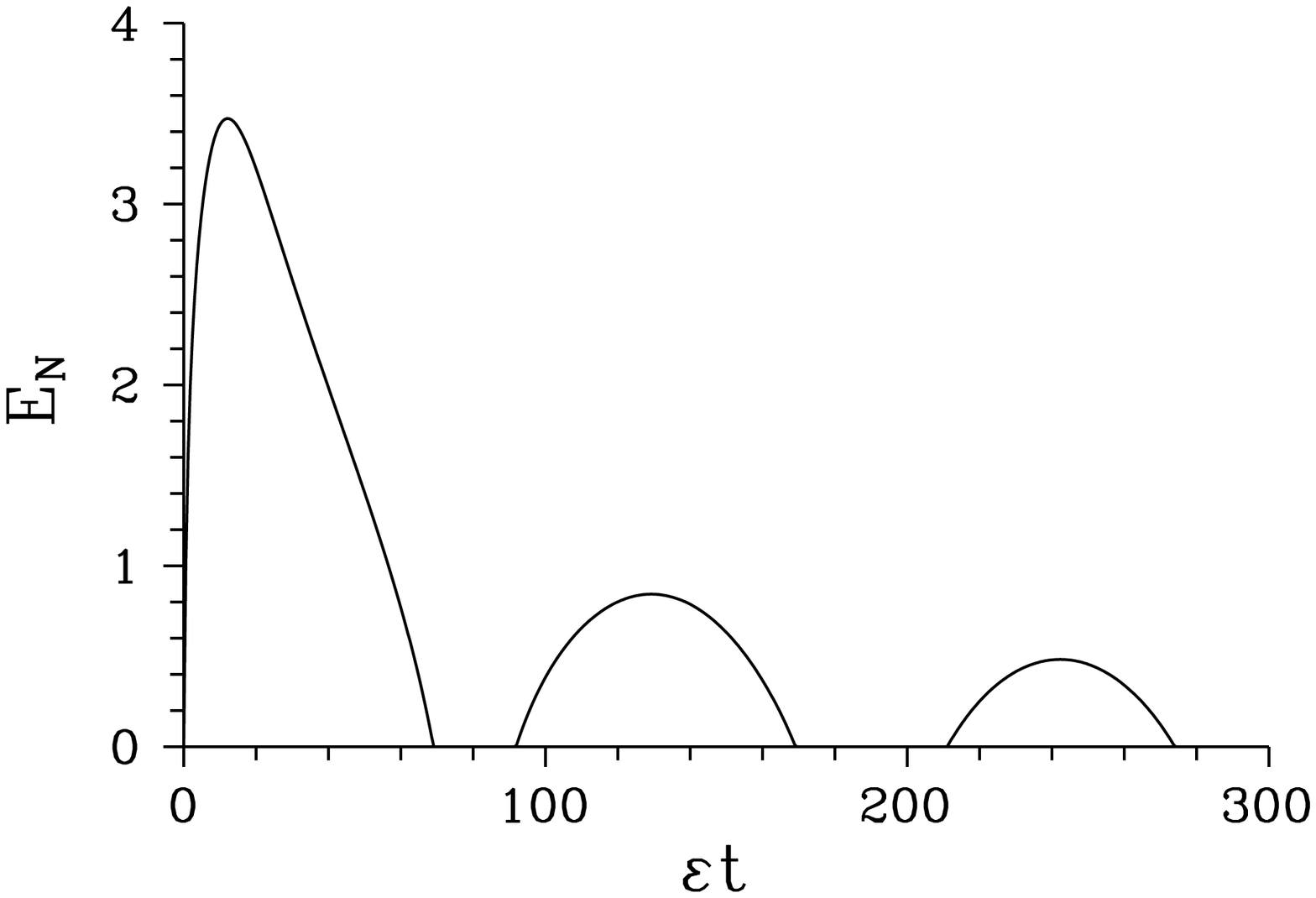}

 \vspace{3mm}
 (b)\includegraphics[width=0.35\textwidth]{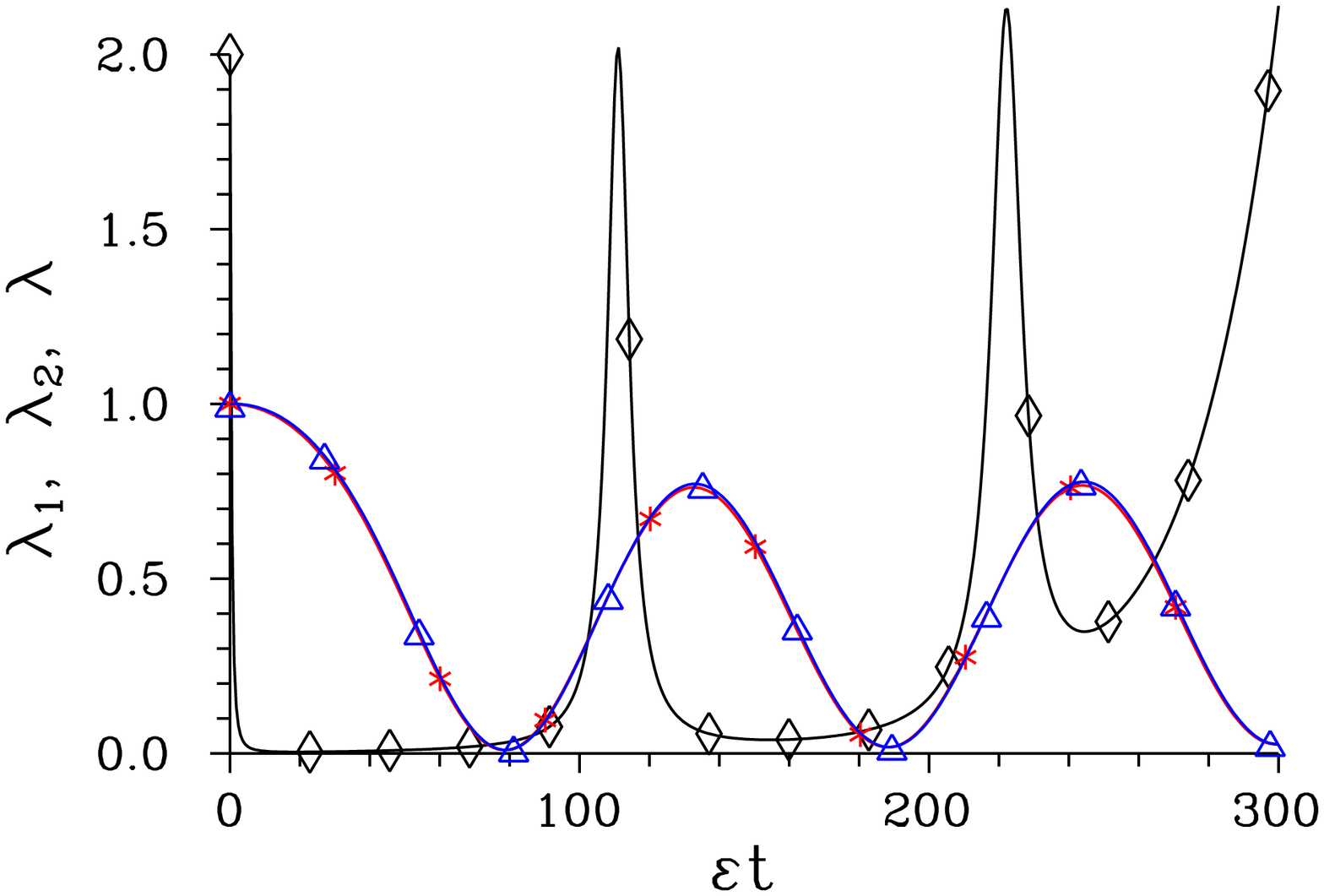}

 \caption{(a) Logarithmic negativity $ E_N $ and (b) principal
  squeezing variances $ \lambda_1 $ (\Red{$ \ast $}), $ \lambda_2 $ (\Blue{$ \triangle $})
  and $ \lambda $ ($ \diamond $) as they depend on dimensionless
  time $ \epsilon t $; $ \gamma = 1\times 10^{-4}\epsilon $,
  $ \kappa= (1- 1\times 10^{-4})\epsilon $, $ \beta_1 =
  \beta_2= 0.05\epsilon $, and $ \beta_c = 0 $. The initial steady
  state in Eq.~(\ref{29}) is assumed.}
\label{fig2}
\end{figure}

Greater values of damping and amplification naturally lead to a
faster loss of non-classicality for two possible reasons. The
first reason is the presence of damping and amplification terms in
the Heisenberg equations in Eqs.~(\ref{9}), the second one is the
presence of fluctuating Langevin forces in Eq.~(\ref{9}) that
constantly add the noise into the state during its evolution.
Although both contributions are intimately related by the
fluctuation-dissipation theorem~\cite{Perina1991}, we may separate
them in our model to judge independently their roles in the state
evolution (see Fig.~\ref{fig3}). Comparison of the solid and
dashed curves in Fig.~\ref{fig3} leads us to the important
conclusion that the Langevin forces are dominantly responsible for
a gradual loss of non-classicality during the evolution.
\begin{figure} 
 \includegraphics[width=0.23\textwidth]{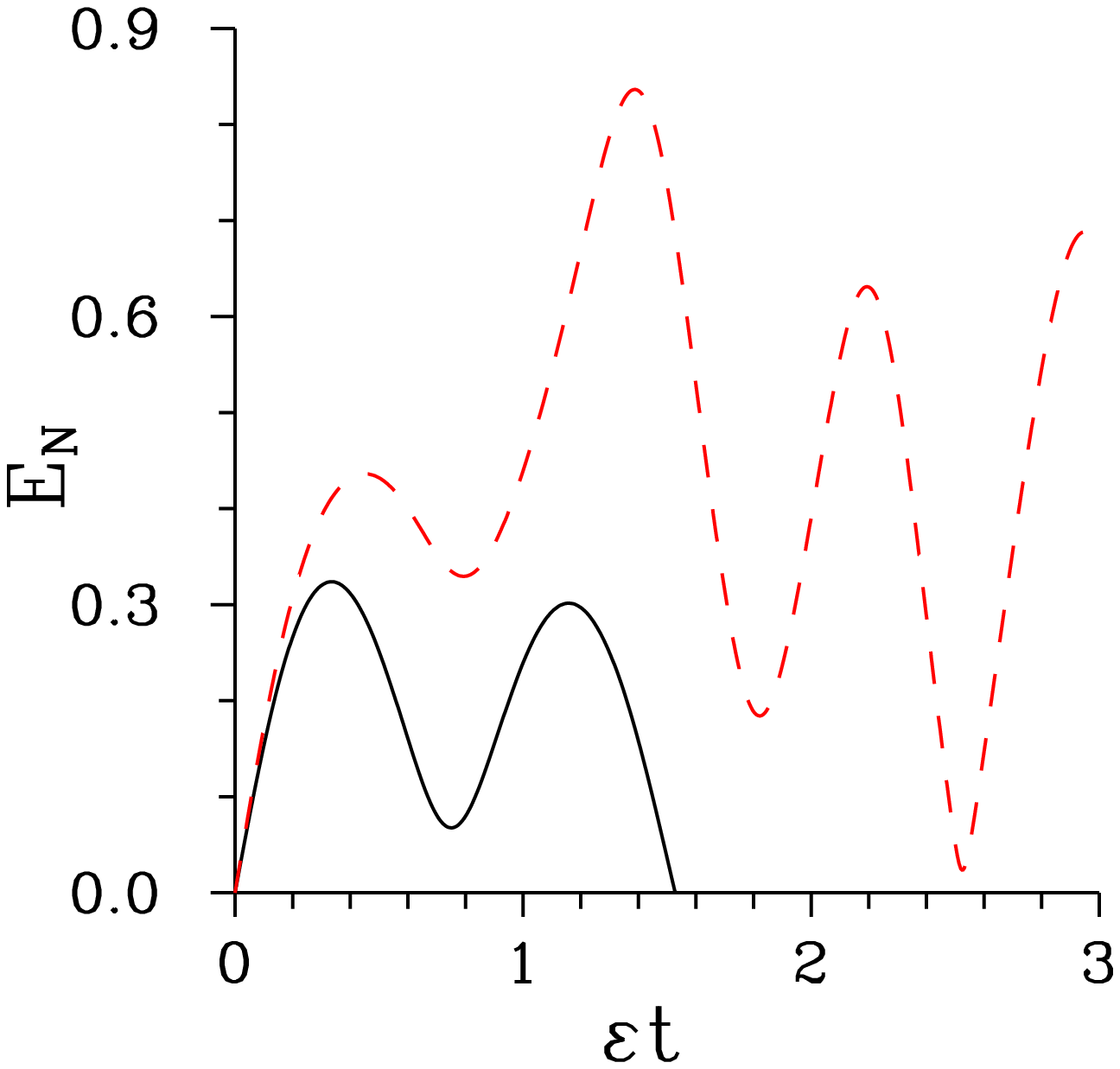}\hspace{2mm}
 \includegraphics[width=0.23\textwidth]{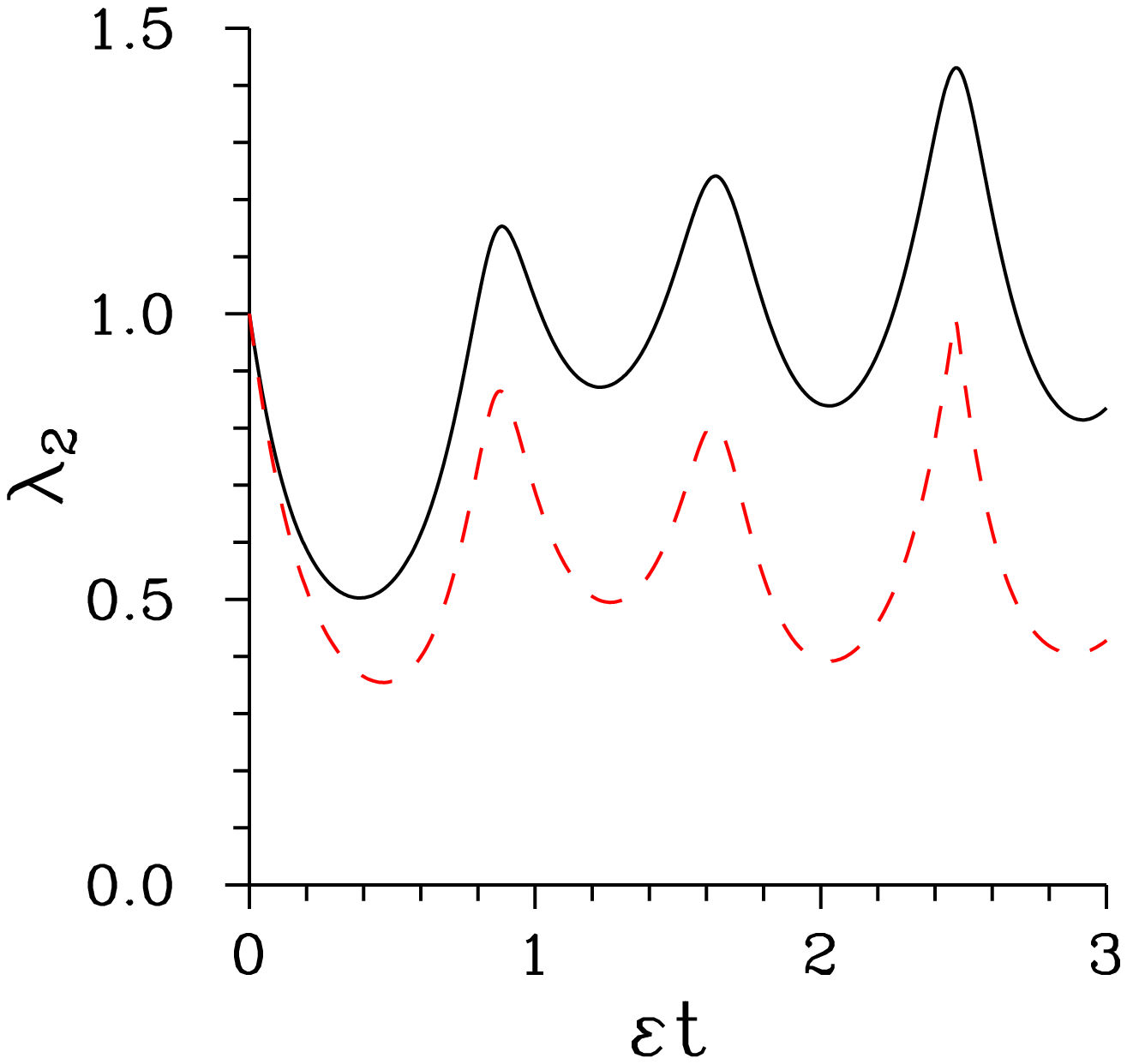}
 \centerline{(a) \hspace{.2\textwidth} (b)}
 \vspace{2mm}
 \includegraphics[width=0.23\textwidth]{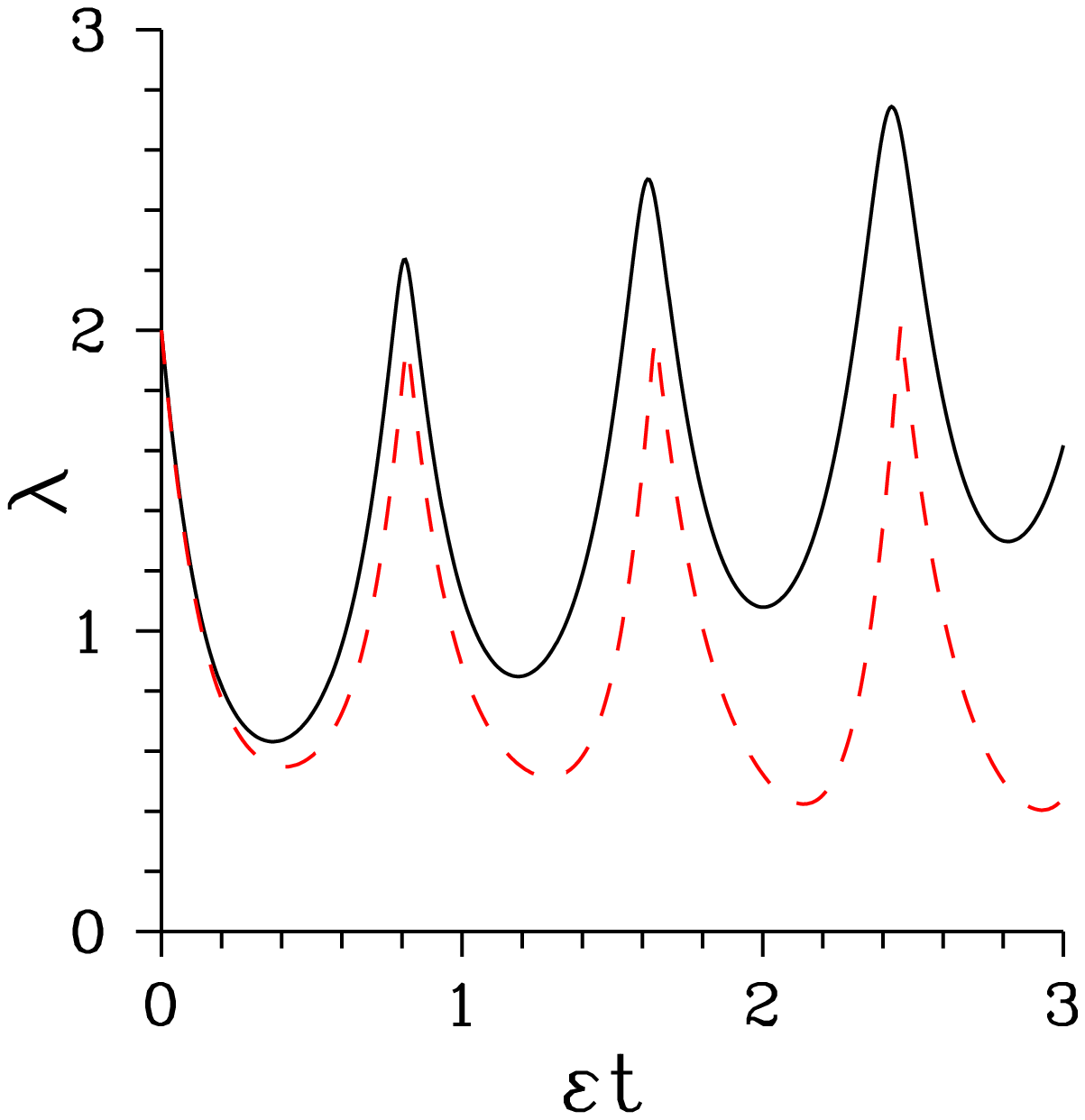}\hspace{2mm}
 {}\includegraphics[width=0.23\textwidth]{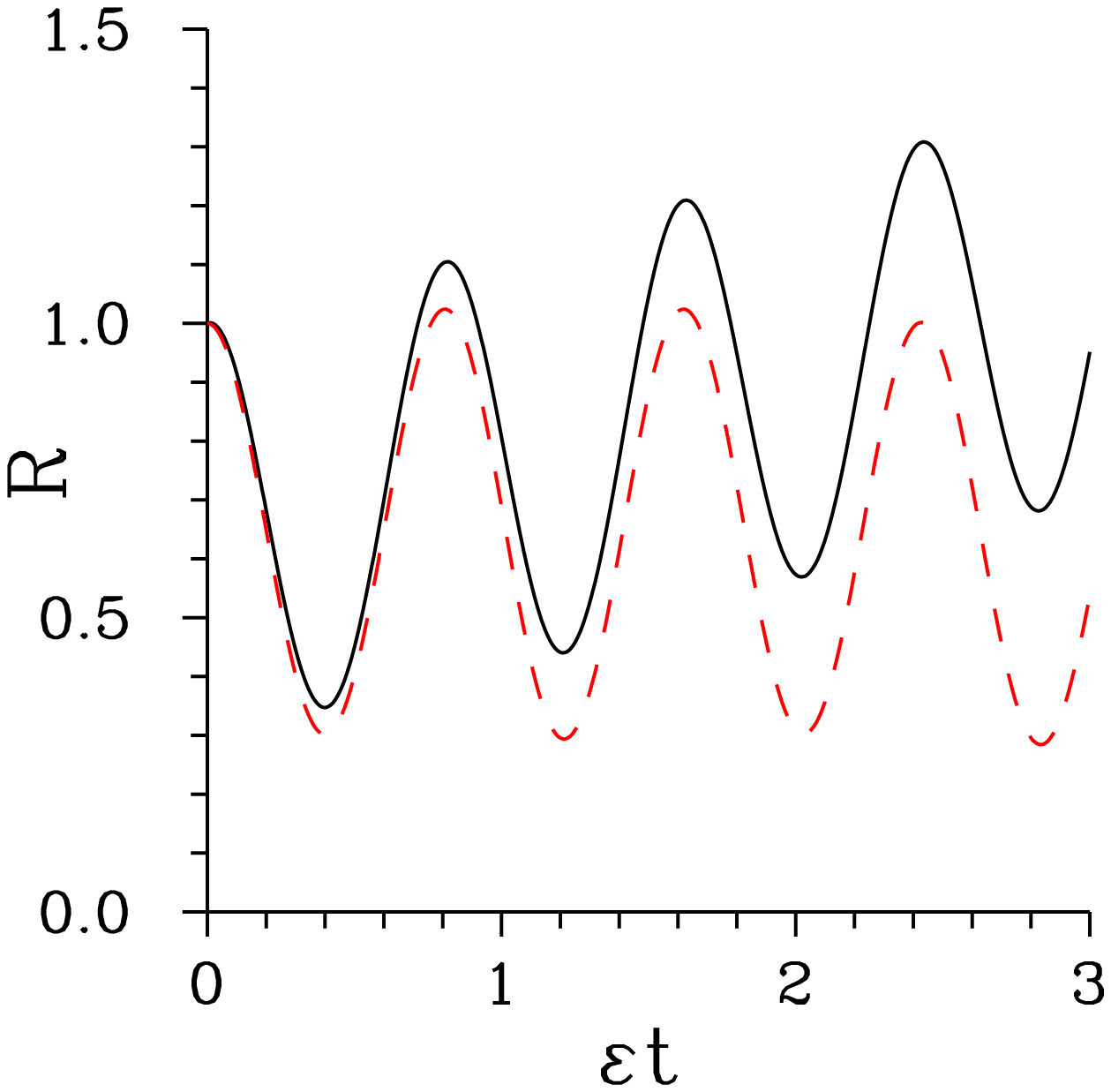}
 \centerline{(c) \hspace{.2\textwidth} (d)}
 \caption{(a) Logarithmic negativity $ E_N $, (b) principal
  squeezing variance $ \lambda_2 $, (c) principal
  squeezing variance $ \lambda $, and (d) sub-shot-noise parameter
  $ R $ as they depend on dimensionless
  time $ \epsilon t $; $ \gamma = 0.1\epsilon $,
  $ \kappa= 0.9\epsilon $, $ \beta_1 =
  \beta_2= 0.05\epsilon $, and $ \beta_c = 0 $. The initial steady
  state in Eq.~(\ref{28}) is assumed, evolution is treated
  with [without] the fluctuating Langevin forces (black solid [\Red{red dashed}] curves).}
\label{fig3}
\end{figure}

Increasing the damping and amplification constants $ \gamma_1 $
and $ \gamma_2 $ in general reduces the system ability to provide
nonclassical states. Greater values of the coupling constant $
\kappa $, that is the source of non-classicality, allow the
observation of highly nonclassical states for greater values of
the damping and amplification constants $ \gamma_1 $ and $
\gamma_2 $. According to Fig.~\ref{fig4}(a), the logarithmic
negativity $ E_N^{\rm max} $ attains its greatest values [the
greatest non-classicality] for $ \kappa/\epsilon $ close to 1 and
$ \gamma $ close to 0 and decreasing (increasing) values of $
\kappa $ ($ \gamma $) systematically reduce its values [reduce the
non-classicality]. From the point of view of non-classicality, the
sub-shot-noise parameter $ R^{\rm min} $ and the two-mode
principal squeezing variance $ \lambda^{\rm min} $ behave
qualitatively similarly as the logarithmic negativity $ E_N^{\rm
max} $. Also the single-mode principal squeezing variances $
\lambda_1^{\rm min} $ and $ \lambda_2^{\rm min} $ reach their
smallest values [the greatest non-classicality] in the area with
greater $ \kappa $ and small $ \gamma $, as shown in
Figs.~\ref{fig4}(b) and \ref{fig4}(c). The comparison of
Figs.~\ref{fig4}(b) and \ref{fig4}(c) confirms our general
observation that the single-mode non-classicality develops more
extensively in the damped mode 1 rather than in the amplified mode
2.
\begin{figure} 
 (a)\includegraphics[width=0.4\textwidth]{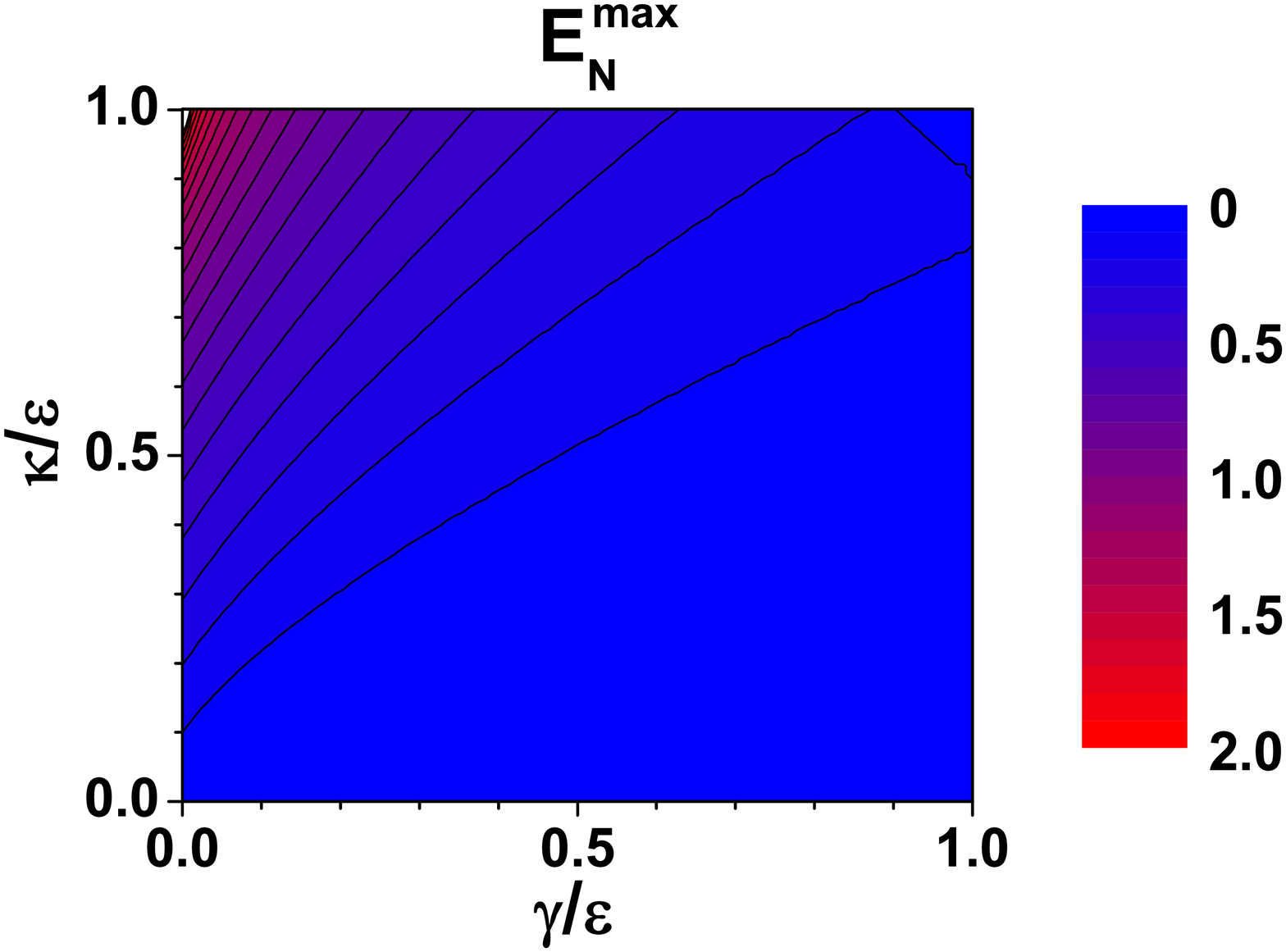}
 \vspace{2mm}
 (b)\includegraphics[width=0.4\textwidth]{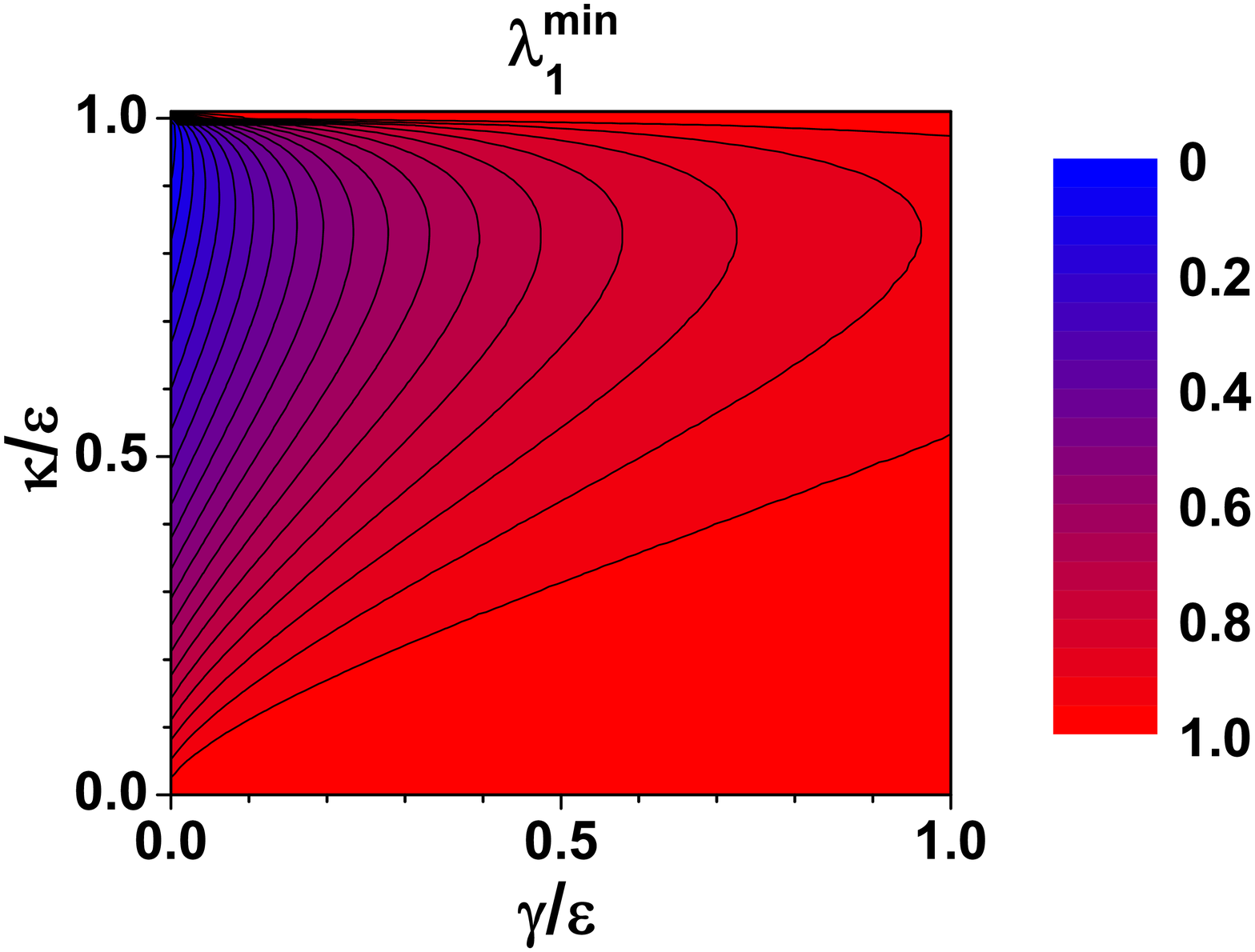}
 \vspace{2mm}
 (c)\includegraphics[width=0.4\textwidth]{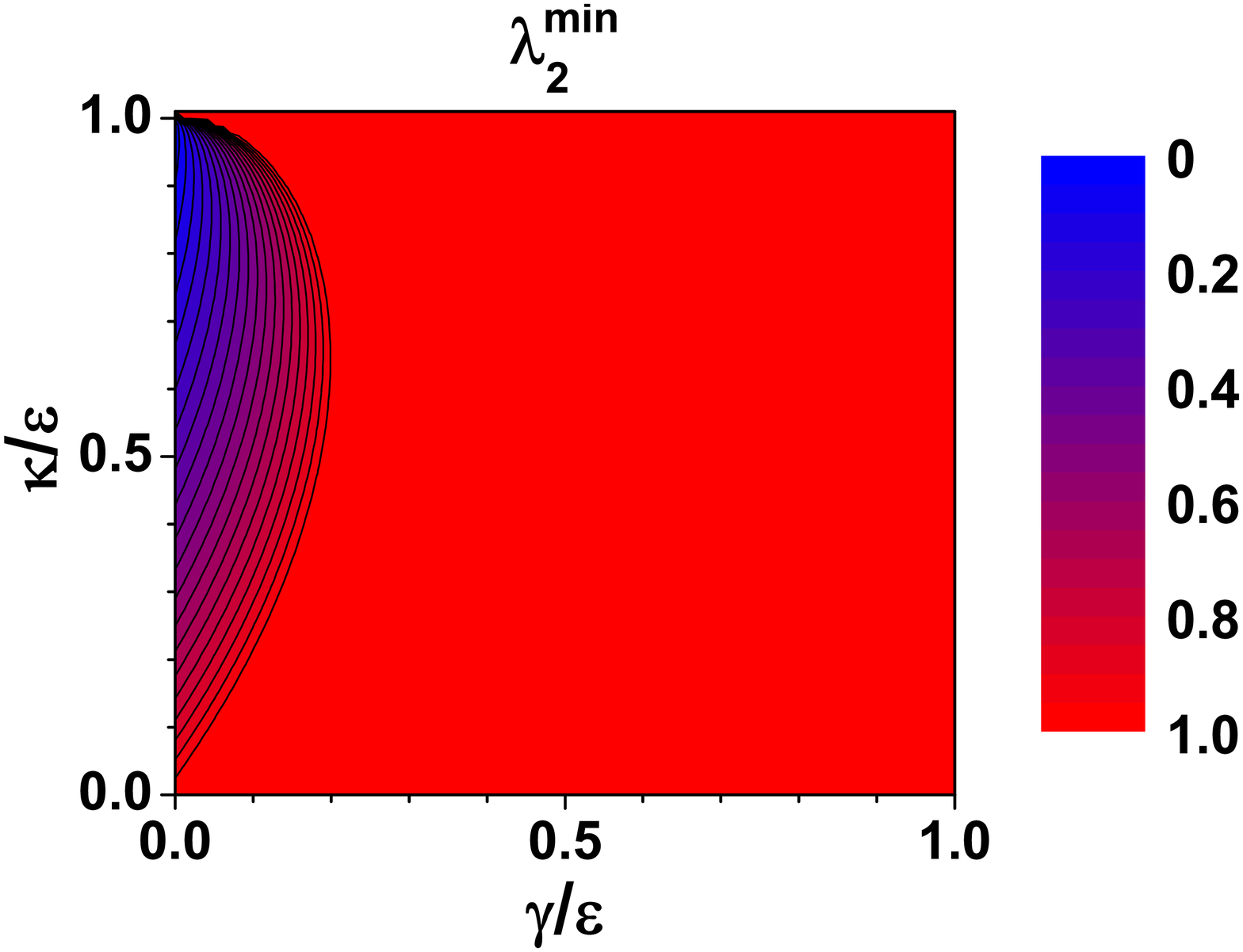}

 \caption{(a) Logarithmic negativity $ E_N $ and principal
  squeezing variances (b) $ \lambda_1 $ and (c) $ \lambda_2 $
  as they depend on dimensionless parameters $
 \gamma/\epsilon $ and $ \kappa/\epsilon $; $ \beta_1=\beta_2 = 0.05\epsilon $
 and $ \beta_c = 0 $. The initial vacuum states are assumed.}
\label{fig4}
\end{figure}

\section{Nonclassical states at exceptional points}

As already discussed in Sec.~III, the most promising conditions
for the nonclassical-light generation are found in EPs in the
parameter space, i.e. if $ \epsilon^2 - \kappa^2 - \gamma^2 = 0 $
[see Eq.~(\ref{6})]. This stems from the fact that the
interactions between modes are enhanced in these
points~\cite{Antonosyan2018}. The ability to generate nonclassical
states decreases with the increasing damping and amplification
constants $ \gamma_1 $ and $ \gamma_2 $, as shown in
Fig.~\ref{fig5}. According to the curves in Fig.~\ref{fig5} drawn
for $ E_N^{\rm max} $, $ \lambda_1^{\rm min} $, $ \lambda^{\rm
min} $, and $ R^{\rm min} $, greater intensities of the incident
coherent states lead to greater values of the logarithmic
negativity $ E_N^{\rm max} $ and smaller values of the principal
squeezing variances $ \lambda_1^{\rm min} $ and $ \lambda^{\rm
min} $. Moreover, they allow for the observation of single- as
well as two-mode squeezing in case of strong damping and
amplification, which is impossible for initial coherent states
close to the vacuum state. The behavior of sub-shot-noise
parameter $ R^{\rm min} $ strongly depends on the relative phase
of the initial coherent states [see Fig.~\ref{fig5}(d)]. If the
initial amplitudes are in-phase, the sub-shot-noise photon-number
correlations gradually vanish with the increasing mode
intensities. On the other hand, if the initial amplitudes are out
of phase, the photon-number correlations become nonclassical ($
R^{\rm min} $) only for greater incident mode intensities.
\begin{figure} 
 \includegraphics[width=0.22\textwidth]{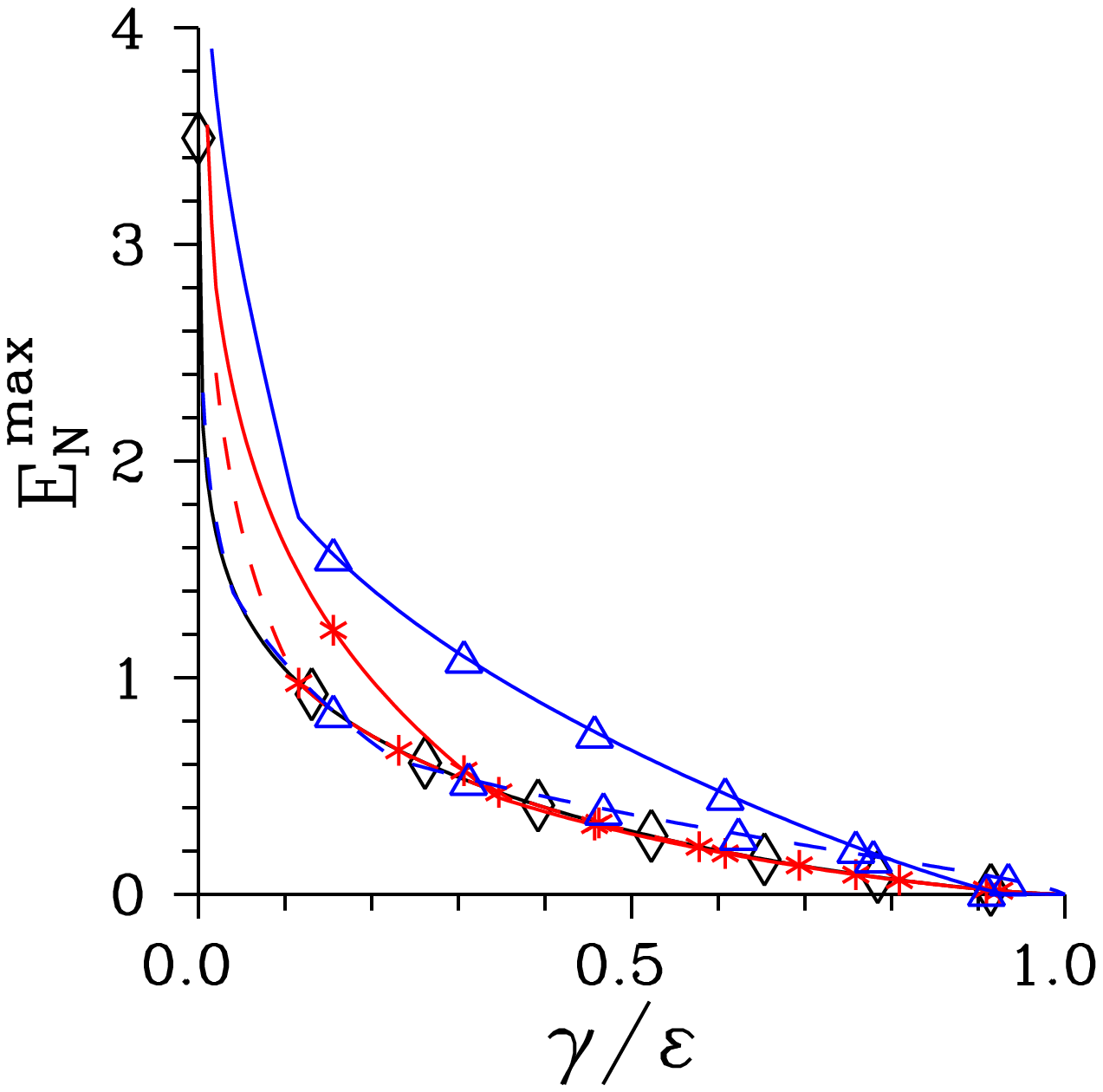}\hspace{2mm}
 \includegraphics[width=0.23\textwidth]{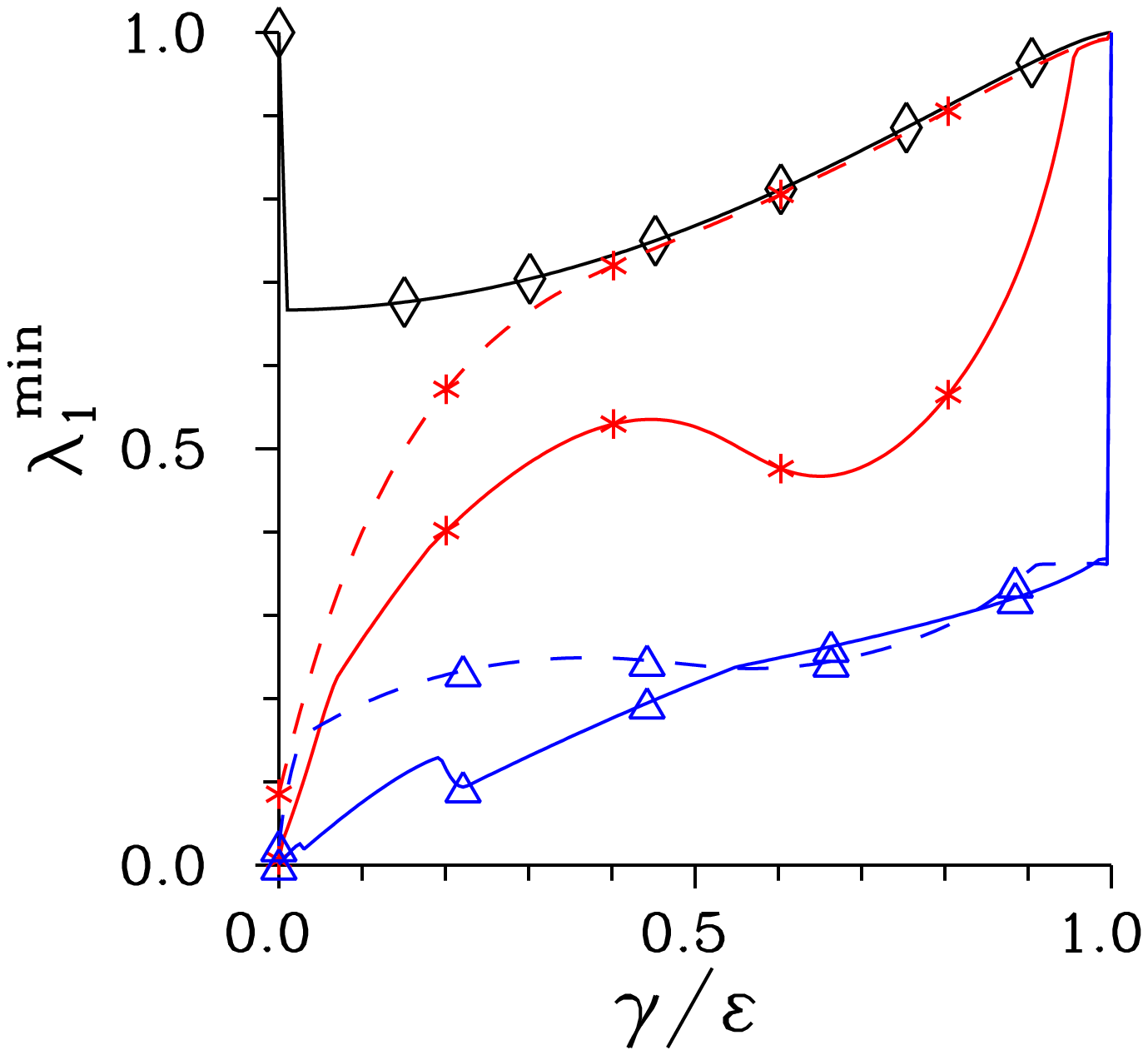}
 \centerline{(a) \hspace{.2\textwidth} (b)}
 \vspace{1mm}

 {\includegraphics[width=0.22\textwidth]{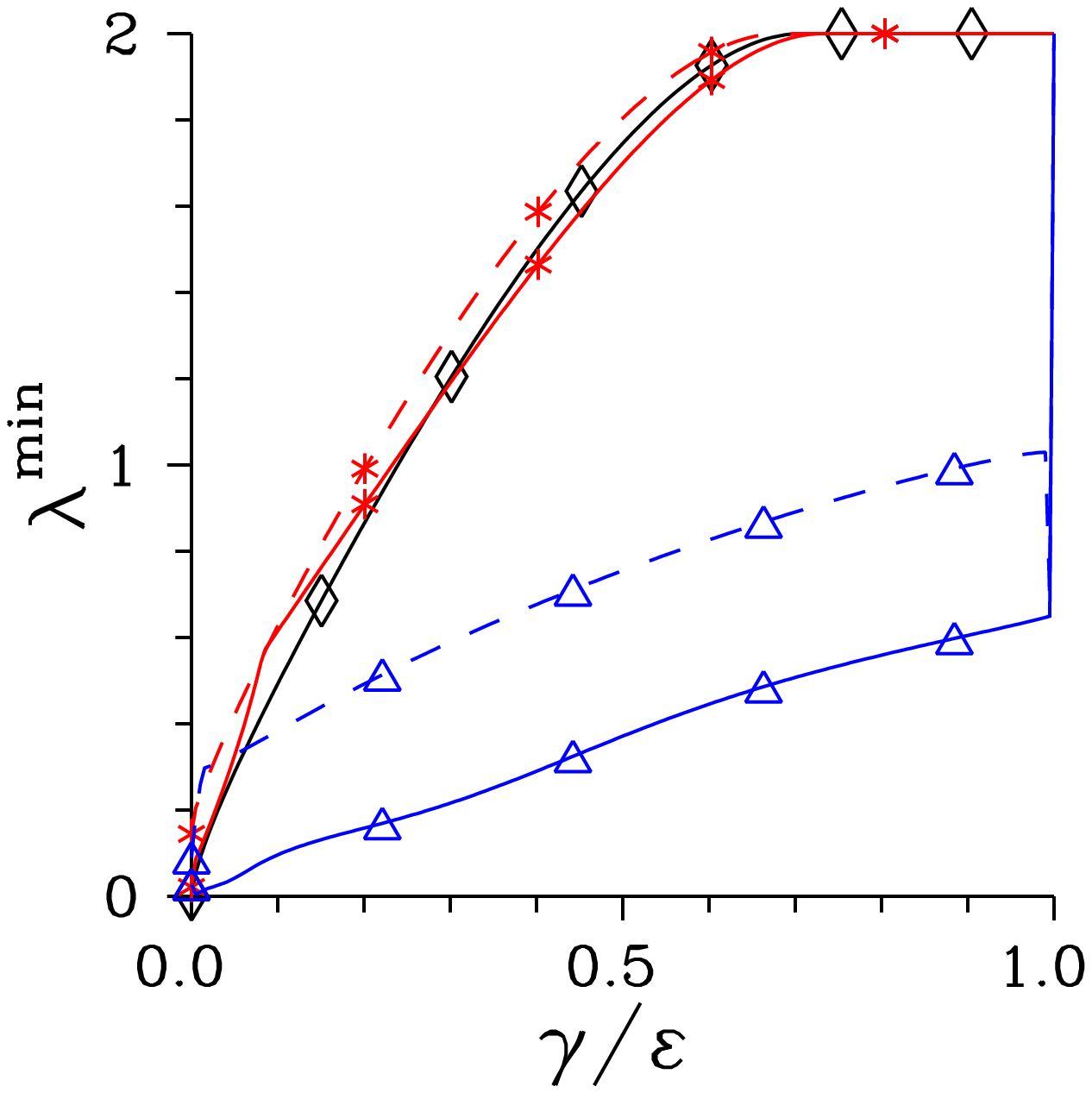}}\hspace{2mm}
 \includegraphics[width=0.23\textwidth]{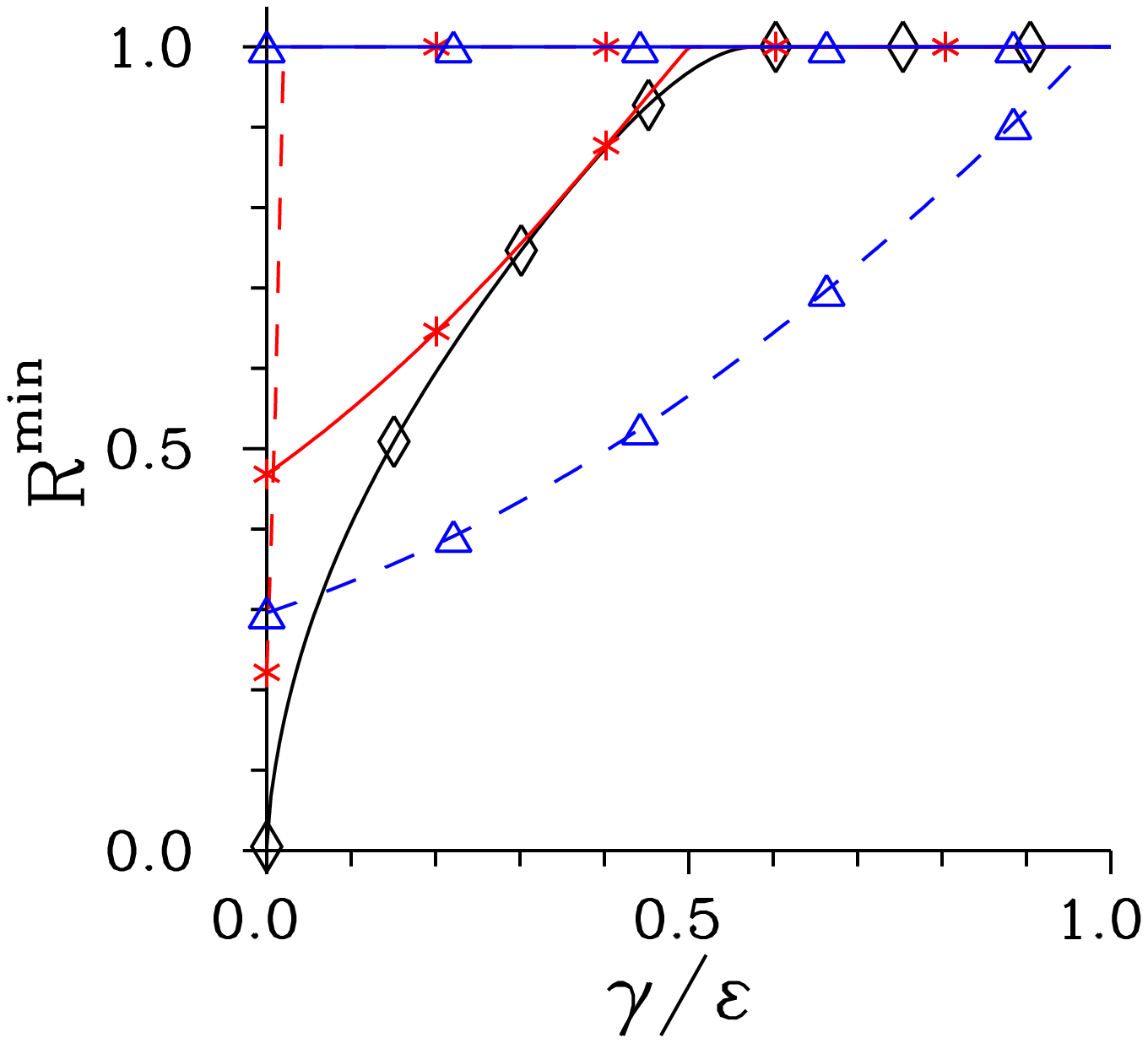}
 \centerline{(c) \hspace{.2\textwidth} (d)}
 \caption{(a) Logarithmic negativity $ E_N^{\rm max} $, (b) principal
  squeezing variance $ \lambda_1^{\rm min} $, (c) principal
  squeezing variance $ \lambda^{\rm min} $, and (d) sub-shot-noise parameter
  $ R^{\rm min} $ as they depend on dimensionless damping and
  amplification constant $ \gamma/\epsilon $ for semiclassical EPs;
  $ \kappa= \sqrt{\epsilon^2-\gamma^2} $, $ \beta_1 =
  \beta_2= 0.05\epsilon $, and $ \beta_c = 0 $. Considered initial states:
  $ \alpha_1^{\rm in} = \alpha_2^{\rm in} = 1\times 10^{-8} $
  (solid curves with $ \diamond $), $ \alpha_1^{\rm in} = \alpha_2^{\rm in} =
  1/\sqrt{2} $ $ [10/\sqrt{2}] $ (solid curves with \Red{$ \ast $} [\Blue{$ \triangle $]}) and
  $ \alpha_1^{\rm in} = -\alpha_2^{\rm in} =
  -i/\sqrt{2} $ $ [-10i/\sqrt{2}] $ (dashed curves with \Red{$ \ast $} [\Blue{$ \triangle $}]).}
\label{fig5}
\end{figure}

Nonclassical properties of the generated states depend strongly on
initial conditions, namely on the overall initial intensity $
I^{\rm in} $ ($ I \equiv |\alpha_1|^2 + |\alpha_2|^2 $), the
initial-intensities unbalance quantified by the angle $
\theta^{\rm in} $ [$ |\alpha_1|^2 = \cos^2(\theta) I $, $
|\alpha_2|^2 = \sin^2(\theta) I $] and the initial relative phase
$ \varphi^{\rm in} $ defined below Eq.~(\ref{22}). We note that
strong dependence of the classical solution of the two-mode Kerr
system on initial conditions was widely discussed
in~\cite{Sukhorukov2010} in relation to $\mathcal{PT}$-symmetry
breaking. As shown in Fig.~\ref{fig6}, the patterns giving the
logarithmic negativity $ E_N^{\rm max} $ in the plane $
(\theta^{\rm in},\varphi^{\rm in}) $ qualitatively depend on the
overall initial intensity $ I^{\rm in} $.
\begin{figure} 
 (a)\includegraphics[width=0.4\textwidth]{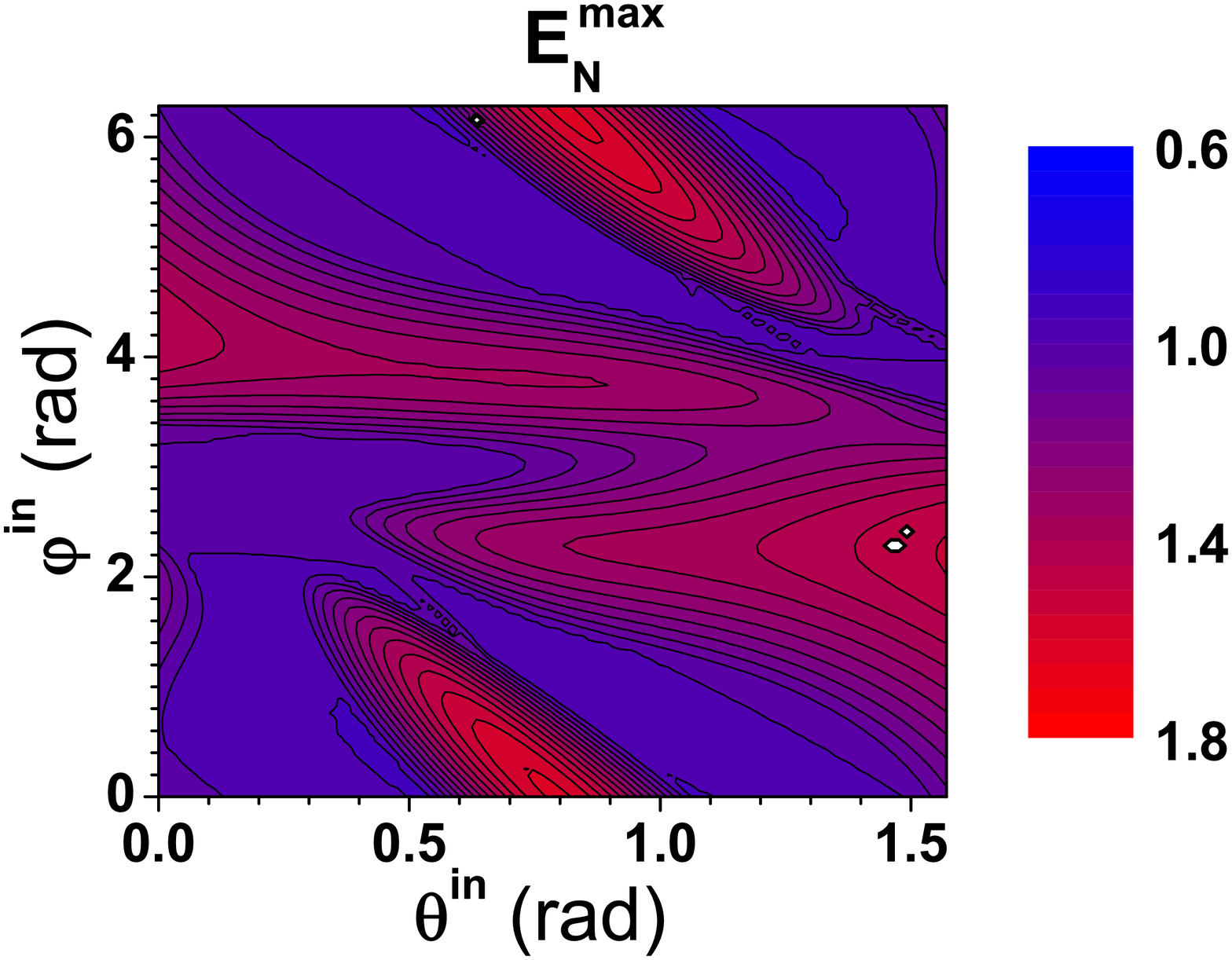}
 \vspace{2mm}
 (b)\includegraphics[width=0.4\textwidth]{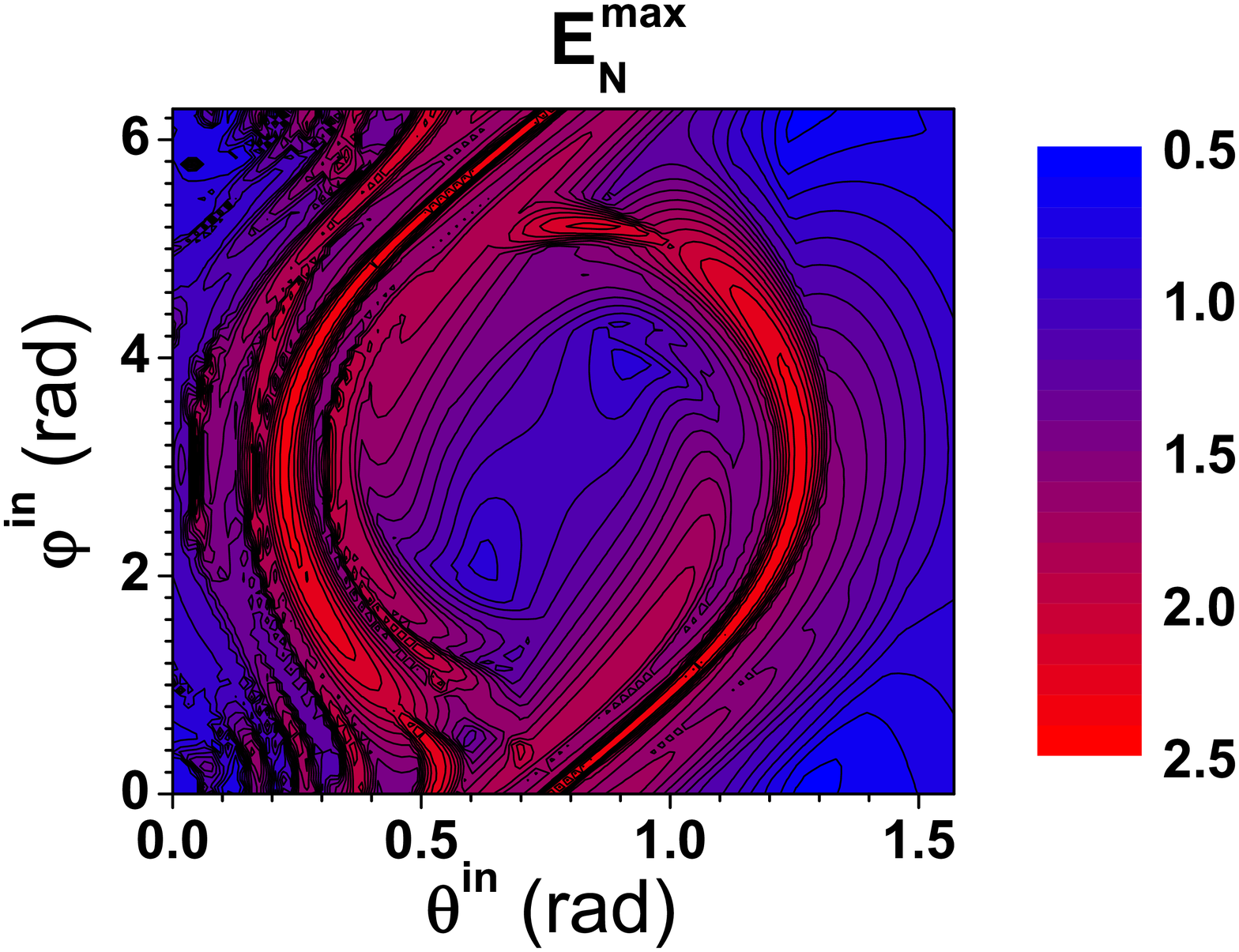}

 \caption{Logarithmic negativity $ E_N^{\rm max} $ for (a) $ I^{\rm in} = 1 $ and
 (b) $ I^{\rm in} = 100 $ as it depends on initial angle $
 \theta^{\rm in} $ giving intensity unbalance and phase difference $ \varphi^{\rm in} $ for
 $ \psi^{\rm in} = 0 $ and the semiclassical EP given as
 $ \gamma/\epsilon = 0.1 $, $ \kappa = \sqrt{\epsilon^2-\gamma^2} $;
 $ \beta_1=\beta_2 = 0.05\epsilon $ and $ \beta_c = 0 $.}
\label{fig6}
\end{figure}
\begin{figure} 
 (a)\includegraphics[width=0.4\textwidth]{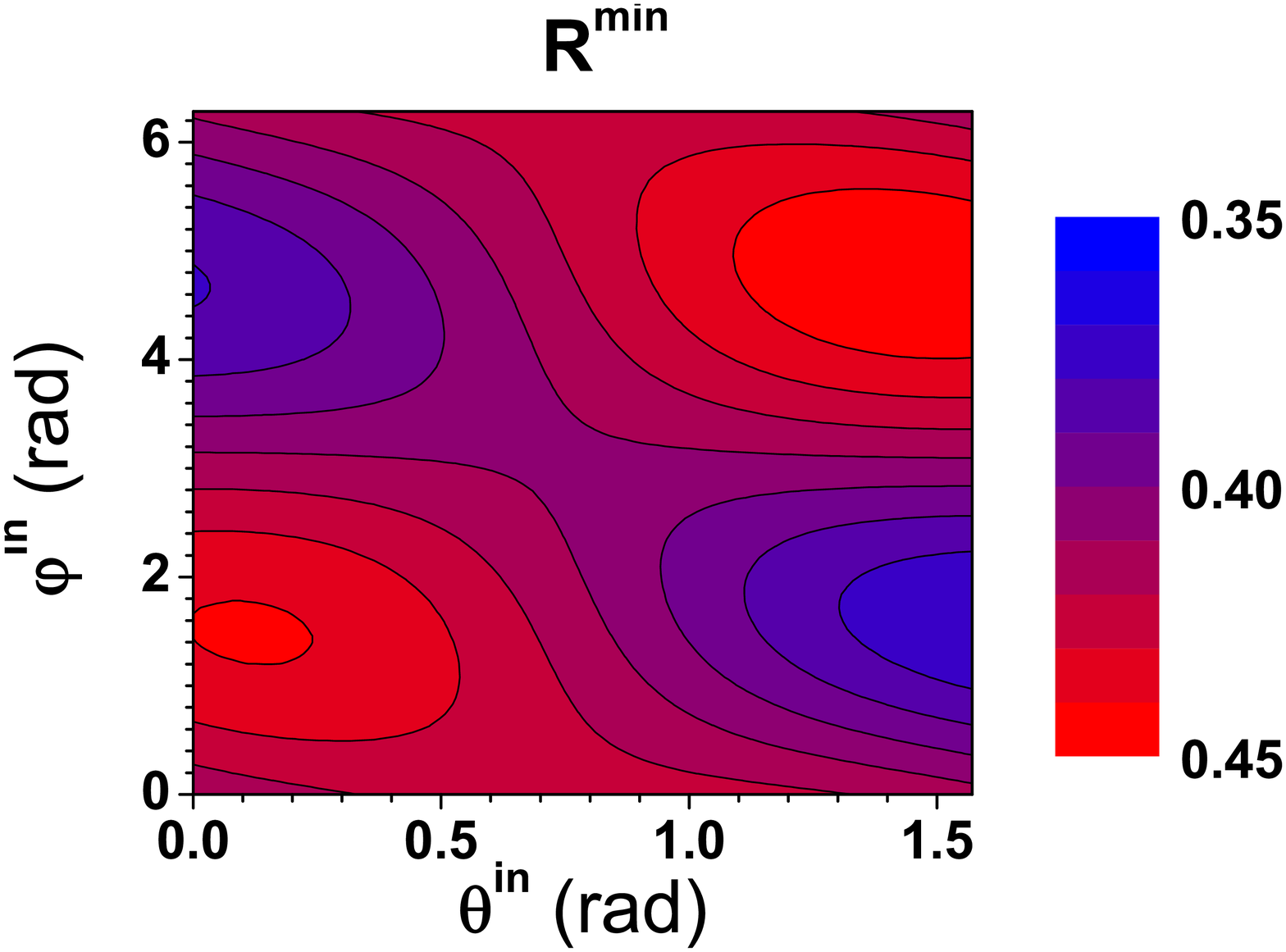}
 \vspace{2mm}
 (b)\includegraphics[width=0.4\textwidth]{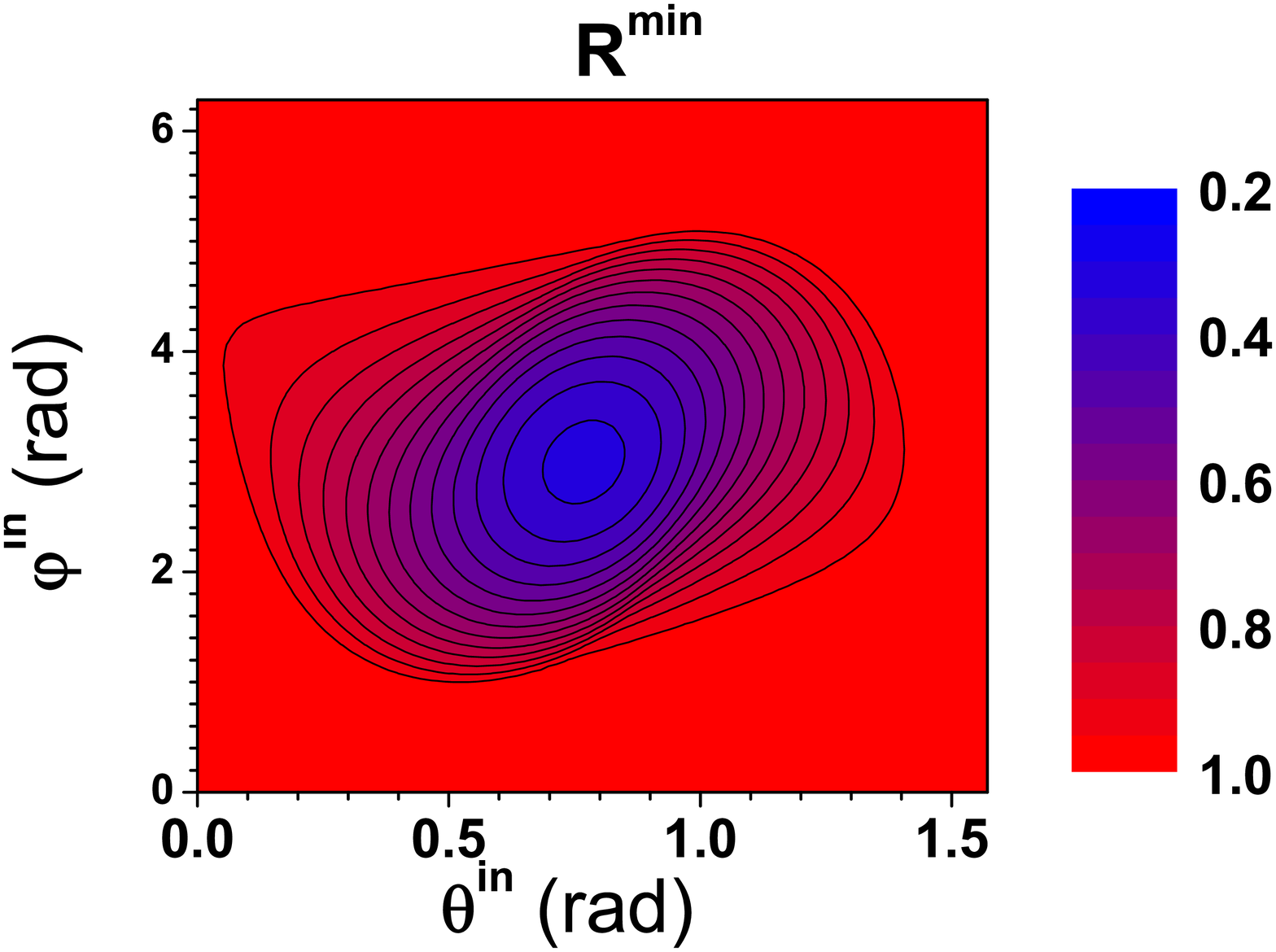}

 \caption{Sub-shot-noise parameter $ R^{\rm min} $ for (a) $ I^{\rm in} = 1 $ and
 (b) $ I^{\rm in} = 100 $ as it depends on initial angle $
 \theta^{\rm in} $ and phase difference $ \varphi^{\rm in} $
 for $ \psi^{\rm in} = 0 $ and the semiclassical EP
 specified in the caption to Fig.~\ref{fig6}.}
\label{fig7}
\end{figure}

Whereas the logarithmic negativity $ E_N^{\rm max} $ depends
smoothly on the initial-intensities unbalance ($ \theta^{\rm in}
$) for small intensities $ I^{\rm in} $, its strong dependence is
observed for greater intensities $ I^{\rm in} $ [compare
Figs.~\ref{fig6}(a) and \ref{fig6}(b)]. The presence of ring-like
structures in Fig.~\ref{fig6}(b) reflects stronger influence of
the nonlinear Kerr terms to the evolution [the terms with $
\beta_1 $, $\beta_2 $ and $ \beta_c $ in Eqs.~(\ref{3})]. These
terms that are weak for small intensities start to play an
important role for greater intensities. They, in parallel to those
depending on $ \kappa $, also serve as a source of
non-classicality of the generated states. Qualitatively different
patterns of sub-shot-noise parameter $ R^{\rm min} $ drawn for
small and large intensities $ I^{\rm in} $ in Figs.~\ref{fig7}(a)
and \ref{fig7}(b) confirm the role of Kerr terms in the state
evolution and the development of its nonclassical properties. The
patterns of the principal squeezing variances $ \lambda_1^{\rm
min}$, $ \lambda_2^{\rm min} $, and $ \lambda^{\rm min} $ are
similarly affected by the initial intensity $ I^{\rm in} $.

\section{Conclusions}

Quantum evolution of a two-mode optical system, described by a quadratic Hamiltonian with additional small nonlinear
Kerr terms and exposed to simultaneous damping and amplification, was discussed in the $\mathcal{PT}$-symmetric
configuration. Two kinds of stationary states were revealed for the classical nonlinear equations corresponding to the
quantum Heisenberg equations of the analyzed system. Whereas the stationary states of the first kind are stable for
damping prevailing the amplification, the stationary states of the second kind are always unstable. The latter states
coincide with the vacuum state in EPs. This allows, in principle, to look for the EPs through seeking for the
stationary states with specific properties. Quantum consistent evolution of fields in the $\mathcal{PT}$-symmetric
model was formulated using linear-operator corrections to the classical solution that was found in general numerically.
In the model, parametric down-conversion with its creation and annihilation of photons in pairs represents the main
source of non-classicality of the emitted fields at lower intensities. These fields can exhibit both two-mode
entanglement and single-mode non-classicalities. Stronger parametric down-conversion together with weak damping and
amplification were shown to be suitable for the generation of highly nonclassical states judged via the values of the
logarithmic negativity, the principal squeezing variances, and the sub-shot-noise parameter. Highly nonclassical states
were especially observed at EPs due to the effective enhancement of nonlinearity. It holds in general that the
evolution gradually conceals the non-classicality of the evolving states for nonzero damping and amplification. Highly
non-classical states are thus obtained for finite time instants. It was shown that the presence of fluctuating Langevin
forces (not the sheer presence of damped and amplified terms in the Heisenberg equations) is the main reason for this
gradual loss of non-classicality. The analyzed $\mathcal{PT}$-symmetric system with parametric down-conversion has been
revealed as the source of highly nonclassical light, especially at EPs with weak damping and amplification. These
nonclassical properties represent a challenge for its practical realization in the form of a nonlinear photonic
structure.

\acknowledgements \noindent{\bf Acknowledgments} J.P.
acknowledges the support by the GA \v{C}R project 18-22102S. 
J.P., J.K.K., and W.L. were supported from ERDF/ESF project
``Nanotechnologies for Future''
(CZ.02.1.01/0.0/0.0/16\_019/0000754). A.L. gratefully acknowledges
the support from the project IGA\_PrF\_2019\_008 of Palack\' y
University. J.K.K. and W.L. acknowledge the financial support from
the program of the Polish Minister of Science and Higher Education
under the name ``Regional Initiative of Excellence'' in 2019-2022,
project no. 003/RID/2018/19, funding amount 11 936 596.10 PLN.

\appendix
\section{Stability analysis of steady states}

To analyze stability of the steady states revealed in Sec.~III and
described in Eqs.~(\ref{28}) and (\ref{29}), we express the used
quantities via their deviations $ \delta\varrho_1 $, $
\delta\varrho_2 $, $ \delta\varphi $, and $ \delta\psi $ from
their steady-state values:
\begin{eqnarray}  
 & \varrho_j = \varrho_j^{\rm st} + \delta\varrho_j, \hspace{5mm}
  j=1,2, & \nonumber \\
 & \varphi = \varphi^{\rm st} + \delta\varphi,
 \hspace{5mm} \psi = \psi^{\rm st} + \delta\psi. &
\label{a1}
\end{eqnarray}

The linearized equations for the deviations are derived as
follows:
\begin{eqnarray}  
 \frac{d}{dt} \left[ \begin{array}{c} \delta\varrho_1 \\
  \delta\varrho_2 \\ \delta\varphi \\ \delta\psi
  \end{array}\right] &=& \left[ \begin{array}{cccc}
   \gamma_1 & s_\epsilon-s_\kappa & c_\epsilon\varrho_2^{\rm st} & - c_\kappa\varrho_2^{\rm st} \\
   -s_\epsilon-s_\kappa & -\gamma_2 & -c_\epsilon\varrho_1^{\rm st} & - c_\kappa\varrho_1^{\rm st} \\
   G^+ & H^+ & I^+ s_\epsilon & I^+ s_\kappa \\
   G^- & H^- & I^- s_\epsilon & I^- s_\kappa \end{array} \right]
   \nonumber \\
  & & \mbox{} \times  \left[ \begin{array}{c} \delta\varrho_1 \\
  \delta\varrho_2 \\ \delta\varphi \\ \delta\psi
  \end{array}\right] .
\label{a2}
\end{eqnarray}
In Eq.~(\ref{a2}), $ s_\epsilon = \epsilon \sin(\varphi^{\rm st})
$, $ s_\kappa = \kappa\sin(\psi^{\rm st}) $, and $ c_\epsilon $
and $ c_\kappa $ are defined in Eqs.~(\ref{28}) and (\ref{29}) for
both steady states. Moreover, the parameters $ G^\pm $, $ H^\pm $,
and $ I^\pm $ introduced in Eq.~(\ref{a2}) are determined as:
\begin{eqnarray}  
 G^\pm &=& -(c_\epsilon+c_\kappa) \left[ 1\pm
  \beta_{12}^2 \right] \frac{1}{\varrho_2^{\rm
  st}} + 2(\pm 2\beta_1-\beta_c)\varrho_1^{\rm st}, \nonumber \\
 H^\pm &=& (c_\epsilon+c_\kappa) \left[ \frac{1}{\beta_{12}^2}\pm 1
  \right] \frac{1}{\varrho_1^{\rm st}} - 2(2\beta_2 \mp\beta_c)\varrho_2^{\rm st}, \nonumber \\
 I^\pm &=& \frac{1}{\beta_{12}} \mp \beta_{12}.
 \label{a3}
\end{eqnarray}

Whereas the dynamical matrix of deviations in Eq.~(\ref{a1}) can
in general be diagonalized only numerically, analytical results
are obtained in the case of $\mathcal{PT}$-symmetry ($
\gamma_1=-\gamma_2\equiv-\gamma $, $ \beta_1=\beta_2\equiv\beta
$):
\begin{eqnarray}  
 & \varrho_1^{\rm st} = \varrho_2^{\rm st} = - (c_\epsilon
  +c_\kappa )/( 2\beta+\beta_c ), & \nonumber \\
 & s_\epsilon = \gamma, \hspace{5mm} s_\kappa = 0, & \nonumber \\
 & |c_\epsilon| = \sqrt{\epsilon^2-\gamma^2}, \hspace{5mm}
 |c_\kappa| = \kappa.
\label{a4}
\end{eqnarray}

Writing the evolution of these deviations in the form $
\exp(-i\tilde\nu_j t) $, the frequencies $ \tilde\nu_j $ are
obtained as follows:
\begin{eqnarray} 
 \tilde\nu_{1,2} &=& \pm 2i\sqrt{-c_\kappa (c_\epsilon+c_\kappa) },
  \nonumber \\
 \tilde\nu_{3,4} &=& \pm 2i\sqrt{-\frac{\beta c_\epsilon (c_\epsilon+c_\kappa)}{\beta+\beta_c/2} },
\label{a5}
\end{eqnarray}
According to Eq.~(\ref{a5}), all $ \tilde\nu_j $ are purely real
for the steady state in Eq.~({\ref{28}) that lies at the border
between the stable and unstable regions. On the other hand, just
one out of four frequencies $ \tilde\nu_j $ has a positive
imaginary part, meaning the instability for the steady state in
Eq.~({\ref{29}). Only if $ c_\epsilon= -c_\kappa $, all four
frequencies $ \tilde\nu_j $ equal zero. This is the case of EPs.


\begin{thebibliography}{100}%
\makeatletter
\providecommand \@ifxundefined [1]{%
 \@ifx{#1\undefined}
}%
\providecommand \@ifnum [1]{%
 \ifnum #1\expandafter \@firstoftwo
 \else \expandafter \@secondoftwo
 \fi
}%
\providecommand \@ifx [1]{%
 \ifx #1\expandafter \@firstoftwo
 \else \expandafter \@secondoftwo
 \fi
}%
\providecommand \natexlab [1]{#1}%
\providecommand \enquote  [1]{``#1''}%
\providecommand \bibnamefont  [1]{#1}%
\providecommand \bibfnamefont [1]{#1}%
\providecommand \citenamefont [1]{#1}%
\providecommand \href@noop [0]{\@secondoftwo}%
\providecommand \href [0]{\begingroup \@sanitize@url \@href}%
\providecommand \@href[1]{\@@startlink{#1}\@@href}%
\providecommand \@@href[1]{\endgroup#1\@@endlink}%
\providecommand \@sanitize@url [0]{\catcode `\\12\catcode `\$12\catcode
  `\&12\catcode `\#12\catcode `\^12\catcode `\_12\catcode `\%12\relax}%
\providecommand \@@startlink[1]{}%
\providecommand \@@endlink[0]{}%
\providecommand \url  [0]{\begingroup\@sanitize@url \@url }%
\providecommand \@url [1]{\endgroup\@href {#1}{\urlprefix }}%
\providecommand \urlprefix  [0]{URL }%
\providecommand \Eprint [0]{\href }%
\providecommand \doibase [0]{http://dx.doi.org/}%
\providecommand \selectlanguage [0]{\@gobble}%
\providecommand \bibinfo  [0]{\@secondoftwo}%
\providecommand \bibfield  [0]{\@secondoftwo}%
\providecommand \translation [1]{[#1]}%
\providecommand \BibitemOpen [0]{}%
\providecommand \bibitemStop [0]{}%
\providecommand \bibitemNoStop [0]{.\EOS\space}%
\providecommand \EOS [0]{\spacefactor3000\relax}%
\providecommand \BibitemShut  [1]{\csname bibitem#1\endcsname}%
\let\auto@bib@innerbib\@empty
\bibitem [{\citenamefont {Bender}\ and\ \citenamefont
  {Boettcher}(1998)}]{Bender1998}%
  \BibitemOpen
  \bibfield  {author} {\bibinfo {author} {\bibfnamefont {C.~M.}\ \bibnamefont
  {Bender}}\ and\ \bibinfo {author} {\bibfnamefont {S.}~\bibnamefont
  {Boettcher}},\ }\bibfield  {title} {\enquote {\bibinfo {title} {Real spectra
  in non-{H}ermitian {H}amiltonians having $\mathcal{PT}$ symmetry},}\ }\href
  {https://doi.org/10.1103/PhysRevLett.80.5243} {\bibfield  {journal} {\bibinfo
   {journal} {Phys. Rev. Lett.}\ }\textbf {\bibinfo {volume} {80}},\ \bibinfo
  {pages} {5243--5246} (\bibinfo {year} {1998})}\BibitemShut {NoStop}%
\bibitem [{\citenamefont {Bender}\ \emph {et~al.}(1999)\citenamefont {Bender},
  \citenamefont {Boettcher},\ and\ \citenamefont {Meisinger}}]{Bender1999}%
  \BibitemOpen
  \bibfield  {author} {\bibinfo {author} {\bibfnamefont {C.~M.}\ \bibnamefont
  {Bender}}, \bibinfo {author} {\bibfnamefont {S.}~\bibnamefont {Boettcher}}, \
  and\ \bibinfo {author} {\bibfnamefont {P.~N.}\ \bibnamefont {Meisinger}},\
  }\bibfield  {title} {\enquote {\bibinfo {title} {{$\mathcal{PT}$-symmetric}
  quantum mechanics},}\ }\href {https://doi.org/10.1063/1.532860} {\bibfield
  {journal} {\bibinfo  {journal} {J. Math. Phys.}\ }\textbf {\bibinfo {volume}
  {40}},\ \bibinfo {pages} {2201--2229} (\bibinfo {year} {1999})}\BibitemShut
  {NoStop}%
\bibitem [{\citenamefont {Bender}\ \emph {et~al.}(2003)\citenamefont {Bender},
  \citenamefont {Brody},\ and\ \citenamefont {Jones}}]{Bender2003}%
  \BibitemOpen
  \bibfield  {author} {\bibinfo {author} {\bibfnamefont {C.~M.}\ \bibnamefont
  {Bender}}, \bibinfo {author} {\bibfnamefont {D.~C.}\ \bibnamefont {Brody}}, \
  and\ \bibinfo {author} {\bibfnamefont {H.~F.}\ \bibnamefont {Jones}},\
  }\bibfield  {title} {\enquote {\bibinfo {title} {Must a {H}amiltonian be
  {H}ermitian?}}\ }\href {https://doi.org/10.1119/1.1574043} {\bibfield
  {journal} {\bibinfo  {journal} {Am. J. Phys.}\ }\textbf {\bibinfo {volume}
  {71}},\ \bibinfo {pages} {1095--1102} (\bibinfo {year} {2003})}\BibitemShut
  {NoStop}%
\bibitem [{\citenamefont {Bender}(2005)}]{Bender2005}%
  \BibitemOpen
  \bibfield  {author} {\bibinfo {author} {\bibfnamefont {C.~M.}\ \bibnamefont
  {Bender}},\ }\bibfield  {title} {\enquote {\bibinfo {title} {Introduction to
  {$\mathcal{PT}$-symmetric} quantum theory},}\ }\href
  {https://doi.org/10.1080/00107500072632} {\bibfield  {journal} {\bibinfo
  {journal} {Contemp. Phys.}\ }\textbf {\bibinfo {volume} {46}},\ \bibinfo
  {pages} {277292} (\bibinfo {year} {2005})}\BibitemShut {NoStop}%
\bibitem [{\citenamefont {El-Ganainy}\ \emph {et~al.}(2007)\citenamefont
  {El-Ganainy}, \citenamefont {Makris}, \citenamefont {Christodoulides},\ and\
  \citenamefont {Musslimani}}]{El-Ganainy2007}%
  \BibitemOpen
  \bibfield  {author} {\bibinfo {author} {\bibfnamefont {R.}~\bibnamefont
  {El-Ganainy}}, \bibinfo {author} {\bibfnamefont {K.~G.}\ \bibnamefont
  {Makris}}, \bibinfo {author} {\bibfnamefont {D.~N.}\ \bibnamefont
  {Christodoulides}}, \ and\ \bibinfo {author} {\bibfnamefont {Ziad~H.}\
  \bibnamefont {Musslimani}},\ }\bibfield  {title} {\enquote {\bibinfo {title}
  {Theory of coupled optical $\mathcal{PT}$-symmetric structures},}\ }\href
  {https://doi.org/10.1364/OL.32.002632} {\bibfield  {journal} {\bibinfo
  {journal} {Opt. Lett.}\ }\textbf {\bibinfo {volume} {32}},\ \bibinfo {pages}
  {2632--2634} (\bibinfo {year} {2007})}\BibitemShut {NoStop}%
\bibitem [{\citenamefont {Ramezani}\ \emph {et~al.}(2010)\citenamefont
  {Ramezani}, \citenamefont {Kottos}, \citenamefont {El-Ganainy},\ and\
  \citenamefont {Christodoulides}}]{Ramezani2010}%
  \BibitemOpen
  \bibfield  {author} {\bibinfo {author} {\bibfnamefont {H.}~\bibnamefont
  {Ramezani}}, \bibinfo {author} {\bibfnamefont {T.}~\bibnamefont {Kottos}},
  \bibinfo {author} {\bibfnamefont {R.}~\bibnamefont {El-Ganainy}}, \ and\
  \bibinfo {author} {\bibfnamefont {D.~N.}\ \bibnamefont {Christodoulides}},\
  }\bibfield  {title} {\enquote {\bibinfo {title} {Unidirectional nonlinear
  $\mathcal{PT}$-symmetric optical structures},}\ }\href
  {https://link.aps.org/doi/10.1103/PhysRevA.82.043803} {\bibfield  {journal}
  {\bibinfo  {journal} {Phys. Rev. A}\ }\textbf {\bibinfo {volume} {82}},\
  \bibinfo {pages} {043803} (\bibinfo {year} {2010})}\BibitemShut {NoStop}%
\bibitem [{\citenamefont {Zyablovsky}\ \emph {et~al.}(2014)\citenamefont
  {Zyablovsky}, \citenamefont {Vinogradov}, \citenamefont {Pukhov},
  \citenamefont {Dorofeenko},\ and\ \citenamefont
  {Lisyansky}}]{Zyablovsky2014}%
  \BibitemOpen
  \bibfield  {author} {\bibinfo {author} {\bibfnamefont {A.~A.}\ \bibnamefont
  {Zyablovsky}}, \bibinfo {author} {\bibfnamefont {A.~P.}\ \bibnamefont
  {Vinogradov}}, \bibinfo {author} {\bibfnamefont {A.~A.}\ \bibnamefont
  {Pukhov}}, \bibinfo {author} {\bibfnamefont {A.~V.}\ \bibnamefont
  {Dorofeenko}}, \ and\ \bibinfo {author} {\bibfnamefont {A.~A.}\ \bibnamefont
  {Lisyansky}},\ }\bibfield  {title} {\enquote {\bibinfo {title}
  {{$\mathcal{PT}$-symmetry} in optics},}\ }\href
  {https://doi.org/10.3367/ufne.0184.201411b.1177} {\bibfield  {journal}
  {\bibinfo  {journal} {Physics-Uspekhi}\ }\textbf {\bibinfo {volume} {57}},\
  \bibinfo {pages} {1063---1082} (\bibinfo {year} {2014})}\BibitemShut
  {NoStop}%
\bibitem [{\citenamefont {\"{O}gren}\ \emph {et~al.}(2017)\citenamefont
  {\"{O}gren}, \citenamefont {Abdullaev},\ and\ \citenamefont
  {Konotop}}]{Ogren2017}%
  \BibitemOpen
  \bibfield  {author} {\bibinfo {author} {\bibfnamefont {M.}~\bibnamefont
  {\"{O}gren}}, \bibinfo {author} {\bibfnamefont {F.~K.}\ \bibnamefont
  {Abdullaev}}, \ and\ \bibinfo {author} {\bibfnamefont {V.~V.}\ \bibnamefont
  {Konotop}},\ }\bibfield  {title} {\enquote {\bibinfo {title} {Solitons in a
  $\mathcal{PT}$-symmetric $\chi^{(2)}$ coupler},}\ }\href
  {https://doi.org/10.1364/OL.42.004079} {\bibfield  {journal} {\bibinfo
  {journal} {Opt. Lett.}\ }\textbf {\bibinfo {volume} {42}},\ \bibinfo {pages}
  {4079--4082} (\bibinfo {year} {2017})}\BibitemShut {NoStop}%
\bibitem [{\citenamefont {Turitsyna}\ \emph {et~al.}(2017)\citenamefont
  {Turitsyna}, \citenamefont {Shadrivov},\ and\ \citenamefont
  {Kivshar}}]{Turitsyna2017}%
  \BibitemOpen
  \bibfield  {author} {\bibinfo {author} {\bibfnamefont {E.~G.}\ \bibnamefont
  {Turitsyna}}, \bibinfo {author} {\bibfnamefont {I.~V.}\ \bibnamefont
  {Shadrivov}}, \ and\ \bibinfo {author} {\bibfnamefont {Y.~S.}\ \bibnamefont
  {Kivshar}},\ }\bibfield  {title} {\enquote {\bibinfo {title} {{Guided modes
  in non-{Hermitian} optical waveguides}},}\ }\href
  {https://doi.org/10.1103/PhysRevA.96.033824} {\bibfield  {journal} {\bibinfo
  {journal} {Phys. Rev. A}\ }\textbf {\bibinfo {volume} {96}},\ \bibinfo
  {pages} {033824} (\bibinfo {year} {2017})}\BibitemShut {NoStop}%
\bibitem [{\citenamefont {Xu}\ \emph {et~al.}(2018)\citenamefont {Xu},
  \citenamefont {Shi}, \citenamefont {Ren},\ and\ \citenamefont
  {Zhang}}]{Xu2018}%
  \BibitemOpen
  \bibfield  {author} {\bibinfo {author} {\bibfnamefont {X.}~\bibnamefont
  {Xu}}, \bibinfo {author} {\bibfnamefont {L.}~\bibnamefont {Shi}}, \bibinfo
  {author} {\bibfnamefont {L.}~\bibnamefont {Ren}}, \ and\ \bibinfo {author}
  {\bibfnamefont {X.}~\bibnamefont {Zhang}},\ }\bibfield  {title} {\enquote
  {\bibinfo {title} {Optical gradient forces in $\mathcal{PT}$-symmetric
  coupled-waveguide structures},}\ }\href
  {https://doi.org/10.1364/OE.26.010220} {\bibfield  {journal} {\bibinfo
  {journal} {Opt. Express}\ }\textbf {\bibinfo {volume} {26}},\ \bibinfo
  {pages} {10220--10229} (\bibinfo {year} {2018})}\BibitemShut {NoStop}%
\bibitem [{\citenamefont {Graefe}\ and\ \citenamefont
  {Jones}(2011)}]{Graefe2011}%
  \BibitemOpen
  \bibfield  {author} {\bibinfo {author} {\bibfnamefont {E.-M.}\ \bibnamefont
  {Graefe}}\ and\ \bibinfo {author} {\bibfnamefont {H.~F.}\ \bibnamefont
  {Jones}},\ }\bibfield  {title} {\enquote {\bibinfo {title}
  {$\mathcal{PT}$-symmetric sinusoidal optical lattices at the
  symmetry-breaking threshold},}\ }\href
  {https://doi.org/10.1103/PhysRevA.84.013818} {\bibfield  {journal} {\bibinfo
  {journal} {Phys. Rev. A}\ }\textbf {\bibinfo {volume} {84}},\ \bibinfo
  {pages} {013818} (\bibinfo {year} {2011})}\BibitemShut {NoStop}%
\bibitem [{\citenamefont {Miri}\ \emph {et~al.}(2012)\citenamefont {Miri},
  \citenamefont {Regensburger}, \citenamefont {Peschel},\ and\ \citenamefont
  {Christodoulides}}]{Miri2012}%
  \BibitemOpen
  \bibfield  {author} {\bibinfo {author} {\bibfnamefont {M.-A.}\ \bibnamefont
  {Miri}}, \bibinfo {author} {\bibfnamefont {A.}~\bibnamefont {Regensburger}},
  \bibinfo {author} {\bibfnamefont {U.}~\bibnamefont {Peschel}}, \ and\
  \bibinfo {author} {\bibfnamefont {D.~N.}\ \bibnamefont {Christodoulides}},\
  }\bibfield  {title} {\enquote {\bibinfo {title} {Optical mesh lattices with
  $\mathcal{PT}$ symmetry},}\ }\href
  {https://doi.org/10.1103/PhysRevA.86.023807} {\bibfield  {journal} {\bibinfo
  {journal} {Phys. Rev. A}\ }\textbf {\bibinfo {volume} {86}},\ \bibinfo
  {pages} {023807} (\bibinfo {year} {2012})}\BibitemShut {NoStop}%
\bibitem [{\citenamefont {Ornigotti}\ and\ \citenamefont
  {Szameit}(2014)}]{Ornigotti2014}%
  \BibitemOpen
  \bibfield  {author} {\bibinfo {author} {\bibfnamefont {M.}~\bibnamefont
  {Ornigotti}}\ and\ \bibinfo {author} {\bibfnamefont {A.}~\bibnamefont
  {Szameit}},\ }\bibfield  {title} {\enquote {\bibinfo {title} {Quasi
  $\mathcal{PT}$-symmetry in passive photonic lattices},}\ }\href
  {https://doi.org/10.1088/2040-8978/16/6/065501} {\bibfield  {journal}
  {\bibinfo  {journal} {J. Opt.}\ }\textbf {\bibinfo {volume} {16}},\ \bibinfo
  {pages} {065501} (\bibinfo {year} {2014})}\BibitemShut {NoStop}%
\bibitem [{\citenamefont {Shui}\ \emph {et~al.}(2019)\citenamefont {Shui},
  \citenamefont {Yang}, \citenamefont {Li},\ and\ \citenamefont
  {Wang}}]{Shui2019}%
  \BibitemOpen
  \bibfield  {author} {\bibinfo {author} {\bibfnamefont {T.}~\bibnamefont
  {Shui}}, \bibinfo {author} {\bibfnamefont {W.-X.}\ \bibnamefont {Yang}},
  \bibinfo {author} {\bibfnamefont {L.}~\bibnamefont {Li}}, \ and\ \bibinfo
  {author} {\bibfnamefont {X.}~\bibnamefont {Wang}},\ }\bibfield  {title}
  {\enquote {\bibinfo {title} {Lop-sided {Raman-Nath} diffraction in
  $\mathcal{PT}$-antisymmetric atomic lattices},}\ }\href
  {https://doi.org/10.1364/OL.44.002089} {\bibfield  {journal} {\bibinfo
  {journal} {Opt. Lett.}\ }\textbf {\bibinfo {volume} {44}},\ \bibinfo {pages}
  {2089--2092} (\bibinfo {year} {2019})}\BibitemShut {NoStop}%
\bibitem [{\citenamefont {Peng}\ \emph
  {et~al.}(2014{\natexlab{a}})\citenamefont {Peng}, \citenamefont {\c{S}.
  K.~{\"O}zdemir}, \citenamefont {Lei}, \citenamefont {Monifi}, \citenamefont
  {Gianfreda}, \citenamefont {Long}, \citenamefont {Fan}, \citenamefont {Nori},
  \citenamefont {Bender},\ and\ \citenamefont {Yang}}]{Peng2014}%
  \BibitemOpen
  \bibfield  {author} {\bibinfo {author} {\bibfnamefont {B.}~\bibnamefont
  {Peng}}, \bibinfo {author} {\bibnamefont {\c{S}. K.~{\"O}zdemir}}, \bibinfo
  {author} {\bibfnamefont {F.}~\bibnamefont {Lei}}, \bibinfo {author}
  {\bibfnamefont {F.}~\bibnamefont {Monifi}}, \bibinfo {author} {\bibfnamefont
  {M.}~\bibnamefont {Gianfreda}}, \bibinfo {author} {\bibfnamefont {G.~L.}\
  \bibnamefont {Long}}, \bibinfo {author} {\bibfnamefont {S.}~\bibnamefont
  {Fan}}, \bibinfo {author} {\bibfnamefont {F.}~\bibnamefont {Nori}}, \bibinfo
  {author} {\bibfnamefont {C.}~\bibnamefont {Bender}}, \ and\ \bibinfo {author}
  {\bibfnamefont {L.}~\bibnamefont {Yang}},\ }\bibfield  {title} {\enquote
  {\bibinfo {title} {Parity{\textendash}time-symmetric whispering-gallery
  microcavities},}\ }\href {https://doi.org/10.1038/nphys2927} {\bibfield
  {journal} {\bibinfo  {journal} {Nat. Phys.}\ }\textbf {\bibinfo {volume}
  {10}},\ \bibinfo {pages} {394--398} (\bibinfo {year}
  {2014}{\natexlab{a}})}\BibitemShut {NoStop}%
\bibitem [{\citenamefont {Peng}\ \emph
  {et~al.}(2014{\natexlab{b}})\citenamefont {Peng}, \citenamefont
  {{\"O}zdemir}, \citenamefont {Rotter}, \citenamefont {Yilmaz}, \citenamefont
  {Liertzer}, \citenamefont {Monifi}, \citenamefont {Bender}, \citenamefont
  {Nori},\ and\ \citenamefont {Yang}}]{Peng2014a}%
  \BibitemOpen
  \bibfield  {author} {\bibinfo {author} {\bibfnamefont {B.}~\bibnamefont
  {Peng}}, \bibinfo {author} {\bibfnamefont {{\c S}.~K.}\ \bibnamefont
  {{\"O}zdemir}}, \bibinfo {author} {\bibfnamefont {S.}~\bibnamefont {Rotter}},
  \bibinfo {author} {\bibfnamefont {H.}~\bibnamefont {Yilmaz}}, \bibinfo
  {author} {\bibfnamefont {M.}~\bibnamefont {Liertzer}}, \bibinfo {author}
  {\bibfnamefont {F.}~\bibnamefont {Monifi}}, \bibinfo {author} {\bibfnamefont
  {C.~M.}\ \bibnamefont {Bender}}, \bibinfo {author} {\bibfnamefont
  {F.}~\bibnamefont {Nori}}, \ and\ \bibinfo {author} {\bibfnamefont
  {L.}~\bibnamefont {Yang}},\ }\bibfield  {title} {\enquote {\bibinfo {title}
  {Loss-induced suppression and revival of lasing},}\ }\href
  {https://doi.org/10.1126/science.1258004} {\bibfield  {journal} {\bibinfo
  {journal} {Science}\ }\textbf {\bibinfo {volume} {346}},\ \bibinfo {pages}
  {328--332} (\bibinfo {year} {2014}{\natexlab{b}})}\BibitemShut {NoStop}%
\bibitem [{\citenamefont {Liu}\ \emph {et~al.}(2016)\citenamefont {Liu},
  \citenamefont {Zhang}, \citenamefont {{\"O}zdemir}, \citenamefont {Peng},
  \citenamefont {Jing}, \citenamefont {L\"u}, \citenamefont {Li}, \citenamefont
  {Yang}, \citenamefont {Nori},\ and\ \citenamefont {Liu}}]{Liu2016}%
  \BibitemOpen
  \bibfield  {author} {\bibinfo {author} {\bibfnamefont {Z.-P.}\ \bibnamefont
  {Liu}}, \bibinfo {author} {\bibfnamefont {J.}~\bibnamefont {Zhang}}, \bibinfo
  {author} {\bibfnamefont {S.~K.}\ \bibnamefont {{\"O}zdemir}}, \bibinfo
  {author} {\bibfnamefont {B.}~\bibnamefont {Peng}}, \bibinfo {author}
  {\bibfnamefont {H.}~\bibnamefont {Jing}}, \bibinfo {author} {\bibfnamefont
  {X.-Y.}\ \bibnamefont {L\"u}}, \bibinfo {author} {\bibfnamefont {C.-W.}\
  \bibnamefont {Li}}, \bibinfo {author} {\bibfnamefont {L.}~\bibnamefont
  {Yang}}, \bibinfo {author} {\bibfnamefont {F.}~\bibnamefont {Nori}}, \ and\
  \bibinfo {author} {\bibfnamefont {Y.-X.}\ \bibnamefont {Liu}},\ }\bibfield
  {title} {\enquote {\bibinfo {title} {Metrology with $\mathcal{PT}$-symmetric
  cavities: Enhanced sensitivity near the $\mathcal{PT}$-phase transition},}\
  }\href {https://doi.org/10.1103/PhysRevLett.117.110802} {\bibfield  {journal}
  {\bibinfo  {journal} {Phys. Rev. Lett.}\ }\textbf {\bibinfo {volume} {117}},\
  \bibinfo {pages} {110802} (\bibinfo {year} {2016})}\BibitemShut {NoStop}%
\bibitem [{\citenamefont {Zhou}\ and\ \citenamefont {Chong}(2016)}]{Zhou2016}%
  \BibitemOpen
  \bibfield  {author} {\bibinfo {author} {\bibfnamefont {X.}~\bibnamefont
  {Zhou}}\ and\ \bibinfo {author} {\bibfnamefont {Y.~D.}\ \bibnamefont
  {Chong}},\ }\bibfield  {title} {\enquote {\bibinfo {title} {$\mathcal{PT}$
  symmetry breaking and nonlinear optical isolation in coupled
  microcavities},}\ }\href {https://doi.org/10.1364/OE.24.006916} {\bibfield
  {journal} {\bibinfo  {journal} {Opt. Express}\ }\textbf {\bibinfo {volume}
  {24}},\ \bibinfo {pages} {6916--6930} (\bibinfo {year} {2016})}\BibitemShut
  {NoStop}%
\bibitem [{\citenamefont {Arkhipov}\ \emph {et~al.}(2019)\citenamefont
  {Arkhipov}, \citenamefont {Miranowicz}, \citenamefont {{Di~Stefano}},
  \citenamefont {Stassi}, \citenamefont {Savasta}, \citenamefont {Nori},\ and\
  \citenamefont {\"Ozdemir}}]{Arkhipov2019}%
  \BibitemOpen
  \bibfield  {author} {\bibinfo {author} {\bibfnamefont {I.~I.}\ \bibnamefont
  {Arkhipov}}, \bibinfo {author} {\bibfnamefont {A.}~\bibnamefont
  {Miranowicz}}, \bibinfo {author} {\bibfnamefont {O.}~\bibnamefont
  {{Di~Stefano}}}, \bibinfo {author} {\bibfnamefont {R.}~\bibnamefont
  {Stassi}}, \bibinfo {author} {\bibfnamefont {S.}~\bibnamefont {Savasta}},
  \bibinfo {author} {\bibfnamefont {F.}~\bibnamefont {Nori}}, \ and\ \bibinfo
  {author} {\bibfnamefont {S.~K.}\ \bibnamefont {\"Ozdemir}},\ }\bibfield
  {title} {\enquote {\bibinfo {title} {{Scully-Lamb} quantum laser model for
  parity-time-symmetric whispering-gallery microcavities: Gain saturation
  effects and nonreciprocity},}\ }\href
  {https://doi.org/10.1103/PhysRevA.99.053806} {\bibfield  {journal} {\bibinfo
  {journal} {Phys. Rev. A}\ }\textbf {\bibinfo {volume} {99}},\ \bibinfo
  {pages} {053806} (\bibinfo {year} {2019})}\BibitemShut {NoStop}%
\bibitem [{\citenamefont {Quijandr\'ia}\ \emph {et~al.}(2018)\citenamefont
  {Quijandr\'ia}, \citenamefont {Naether}, \citenamefont {{\"Ozdemir}},
  \citenamefont {Nori},\ and\ \citenamefont {Zueco}}]{Quijandria2018}%
  \BibitemOpen
  \bibfield  {author} {\bibinfo {author} {\bibfnamefont {F.}~\bibnamefont
  {Quijandr\'ia}}, \bibinfo {author} {\bibfnamefont {U.}~\bibnamefont
  {Naether}}, \bibinfo {author} {\bibfnamefont {S.~K.}\ \bibnamefont
  {{\"Ozdemir}}}, \bibinfo {author} {\bibfnamefont {F.}~\bibnamefont {Nori}}, \
  and\ \bibinfo {author} {\bibfnamefont {D.}~\bibnamefont {Zueco}},\ }\bibfield
   {title} {\enquote {\bibinfo {title} {{$\mathcal{PT}$-symmetric} circuit
  {QED}},}\ }\href {https://doi.org/10.1103/PhysRevA.97.053846} {\bibfield
  {journal} {\bibinfo  {journal} {Phys. Rev. A}\ }\textbf {\bibinfo {volume}
  {97}},\ \bibinfo {pages} {053846} (\bibinfo {year} {2018})}\BibitemShut
  {NoStop}%
\bibitem [{\citenamefont {Tchodimou}\ \emph {et~al.}(2017)\citenamefont
  {Tchodimou}, \citenamefont {Djorwe},\ and\ \citenamefont
  {{Nana~Engo}}}]{Tchodimou2017}%
  \BibitemOpen
  \bibfield  {author} {\bibinfo {author} {\bibfnamefont {C.}~\bibnamefont
  {Tchodimou}}, \bibinfo {author} {\bibfnamefont {P.}~\bibnamefont {Djorwe}}, \
  and\ \bibinfo {author} {\bibfnamefont {S.~G.}\ \bibnamefont {{Nana~Engo}}},\
  }\bibfield  {title} {\enquote {\bibinfo {title} {Distant entanglement
  enhanced in $\mathcal{PT}$-symmetric optomechanics},}\ }\href
  {https://doi.org/10.1103/PhysRevA.96.033856} {\bibfield  {journal} {\bibinfo
  {journal} {Phys. Rev. A}\ }\textbf {\bibinfo {volume} {96}},\ \bibinfo
  {pages} {033856} (\bibinfo {year} {2017})}\BibitemShut {NoStop}%
\bibitem [{\citenamefont {Wang}\ \emph {et~al.}(2019)\citenamefont {Wang},
  \citenamefont {Bai}, \citenamefont {Liu}, \citenamefont {Zhang},\ and\
  \citenamefont {Wang}}]{Wang2019}%
  \BibitemOpen
  \bibfield  {author} {\bibinfo {author} {\bibfnamefont {D.-Y.}\ \bibnamefont
  {Wang}}, \bibinfo {author} {\bibfnamefont {C.-H.}\ \bibnamefont {Bai}},
  \bibinfo {author} {\bibfnamefont {S.}~\bibnamefont {Liu}}, \bibinfo {author}
  {\bibfnamefont {S.}~\bibnamefont {Zhang}}, \ and\ \bibinfo {author}
  {\bibfnamefont {H.-F.}\ \bibnamefont {Wang}},\ }\bibfield  {title} {\enquote
  {\bibinfo {title} {Distinguishing photon blockade in a
  $\mathcal{PT}$-symmetric optomechanical system},}\ }\href
  {https://doi.org/10.1103/PhysRevA.99.043818} {\bibfield  {journal} {\bibinfo
  {journal} {Phys. Rev. A}\ }\textbf {\bibinfo {volume} {99}},\ \bibinfo
  {pages} {043818} (\bibinfo {year} {2019})}\BibitemShut {NoStop}%
\bibitem [{\citenamefont {Guo}\ \emph {et~al.}(2009)\citenamefont {Guo},
  \citenamefont {Salamo}, \citenamefont {Duchesne}, \citenamefont {Morandotti},
  \citenamefont {Volatier-Ravat}, \citenamefont {Aimez}, \citenamefont
  {Siviloglou},\ and\ \citenamefont {Christodoulides}}]{Guo2009}%
  \BibitemOpen
  \bibfield  {author} {\bibinfo {author} {\bibfnamefont {A.}~\bibnamefont
  {Guo}}, \bibinfo {author} {\bibfnamefont {G.~J.}\ \bibnamefont {Salamo}},
  \bibinfo {author} {\bibfnamefont {D.}~\bibnamefont {Duchesne}}, \bibinfo
  {author} {\bibfnamefont {R.}~\bibnamefont {Morandotti}}, \bibinfo {author}
  {\bibfnamefont {M.}~\bibnamefont {Volatier-Ravat}}, \bibinfo {author}
  {\bibfnamefont {V.}~\bibnamefont {Aimez}}, \bibinfo {author} {\bibfnamefont
  {G.~A.}\ \bibnamefont {Siviloglou}}, \ and\ \bibinfo {author} {\bibfnamefont
  {D.~N.}\ \bibnamefont {Christodoulides}},\ }\bibfield  {title} {\enquote
  {\bibinfo {title} {Observation of $\mathcal{PT}$-symmetry breaking in complex
  optical potentials},}\ }\href
  {https://doi.org/10.1103/PhysRevLett.103.093902} {\bibfield  {journal}
  {\bibinfo  {journal} {Phys. Rev. Lett.}\ }\textbf {\bibinfo {volume} {103}},\
  \bibinfo {pages} {093902} (\bibinfo {year} {2009})}\BibitemShut {NoStop}%
\bibitem [{\citenamefont {El-Ganainy}\ \emph {et~al.}(2014)\citenamefont
  {El-Ganainy}, \citenamefont {Khajavikhan},\ and\ \citenamefont
  {Ge}}]{El-Ganainy2014}%
  \BibitemOpen
  \bibfield  {author} {\bibinfo {author} {\bibfnamefont {R.}~\bibnamefont
  {El-Ganainy}}, \bibinfo {author} {\bibfnamefont {M.}~\bibnamefont
  {Khajavikhan}}, \ and\ \bibinfo {author} {\bibfnamefont {L.}~\bibnamefont
  {Ge}},\ }\bibfield  {title} {\enquote {\bibinfo {title} {Exceptional points
  and lasing self-termination in photonic molecules},}\ }\href
  {https://doi.org/10.1103/PhysRevA.90.013802} {\bibfield  {journal} {\bibinfo
  {journal} {Phys. Rev. A}\ }\textbf {\bibinfo {volume} {90}},\ \bibinfo
  {pages} {013802} (\bibinfo {year} {2014})}\BibitemShut {NoStop}%
\bibitem [{\citenamefont {Chen}\ \emph {et~al.}(2017)\citenamefont {Chen},
  \citenamefont {\"{O}zdemir}, \citenamefont {Zhao}, \citenamefont {Wiersig},\
  and\ \citenamefont {Yang}}]{Chen2017}%
  \BibitemOpen
  \bibfield  {author} {\bibinfo {author} {\bibfnamefont {W.}~\bibnamefont
  {Chen}}, \bibinfo {author} {\bibfnamefont {{\c{S}}.~Kaya}\ \bibnamefont
  {\"{O}zdemir}}, \bibinfo {author} {\bibfnamefont {G.}~\bibnamefont {Zhao}},
  \bibinfo {author} {\bibfnamefont {J.}~\bibnamefont {Wiersig}}, \ and\
  \bibinfo {author} {\bibfnamefont {L.}~\bibnamefont {Yang}},\ }\bibfield
  {title} {\enquote {\bibinfo {title} {Exceptional points enhance sensing in an
  optical microcavity},}\ }\href {https://doi.org/10.1038/nature23281}
  {\bibfield  {journal} {\bibinfo  {journal} {Nature}\ }\textbf {\bibinfo
  {volume} {548}},\ \bibinfo {pages} {192--196} (\bibinfo {year}
  {2017})}\BibitemShut {NoStop}%
\bibitem [{\citenamefont {Rubinstein}\ \emph {et~al.}(2007)\citenamefont
  {Rubinstein}, \citenamefont {Sternberg},\ and\ \citenamefont
  {Ma}}]{Rubinstein2007}%
  \BibitemOpen
  \bibfield  {author} {\bibinfo {author} {\bibfnamefont {J.}~\bibnamefont
  {Rubinstein}}, \bibinfo {author} {\bibfnamefont {P.}~\bibnamefont
  {Sternberg}}, \ and\ \bibinfo {author} {\bibfnamefont {Q.}~\bibnamefont
  {Ma}},\ }\bibfield  {title} {\enquote {\bibinfo {title} {Bifurcation diagram
  and pattern formation of phase slip centers in superconducting wires driven
  with electric currents},}\ }\href
  {https://doi.org/10.1103/PhysRevLett.99.167003} {\bibfield  {journal}
  {\bibinfo  {journal} {Phys. Rev. Lett.}\ }\textbf {\bibinfo {volume} {99}},\
  \bibinfo {pages} {167003} (\bibinfo {year} {2007})}\BibitemShut {NoStop}%
\bibitem [{\citenamefont {Bittner}\ \emph {et~al.}(2012)\citenamefont
  {Bittner}, \citenamefont {Dietz}, \citenamefont {G\"unther}, \citenamefont
  {Harney}, \citenamefont {Miski-Oglu}, \citenamefont {Richter},\ and\
  \citenamefont {Sch\"afer}}]{Bittner2012}%
  \BibitemOpen
  \bibfield  {author} {\bibinfo {author} {\bibfnamefont {S.}~\bibnamefont
  {Bittner}}, \bibinfo {author} {\bibfnamefont {B.}~\bibnamefont {Dietz}},
  \bibinfo {author} {\bibfnamefont {U.}~\bibnamefont {G\"unther}}, \bibinfo
  {author} {\bibfnamefont {H.~L.}\ \bibnamefont {Harney}}, \bibinfo {author}
  {\bibfnamefont {M.}~\bibnamefont {Miski-Oglu}}, \bibinfo {author}
  {\bibfnamefont {A.}~\bibnamefont {Richter}}, \ and\ \bibinfo {author}
  {\bibfnamefont {F.}~\bibnamefont {Sch\"afer}},\ }\bibfield  {title} {\enquote
  {\bibinfo {title} {$\mathcal{PT}$ symmetry and spontaneous symmetry breaking
  in a microwave billiard},}\ }\href
  {https://doi.org/10.1103/PhysRevLett.108.024101} {\bibfield  {journal}
  {\bibinfo  {journal} {Phys. Rev. Lett.}\ }\textbf {\bibinfo {volume} {108}},\
  \bibinfo {pages} {024101} (\bibinfo {year} {2012})}\BibitemShut {NoStop}%
\bibitem [{\citenamefont {L\"u}\ \emph {et~al.}(2015)\citenamefont {L\"u},
  \citenamefont {Jing}, \citenamefont {Ma},\ and\ \citenamefont {Wu}}]{Lu2015}%
  \BibitemOpen
  \bibfield  {author} {\bibinfo {author} {\bibfnamefont {X.-Y.}\ \bibnamefont
  {L\"u}}, \bibinfo {author} {\bibfnamefont {H.}~\bibnamefont {Jing}}, \bibinfo
  {author} {\bibfnamefont {J.-Y.}\ \bibnamefont {Ma}}, \ and\ \bibinfo {author}
  {\bibfnamefont {Y.}~\bibnamefont {Wu}},\ }\bibfield  {title} {\enquote
  {\bibinfo {title} {$\mathcal{PT}$-symmetry-breaking chaos in
  optomechanics},}\ }\href
  {https://link.aps.org/doi/10.1103/PhysRevLett.114.253601} {\bibfield
  {journal} {\bibinfo  {journal} {Phys. Rev. Lett.}\ }\textbf {\bibinfo
  {volume} {114}},\ \bibinfo {pages} {253601} (\bibinfo {year}
  {2015})}\BibitemShut {NoStop}%
\bibitem [{\citenamefont {Koutserimpas}\ \emph {et~al.}(2018)\citenamefont
  {Koutserimpas}, \citenamefont {Al\`u},\ and\ \citenamefont
  {Fleury}}]{Koutserimpas2018}%
  \BibitemOpen
  \bibfield  {author} {\bibinfo {author} {\bibfnamefont {T.~T.}\ \bibnamefont
  {Koutserimpas}}, \bibinfo {author} {\bibfnamefont {A.}~\bibnamefont {Al\`u}},
  \ and\ \bibinfo {author} {\bibfnamefont {R.}~\bibnamefont {Fleury}},\
  }\bibfield  {title} {\enquote {\bibinfo {title} {Parametric amplification and
  bidirectional invisibility in $\mathcal{PT}$-symmetric time-{Floquet}
  systems},}\ }\href {https://doi.org/10.1103/PhysRevA.97.013839} {\bibfield
  {journal} {\bibinfo  {journal} {Phys. Rev. A}\ }\textbf {\bibinfo {volume}
  {97}},\ \bibinfo {pages} {013839} (\bibinfo {year} {2018})}\BibitemShut
  {NoStop}%
\bibitem [{\citenamefont {\"{O}zdemir}\ \emph {et~al.}(2019)\citenamefont
  {\"{O}zdemir}, \citenamefont {Rotter}, \citenamefont {Nori},\ and\
  \citenamefont {Yang}}]{Ozdemir2019}%
  \BibitemOpen
  \bibfield  {author} {\bibinfo {author} {\bibfnamefont {{\c{S}}.~K.}\
  \bibnamefont {\"{O}zdemir}}, \bibinfo {author} {\bibfnamefont
  {S.}~\bibnamefont {Rotter}}, \bibinfo {author} {\bibfnamefont
  {F.}~\bibnamefont {Nori}}, \ and\ \bibinfo {author} {\bibfnamefont
  {L.}~\bibnamefont {Yang}},\ }\bibfield  {title} {\enquote {\bibinfo {title}
  {Parity{\textendash}time symmetry and exceptional points in photonics},}\
  }\href {https://doi.org/10.1038/s41563-019-0304-9} {\bibfield  {journal}
  {\bibinfo  {journal} {Nat. Mat.}\ }\textbf {\bibinfo {volume} {18}},\
  \bibinfo {pages} {783--798} (\bibinfo {year} {2019})}\BibitemShut {NoStop}%
\bibitem [{\citenamefont {Miri}\ and\ \citenamefont {Al\`u}(2019)}]{Miri2019}%
  \BibitemOpen
  \bibfield  {author} {\bibinfo {author} {\bibfnamefont {M.}~\bibnamefont
  {Miri}}\ and\ \bibinfo {author} {\bibfnamefont {A.}~\bibnamefont {Al\`u}},\
  }\bibfield  {title} {\enquote {\bibinfo {title} {Exceptional points in optics
  and photonics},}\ }\href {https://doi.org/10.1126/science.aar7709} {\bibfield
   {journal} {\bibinfo  {journal} {Science}\ }\textbf {\bibinfo {volume}
  {363}},\ \bibinfo {pages} {7709} (\bibinfo {year} {2019})}\BibitemShut
  {NoStop}%
\bibitem [{\citenamefont {Agarwal}\ and\ \citenamefont
  {Qu}(2012)}]{Agarwal2012}%
  \BibitemOpen
  \bibfield  {author} {\bibinfo {author} {\bibfnamefont {G.~S.}\ \bibnamefont
  {Agarwal}}\ and\ \bibinfo {author} {\bibfnamefont {K.}~\bibnamefont {Qu}},\
  }\bibfield  {title} {\enquote {\bibinfo {title} {Spontaneous generation of
  photons in transmission of quantum fields in {$\mathcal{PT}$-symmetric}
  optical systems},}\ }\href {https://doi.org/10.1103/PhysRevA.85.031802}
  {\bibfield  {journal} {\bibinfo  {journal} {Phys. Rev. A}\ }\textbf {\bibinfo
  {volume} {85}},\ \bibinfo {pages} {031802(R)} (\bibinfo {year}
  {2012})}\BibitemShut {NoStop}%
\bibitem [{\citenamefont {He}\ \emph {et~al.}(2015)\citenamefont {He},
  \citenamefont {Yan}, \citenamefont {Wang},\ and\ \citenamefont
  {Xiao}}]{He2015}%
  \BibitemOpen
  \bibfield  {author} {\bibinfo {author} {\bibfnamefont {B.}~\bibnamefont
  {He}}, \bibinfo {author} {\bibfnamefont {S.-B.}\ \bibnamefont {Yan}},
  \bibinfo {author} {\bibfnamefont {J.}~\bibnamefont {Wang}}, \ and\ \bibinfo
  {author} {\bibfnamefont {M.}~\bibnamefont {Xiao}},\ }\bibfield  {title}
  {\enquote {\bibinfo {title} {Quantum noise effects with {K}err-nonlinearity
  enhancement in coupled gain-loss waveguides},}\ }\href
  {https://doi.org/10.1103/PhysRevA.91.053832} {\bibfield  {journal} {\bibinfo
  {journal} {Phys. Rev. A}\ }\textbf {\bibinfo {volume} {91}},\ \bibinfo
  {pages} {053832} (\bibinfo {year} {2015})}\BibitemShut {NoStop}%
\bibitem [{\citenamefont {Vashahri-Ghamsari}\ \emph {et~al.}(2017)\citenamefont
  {Vashahri-Ghamsari}, \citenamefont {He},\ and\ \citenamefont
  {Xiao}}]{Vashahri2017}%
  \BibitemOpen
  \bibfield  {author} {\bibinfo {author} {\bibfnamefont {S.}~\bibnamefont
  {Vashahri-Ghamsari}}, \bibinfo {author} {\bibfnamefont {B.}~\bibnamefont
  {He}}, \ and\ \bibinfo {author} {\bibfnamefont {M.}~\bibnamefont {Xiao}},\
  }\bibfield  {title} {\enquote {\bibinfo {title} {Continuous-variable
  entanglement generation using a hybrid {$\mathcal{PT}$-symmetric} system},}\
  }\href {https://doi.org/10.1103/PhysRevA.96.033806} {\bibfield  {journal}
  {\bibinfo  {journal} {Phys. Rev. A}\ }\textbf {\bibinfo {volume} {96}},\
  \bibinfo {pages} {033806} (\bibinfo {year} {2017})}\BibitemShut {NoStop}%
\bibitem [{\citenamefont {Scheel}\ and\ \citenamefont
  {Szameit}(2018)}]{Scheel2018}%
  \BibitemOpen
  \bibfield  {author} {\bibinfo {author} {\bibfnamefont {S.}~\bibnamefont
  {Scheel}}\ and\ \bibinfo {author} {\bibfnamefont {A.}~\bibnamefont
  {Szameit}},\ }\bibfield  {title} {\enquote {\bibinfo {title}
  {$\mathcal{PT}$-symmetric photonic quantum systems with gain and loss do not
  exist},}\ }\href
  {https://iopscience.iop.org/article/10.1209/0295-5075/122/34001/meta}
  {\bibfield  {journal} {\bibinfo  {journal} {Eur. Phys. Lett.}\ }\textbf
  {\bibinfo {volume} {122}},\ \bibinfo {pages} {34001} (\bibinfo {year}
  {2018})}\BibitemShut {NoStop}%
\bibitem [{\citenamefont {Pe\v{r}inov\'a}\ \emph {et~al.}(2019)\citenamefont
  {Pe\v{r}inov\'a}, \citenamefont {Luk\v{s}},\ and\ \citenamefont
  {K\v{r}epelka}}]{Perinova2019}%
  \BibitemOpen
  \bibfield  {author} {\bibinfo {author} {\bibfnamefont {V.}~\bibnamefont
  {Pe\v{r}inov\'a}}, \bibinfo {author} {\bibfnamefont {A.}~\bibnamefont
  {Luk\v{s}}}, \ and\ \bibinfo {author} {\bibfnamefont {J.}~\bibnamefont
  {K\v{r}epelka}},\ }\bibfield  {title} {\enquote {\bibinfo {title} {Quantum
  description of a {$\mathcal{PT}$-symmetric} nonlinear directional coupler},}\
  }\href {https://doi.org/10.1364/JOSAB.36.000855} {\bibfield  {journal}
  {\bibinfo  {journal} {J. Opt. Soc. Am. B}\ }\textbf {\bibinfo {volume}
  {36}},\ \bibinfo {pages} {855---861} (\bibinfo {year} {2019})}\BibitemShut
  {NoStop}%
\bibitem [{\citenamefont {Minganti}\ \emph {et~al.}(2019)\citenamefont
  {Minganti}, \citenamefont {Miranowicz}, \citenamefont {Chhajlany},\ and\
  \citenamefont {Nori}}]{Minganti2019}%
  \BibitemOpen
  \bibfield  {author} {\bibinfo {author} {\bibfnamefont {F.}~\bibnamefont
  {Minganti}}, \bibinfo {author} {\bibfnamefont {A.}~\bibnamefont
  {Miranowicz}}, \bibinfo {author} {\bibfnamefont {R.}~\bibnamefont
  {Chhajlany}}, \ and\ \bibinfo {author} {\bibfnamefont {F.}~\bibnamefont
  {Nori}},\ }\bibfield  {title} {\enquote {\bibinfo {title} {Quantum
  exceptional points of non-{H}ermitian {H}amiltonians and {L}iouvillians:
  {T}he effects of quantum jumps},}\ }\href@noop {} {\bibfield  {journal}
  {\bibinfo  {journal} {e-print arXiv:1909.11619}\ } (\bibinfo {year}
  {2019})}\BibitemShut {NoStop}%
\bibitem [{\citenamefont {Zhong}\ \emph {et~al.}(2016)\citenamefont {Zhong},
  \citenamefont {Ahmed}, \citenamefont {Dadap}, \citenamefont {{Osgood~Jr.}},\
  and\ \citenamefont {El-Ganainy}}]{Zhong2016}%
  \BibitemOpen
  \bibfield  {author} {\bibinfo {author} {\bibfnamefont {Q.}~\bibnamefont
  {Zhong}}, \bibinfo {author} {\bibfnamefont {A.}~\bibnamefont {Ahmed}},
  \bibinfo {author} {\bibfnamefont {J.~I.}\ \bibnamefont {Dadap}}, \bibinfo
  {author} {\bibfnamefont {R.~M.}\ \bibnamefont {{Osgood~Jr.}}}, \ and\
  \bibinfo {author} {\bibfnamefont {R.}~\bibnamefont {El-Ganainy}},\ }\bibfield
   {title} {\enquote {\bibinfo {title} {Parametric amplification in
  {quasi-$\mathcal{PT}$} symmetric coupled waveguide structures},}\ }\href
  {https://doi.org/10.1088/1367-2630/18/12/125006} {\bibfield  {journal}
  {\bibinfo  {journal} {New J. Phys.}\ }\textbf {\bibinfo {volume} {18}},\
  \bibinfo {pages} {125006} (\bibinfo {year} {2016})}\BibitemShut {NoStop}%
\bibitem [{\citenamefont {Antonosyan}\ \emph {et~al.}(2018)\citenamefont
  {Antonosyan}, \citenamefont {Solntsev},\ and\ \citenamefont
  {Sukhorukov}}]{Antonosyan2018}%
  \BibitemOpen
  \bibfield  {author} {\bibinfo {author} {\bibfnamefont {D.~A.}\ \bibnamefont
  {Antonosyan}}, \bibinfo {author} {\bibfnamefont {A.~S.}\ \bibnamefont
  {Solntsev}}, \ and\ \bibinfo {author} {\bibfnamefont {A.~A.}\ \bibnamefont
  {Sukhorukov}},\ }\bibfield  {title} {\enquote {\bibinfo {title} {Photon-pair
  generation in a quadratically nonlinear parity-time symmetric coupler},}\
  }\href {https://doi.org/10.1364/PRJ.6.0000A6} {\bibfield  {journal} {\bibinfo
   {journal} {Phot. Res.}\ }\textbf {\bibinfo {volume} {6}},\ \bibinfo {pages}
  {A6--A9} (\bibinfo {year} {2018})}\BibitemShut {NoStop}%
\bibitem [{\citenamefont {Naikoo}\ \emph {et~al.}(2019)\citenamefont {Naikoo},
  \citenamefont {Thapliyal}, \citenamefont {Banerjee},\ and\ \citenamefont
  {Pathak}}]{Naikoo2019}%
  \BibitemOpen
  \bibfield  {author} {\bibinfo {author} {\bibfnamefont {J.}~\bibnamefont
  {Naikoo}}, \bibinfo {author} {\bibfnamefont {K.}~\bibnamefont {Thapliyal}},
  \bibinfo {author} {\bibfnamefont {S.}~\bibnamefont {Banerjee}}, \ and\
  \bibinfo {author} {\bibfnamefont {A.}~\bibnamefont {Pathak}},\ }\bibfield
  {title} {\enquote {\bibinfo {title} {Quantum {Zeno} effect and
  nonclassicality in a $\mathcal{PT}$-symmetric system of coupled cavities},}\
  }\href {https://doi.org/10.1103/PhysRevA.99.023820} {\bibfield  {journal}
  {\bibinfo  {journal} {Phys. Rev. A}\ }\textbf {\bibinfo {volume} {99}},\
  \bibinfo {pages} {023820} (\bibinfo {year} {2019})}\BibitemShut {NoStop}%
\bibitem [{\citenamefont {Croke}(2015)}]{Croke2015}%
  \BibitemOpen
  \bibfield  {author} {\bibinfo {author} {\bibfnamefont {S.}~\bibnamefont
  {Croke}},\ }\bibfield  {title} {\enquote {\bibinfo {title}
  {$\mathcal{PT}$-symmetric {Hamiltonians} and their application in quantum
  information},}\ }\href {https://doi.org/10.1103/PhysRevA.91.052113}
  {\bibfield  {journal} {\bibinfo  {journal} {Phys. Rev. A}\ }\textbf {\bibinfo
  {volume} {91}},\ \bibinfo {pages} {052113} (\bibinfo {year}
  {2015})}\BibitemShut {NoStop}%
\bibitem [{\citenamefont {Hodaei}\ \emph {et~al.}(2017)\citenamefont {Hodaei},
  \citenamefont {Absar}, \citenamefont {Wittek}, \citenamefont {Garcia-Gracia},
  \citenamefont {El-Ganainy}, \citenamefont {Christodoulides},\ and\
  \citenamefont {Khajavikhan}}]{Hodaei2017}%
  \BibitemOpen
  \bibfield  {author} {\bibinfo {author} {\bibfnamefont {H.}~\bibnamefont
  {Hodaei}}, \bibinfo {author} {\bibfnamefont {U.~H.}\ \bibnamefont {Absar}},
  \bibinfo {author} {\bibfnamefont {S.}~\bibnamefont {Wittek}}, \bibinfo
  {author} {\bibfnamefont {H.}~\bibnamefont {Garcia-Gracia}}, \bibinfo {author}
  {\bibfnamefont {R.}~\bibnamefont {El-Ganainy}}, \bibinfo {author}
  {\bibfnamefont {D.~N.}\ \bibnamefont {Christodoulides}}, \ and\ \bibinfo
  {author} {\bibfnamefont {M.}~\bibnamefont {Khajavikhan}},\ }\bibfield
  {title} {\enquote {\bibinfo {title} {Enhanced sensitivity at higher-order
  exceptional points},}\ }\href {https://doi.org/10.1038/nature23280}
  {\bibfield  {journal} {\bibinfo  {journal} {Nature}\ }\textbf {\bibinfo
  {volume} {548}},\ \bibinfo {pages} {187} (\bibinfo {year}
  {2017})}\BibitemShut {NoStop}%
\bibitem [{\citenamefont {Brandstetter}\ \emph {et~al.}(2014)\citenamefont
  {Brandstetter}, \citenamefont {Liertzer}, \citenamefont {Deutsch},
  \citenamefont {Klang}, \citenamefont {Schoberl}, \citenamefont {Tureci},
  \citenamefont {Strasser}, \citenamefont {Unterrainer},\ and\ \citenamefont
  {Rotter}}]{Brands2014a}%
  \BibitemOpen
  \bibfield  {author} {\bibinfo {author} {\bibfnamefont {M.}~\bibnamefont
  {Brandstetter}}, \bibinfo {author} {\bibfnamefont {M.}~\bibnamefont
  {Liertzer}}, \bibinfo {author} {\bibfnamefont {C.}~\bibnamefont {Deutsch}},
  \bibinfo {author} {\bibfnamefont {P.}~\bibnamefont {Klang}}, \bibinfo
  {author} {\bibfnamefont {J.}~\bibnamefont {Schoberl}}, \bibinfo {author}
  {\bibfnamefont {H.~E.}\ \bibnamefont {Tureci}}, \bibinfo {author}
  {\bibfnamefont {G.}~\bibnamefont {Strasser}}, \bibinfo {author}
  {\bibfnamefont {K.}~\bibnamefont {Unterrainer}}, \ and\ \bibinfo {author}
  {\bibfnamefont {S.}~\bibnamefont {Rotter}},\ }\bibfield  {title} {\enquote
  {\bibinfo {title} {Reversing the pump dependence of a laser at an exceptional
  point},}\ }\href {https://doi.org/10.1038/ncomms5034} {\bibfield  {journal}
  {\bibinfo  {journal} {Nat. Commun.}\ }\textbf {\bibinfo {volume} {5}},\
  \bibinfo {pages} {4034} (\bibinfo {year} {2014})}\BibitemShut {NoStop}%
\bibitem [{\citenamefont {Chang}\ \emph {et~al.}(2014)\citenamefont {Chang},
  \citenamefont {Jiang}, \citenamefont {Hua}, \citenamefont {Yang},
  \citenamefont {Wen}, \citenamefont {Jiang}, \citenamefont {Li}, \citenamefont
  {Wang},\ and\ \citenamefont {Xiao}}]{Chang2014}%
  \BibitemOpen
  \bibfield  {author} {\bibinfo {author} {\bibfnamefont {L.}~\bibnamefont
  {Chang}}, \bibinfo {author} {\bibfnamefont {X.}~\bibnamefont {Jiang}},
  \bibinfo {author} {\bibfnamefont {S.}~\bibnamefont {Hua}}, \bibinfo {author}
  {\bibfnamefont {C.}~\bibnamefont {Yang}}, \bibinfo {author} {\bibfnamefont
  {J.}~\bibnamefont {Wen}}, \bibinfo {author} {\bibfnamefont {L.}~\bibnamefont
  {Jiang}}, \bibinfo {author} {\bibfnamefont {G.}~\bibnamefont {Li}}, \bibinfo
  {author} {\bibfnamefont {G.}~\bibnamefont {Wang}}, \ and\ \bibinfo {author}
  {\bibfnamefont {M.}~\bibnamefont {Xiao}},\ }\bibfield  {title} {\enquote
  {\bibinfo {title} {Parity-time symmetry and variable optical isolation in
  active-passive-coupled microresonators},}\ }\href
  {https://doi.org/10.1038/nphoton.2014.133} {\bibfield  {journal} {\bibinfo
  {journal} {Nat. Photon.}\ }\textbf {\bibinfo {volume} {8}},\ \bibinfo {pages}
  {524} (\bibinfo {year} {2014})}\BibitemShut {NoStop}%
\bibitem [{\citenamefont {Jing}\ \emph {et~al.}(2014)\citenamefont {Jing},
  \citenamefont {\c{S}. K.~{\"O}zdemir}, \citenamefont {L{\"u}}, \citenamefont
  {Zhang}, \citenamefont {Yang},\ and\ \citenamefont {Nori}}]{Jing2014}%
  \BibitemOpen
  \bibfield  {author} {\bibinfo {author} {\bibfnamefont {H.}~\bibnamefont
  {Jing}}, \bibinfo {author} {\bibnamefont {\c{S}. K.~{\"O}zdemir}}, \bibinfo
  {author} {\bibfnamefont {X.-Y.}\ \bibnamefont {L{\"u}}}, \bibinfo {author}
  {\bibfnamefont {J.}~\bibnamefont {Zhang}}, \bibinfo {author} {\bibfnamefont
  {L.}~\bibnamefont {Yang}}, \ and\ \bibinfo {author} {\bibfnamefont
  {F.}~\bibnamefont {Nori}},\ }\bibfield  {title} {\enquote {\bibinfo {title}
  {$\mathcal{PT}$-symmetric phonon laser},}\ }\href
  {https://doi.org/10.1103/PhysRevLett.113.053604} {\bibfield  {journal}
  {\bibinfo  {journal} {Phys. Rev. Lett.}\ }\textbf {\bibinfo {volume} {113}},\
  \bibinfo {pages} {053604} (\bibinfo {year} {2014})}\BibitemShut {NoStop}%
\bibitem [{\citenamefont {L{\"u}}\ \emph {et~al.}(2017)\citenamefont {L{\"u}},
  \citenamefont {\c{S}. K.~{\"O}zdemir}, \citenamefont {Kuang}, \citenamefont
  {Nori},\ and\ \citenamefont {Jing}}]{Lu2017}%
  \BibitemOpen
  \bibfield  {author} {\bibinfo {author} {\bibfnamefont {H.}~\bibnamefont
  {L{\"u}}}, \bibinfo {author} {\bibnamefont {\c{S}. K.~{\"O}zdemir}}, \bibinfo
  {author} {\bibfnamefont {L.~M.}\ \bibnamefont {Kuang}}, \bibinfo {author}
  {\bibfnamefont {F.}~\bibnamefont {Nori}}, \ and\ \bibinfo {author}
  {\bibfnamefont {H.}~\bibnamefont {Jing}},\ }\bibfield  {title} {\enquote
  {\bibinfo {title} {Exceptional points in random-defect phonon lasers},}\
  }\href {https://doi.org/10.1103/PhysRevApplied.8.044020} {\bibfield
  {journal} {\bibinfo  {journal} {Phys. Rev. App.}\ }\textbf {\bibinfo {volume}
  {8}},\ \bibinfo {pages} {044020} (\bibinfo {year} {2017})}\BibitemShut
  {NoStop}%
\bibitem [{\citenamefont {Lin}\ \emph {et~al.}(2011)\citenamefont {Lin},
  \citenamefont {Ramezani}, \citenamefont {Eichelkraut}, \citenamefont
  {Kottos}, \citenamefont {Cao},\ and\ \citenamefont
  {Christodoulides}}]{Lin2011}%
  \BibitemOpen
  \bibfield  {author} {\bibinfo {author} {\bibfnamefont {Z.}~\bibnamefont
  {Lin}}, \bibinfo {author} {\bibfnamefont {H.}~\bibnamefont {Ramezani}},
  \bibinfo {author} {\bibfnamefont {T.}~\bibnamefont {Eichelkraut}}, \bibinfo
  {author} {\bibfnamefont {T.}~\bibnamefont {Kottos}}, \bibinfo {author}
  {\bibfnamefont {H.}~\bibnamefont {Cao}}, \ and\ \bibinfo {author}
  {\bibfnamefont {D.~N.}\ \bibnamefont {Christodoulides}},\ }\bibfield  {title}
  {\enquote {\bibinfo {title} {Unidirectional invisibility induced by
  $\mathcal{PT}$-symmetric periodic structures},}\ }\href
  {https://doi.org/10.1103/PhysRevLett.106.213901} {\bibfield  {journal}
  {\bibinfo  {journal} {Phys. Rev. Lett.}\ }\textbf {\bibinfo {volume} {106}},\
  \bibinfo {pages} {213901} (\bibinfo {year} {2011})}\BibitemShut {NoStop}%
\bibitem [{\citenamefont {Regensburger}\ \emph {et~al.}(2012)\citenamefont
  {Regensburger}, \citenamefont {Bersch}, \citenamefont {Miri}, \citenamefont
  {Onishchukov}, \citenamefont {Christodoulides},\ and\ \citenamefont
  {Peschel}}]{Regen2012}%
  \BibitemOpen
  \bibfield  {author} {\bibinfo {author} {\bibfnamefont {A.}~\bibnamefont
  {Regensburger}}, \bibinfo {author} {\bibfnamefont {C.}~\bibnamefont
  {Bersch}}, \bibinfo {author} {\bibfnamefont {M.-A.}\ \bibnamefont {Miri}},
  \bibinfo {author} {\bibfnamefont {G.}~\bibnamefont {Onishchukov}}, \bibinfo
  {author} {\bibfnamefont {D.~N.}\ \bibnamefont {Christodoulides}}, \ and\
  \bibinfo {author} {\bibfnamefont {U.}~\bibnamefont {Peschel}},\ }\bibfield
  {title} {\enquote {\bibinfo {title} {Parity-time synthetic photonic
  lattices},}\ }\href {https://doi.org/10.1038/nature11298} {\bibfield
  {journal} {\bibinfo  {journal} {Nature}\ }\textbf {\bibinfo {volume} {488}},\
  \bibinfo {pages} {167} (\bibinfo {year} {2012})}\BibitemShut {NoStop}%
\bibitem [{\citenamefont {Peng}\ \emph {et~al.}(2016)\citenamefont {Peng},
  \citenamefont {\"{O}zdemir}, \citenamefont {Liertzer}, \citenamefont {Chen},
  \citenamefont {Kramer}, \citenamefont {Y{\i}lmaz}, \citenamefont {Wiersig},
  \citenamefont {Rotter},\ and\ \citenamefont {Yang}}]{Peng2016}%
  \BibitemOpen
  \bibfield  {author} {\bibinfo {author} {\bibfnamefont {B.}~\bibnamefont
  {Peng}}, \bibinfo {author} {\bibfnamefont {{\c{S}}.~K.}\ \bibnamefont
  {\"{O}zdemir}}, \bibinfo {author} {\bibfnamefont {M.}~\bibnamefont
  {Liertzer}}, \bibinfo {author} {\bibfnamefont {W.}~\bibnamefont {Chen}},
  \bibinfo {author} {\bibfnamefont {J.}~\bibnamefont {Kramer}}, \bibinfo
  {author} {\bibfnamefont {H.}~\bibnamefont {Y{\i}lmaz}}, \bibinfo {author}
  {\bibfnamefont {J.}~\bibnamefont {Wiersig}}, \bibinfo {author} {\bibfnamefont
  {S.}~\bibnamefont {Rotter}}, \ and\ \bibinfo {author} {\bibfnamefont
  {L.}~\bibnamefont {Yang}},\ }\bibfield  {title} {\enquote {\bibinfo {title}
  {Chiral modes and directional lasing at exceptional points},}\ }\href
  {\doibase 10.1073/pnas.1603318113} {\bibfield  {journal} {\bibinfo  {journal}
  {PNAS}\ }\textbf {\bibinfo {volume} {113}},\ \bibinfo {pages} {6845--6850}
  (\bibinfo {year} {2016})}\BibitemShut {NoStop}%
\bibitem [{\citenamefont {Feng}\ \emph {et~al.}(2014)\citenamefont {Feng},
  \citenamefont {Wong}, \citenamefont {Ma}, \citenamefont {Wang},\ and\
  \citenamefont {Zhang}}]{Feng2014}%
  \BibitemOpen
  \bibfield  {author} {\bibinfo {author} {\bibfnamefont {L.}~\bibnamefont
  {Feng}}, \bibinfo {author} {\bibfnamefont {Z.~J.}\ \bibnamefont {Wong}},
  \bibinfo {author} {\bibfnamefont {R.-M.}\ \bibnamefont {Ma}}, \bibinfo
  {author} {\bibfnamefont {Y.}~\bibnamefont {Wang}}, \ and\ \bibinfo {author}
  {\bibfnamefont {X.}~\bibnamefont {Zhang}},\ }\bibfield  {title} {\enquote
  {\bibinfo {title} {Single-mode laser by parity-time symmetry breaking},}\
  }\href {https://doi.org/10.1126/science.1258479} {\bibfield  {journal}
  {\bibinfo  {journal} {Science}\ }\textbf {\bibinfo {volume} {346}},\ \bibinfo
  {pages} {972} (\bibinfo {year} {2014})}\BibitemShut {NoStop}%
\bibitem [{\citenamefont {Hodaei}\ \emph {et~al.}(2014)\citenamefont {Hodaei},
  \citenamefont {Miri}, \citenamefont {Heinrich}, \citenamefont
  {Christodoulides},\ and\ \citenamefont {Khajavikhan}}]{Hodaei2014}%
  \BibitemOpen
  \bibfield  {author} {\bibinfo {author} {\bibfnamefont {H.}~\bibnamefont
  {Hodaei}}, \bibinfo {author} {\bibfnamefont {M.-A.}\ \bibnamefont {Miri}},
  \bibinfo {author} {\bibfnamefont {M.}~\bibnamefont {Heinrich}}, \bibinfo
  {author} {\bibfnamefont {D.~N.}\ \bibnamefont {Christodoulides}}, \ and\
  \bibinfo {author} {\bibfnamefont {M.}~\bibnamefont {Khajavikhan}},\
  }\bibfield  {title} {\enquote {\bibinfo {title} {Parity-time-symmetric
  microring lasers},}\ }\href {https://doi.org/10.1126/science.1258480}
  {\bibfield  {journal} {\bibinfo  {journal} {Science}\ }\textbf {\bibinfo
  {volume} {346}},\ \bibinfo {pages} {975} (\bibinfo {year}
  {2014})}\BibitemShut {NoStop}%
\bibitem [{\citenamefont {Doppler}\ \emph {et~al.}(2016)\citenamefont
  {Doppler}, \citenamefont {Mailybaev}, \citenamefont {B\"{o}hm}, \citenamefont
  {Kuhl}, \citenamefont {Girschik}, \citenamefont {Libisch}, \citenamefont
  {Milburn}, \citenamefont {Rabl}, \citenamefont {Moiseyev},\ and\
  \citenamefont {Rotter}}]{Doppler2016}%
  \BibitemOpen
  \bibfield  {author} {\bibinfo {author} {\bibfnamefont {J.}~\bibnamefont
  {Doppler}}, \bibinfo {author} {\bibfnamefont {A.~A.}\ \bibnamefont
  {Mailybaev}}, \bibinfo {author} {\bibfnamefont {J.}~\bibnamefont {B\"{o}hm}},
  \bibinfo {author} {\bibfnamefont {U.}~\bibnamefont {Kuhl}}, \bibinfo {author}
  {\bibfnamefont {A.}~\bibnamefont {Girschik}}, \bibinfo {author}
  {\bibfnamefont {F.}~\bibnamefont {Libisch}}, \bibinfo {author} {\bibfnamefont
  {T.~J.}\ \bibnamefont {Milburn}}, \bibinfo {author} {\bibfnamefont
  {P.}~\bibnamefont {Rabl}}, \bibinfo {author} {\bibfnamefont {N.}~\bibnamefont
  {Moiseyev}}, \ and\ \bibinfo {author} {\bibfnamefont {S.}~\bibnamefont
  {Rotter}},\ }\bibfield  {title} {\enquote {\bibinfo {title} {Dynamically
  encircling an exceptional point for asymmetric mode switching},}\ }\href
  {https://doi.org/10.1038/nature18605} {\bibfield  {journal} {\bibinfo
  {journal} {Nature}\ }\textbf {\bibinfo {volume} {537}},\ \bibinfo {pages}
  {76--79} (\bibinfo {year} {2016})}\BibitemShut {NoStop}%
\bibitem [{\citenamefont {Jing}\ \emph {et~al.}(2015)\citenamefont {Jing},
  \citenamefont {\"{O}zdemir}, \citenamefont {Geng}, \citenamefont {Zhang},
  \citenamefont {L\"{u}}, \citenamefont {Peng}, \citenamefont {Yang},\ and\
  \citenamefont {Nori}}]{Jing2015}%
  \BibitemOpen
  \bibfield  {author} {\bibinfo {author} {\bibfnamefont {H.}~\bibnamefont
  {Jing}}, \bibinfo {author} {\bibfnamefont {{\c{S}}ahin~K.}\ \bibnamefont
  {\"{O}zdemir}}, \bibinfo {author} {\bibfnamefont {Z.}~\bibnamefont {Geng}},
  \bibinfo {author} {\bibfnamefont {Jing}\ \bibnamefont {Zhang}}, \bibinfo
  {author} {\bibfnamefont {Xin-You}\ \bibnamefont {L\"{u}}}, \bibinfo {author}
  {\bibfnamefont {Bo}~\bibnamefont {Peng}}, \bibinfo {author} {\bibfnamefont
  {Lan}\ \bibnamefont {Yang}}, \ and\ \bibinfo {author} {\bibfnamefont
  {Franco}\ \bibnamefont {Nori}},\ }\bibfield  {title} {\enquote {\bibinfo
  {title} {Optomechanically-induced transparency in parity-time-symmetric
  microresonators},}\ }\href {https://doi.org/10.1038/srep09663} {\bibfield
  {journal} {\bibinfo  {journal} {Sci. Rep.}\ }\textbf {\bibinfo {volume}
  {5}},\ \bibinfo {pages} {9663} (\bibinfo {year} {2015})}\BibitemShut
  {NoStop}%
\bibitem [{\citenamefont {Jing}\ \emph {et~al.}(2017)\citenamefont {Jing},
  \citenamefont {\c{S}. K.~{\"O}zdemir}, \citenamefont {L{\"u}},\ and\
  \citenamefont {Nori}}]{Jing2017}%
  \BibitemOpen
  \bibfield  {author} {\bibinfo {author} {\bibfnamefont {H.}~\bibnamefont
  {Jing}}, \bibinfo {author} {\bibnamefont {\c{S}. K.~{\"O}zdemir}}, \bibinfo
  {author} {\bibfnamefont {H.}~\bibnamefont {L{\"u}}}, \ and\ \bibinfo {author}
  {\bibfnamefont {F.}~\bibnamefont {Nori}},\ }\bibfield  {title} {\enquote
  {\bibinfo {title} {High-order exceptional points in optomechanics},}\ }\href
  {https://doi.org/10.1038/s41598-017-03546-7} {\bibfield  {journal} {\bibinfo
  {journal} {Sci. Rep.}\ }\textbf {\bibinfo {volume} {7}},\ \bibinfo {pages}
  {3386} (\bibinfo {year} {2017})}\BibitemShut {NoStop}%
\bibitem [{\citenamefont {Benisty}\ \emph {et~al.}(2011)\citenamefont
  {Benisty}, \citenamefont {Degiron}, \citenamefont {Lupu}, \citenamefont
  {Lustrac}, \citenamefont {Chenais}, \citenamefont {Forget}, \citenamefont
  {Besbes}, \citenamefont {Barbillon}, \citenamefont {Bruyant}, \citenamefont
  {Blaize},\ and\ \citenamefont {Lerondel}}]{Benisty2011}%
  \BibitemOpen
  \bibfield  {author} {\bibinfo {author} {\bibfnamefont {H.}~\bibnamefont
  {Benisty}}, \bibinfo {author} {\bibfnamefont {A.}~\bibnamefont {Degiron}},
  \bibinfo {author} {\bibfnamefont {A.}~\bibnamefont {Lupu}}, \bibinfo {author}
  {\bibfnamefont {A.~De}\ \bibnamefont {Lustrac}}, \bibinfo {author}
  {\bibfnamefont {S.}~\bibnamefont {Chenais}}, \bibinfo {author} {\bibfnamefont
  {S.}~\bibnamefont {Forget}}, \bibinfo {author} {\bibfnamefont
  {M.}~\bibnamefont {Besbes}}, \bibinfo {author} {\bibfnamefont
  {G.}~\bibnamefont {Barbillon}}, \bibinfo {author} {\bibfnamefont
  {A.}~\bibnamefont {Bruyant}}, \bibinfo {author} {\bibfnamefont
  {S.}~\bibnamefont {Blaize}}, \ and\ \bibinfo {author} {\bibfnamefont
  {G.}~\bibnamefont {Lerondel}},\ }\bibfield  {title} {\enquote {\bibinfo
  {title} {Implementation of $\mathcal{PT}$ symmetric devices using plasmonics:
  principle and applications},}\ }\href {https://doi.org/10.1364/OE.19.018004}
  {\bibfield  {journal} {\bibinfo  {journal} {Opt. Express}\ }\textbf {\bibinfo
  {volume} {19}},\ \bibinfo {pages} {18004} (\bibinfo {year}
  {2011})}\BibitemShut {NoStop}%
\bibitem [{\citenamefont {Tame}\ \emph {et~al.}(2013)\citenamefont {Tame},
  \citenamefont {McEnery}, \citenamefont {\"{O}zdemir}, \citenamefont {Lee},
  \citenamefont {Maier},\ and\ \citenamefont {Kim}}]{Tame2013}%
  \BibitemOpen
  \bibfield  {author} {\bibinfo {author} {\bibfnamefont {M.~S.}\ \bibnamefont
  {Tame}}, \bibinfo {author} {\bibfnamefont {K.~R.}\ \bibnamefont {McEnery}},
  \bibinfo {author} {\bibfnamefont {{\c{S}}.~K.}\ \bibnamefont {\"{O}zdemir}},
  \bibinfo {author} {\bibfnamefont {J.}~\bibnamefont {Lee}}, \bibinfo {author}
  {\bibfnamefont {S.~A.}\ \bibnamefont {Maier}}, \ and\ \bibinfo {author}
  {\bibfnamefont {M.~S.}\ \bibnamefont {Kim}},\ }\bibfield  {title} {\enquote
  {\bibinfo {title} {Quantum plasmonics},}\ }\href
  {https://doi.org/10.1038/nphys2615} {\bibfield  {journal} {\bibinfo
  {journal} {Nat. Phys.}\ }\textbf {\bibinfo {volume} {9}},\ \bibinfo {pages}
  {329--340} (\bibinfo {year} {2013})}\BibitemShut {NoStop}%
\bibitem [{\citenamefont {Schindler}\ \emph {et~al.}(2011)\citenamefont
  {Schindler}, \citenamefont {Li}, \citenamefont {Zheng}, \citenamefont
  {Ellis},\ and\ \citenamefont {Kottos}}]{Schindler2011}%
  \BibitemOpen
  \bibfield  {author} {\bibinfo {author} {\bibfnamefont {J.}~\bibnamefont
  {Schindler}}, \bibinfo {author} {\bibfnamefont {A.}~\bibnamefont {Li}},
  \bibinfo {author} {\bibfnamefont {M.C.}\ \bibnamefont {Zheng}}, \bibinfo
  {author} {\bibfnamefont {F.~M.}\ \bibnamefont {Ellis}}, \ and\ \bibinfo
  {author} {\bibfnamefont {T.}~\bibnamefont {Kottos}},\ }\bibfield  {title}
  {\enquote {\bibinfo {title} {Experimental study of active {LRC} circuits with
  $\mathcal{PT}$ symmetries},}\ }\href
  {https://doi.org/10.1103/PhysRevA.84.040101} {\bibfield  {journal} {\bibinfo
  {journal} {Phys. Rev. A}\ }\textbf {\bibinfo {volume} {84}},\ \bibinfo
  {pages} {040101(R)} (\bibinfo {year} {2011})}\BibitemShut {NoStop}%
\bibitem [{\citenamefont {Chen}\ and\ \citenamefont
  {El-Ganainy}(2019)}]{Chen2019}%
  \BibitemOpen
  \bibfield  {author} {\bibinfo {author} {\bibfnamefont {P.-Y.}\ \bibnamefont
  {Chen}}\ and\ \bibinfo {author} {\bibfnamefont {R.}~\bibnamefont
  {El-Ganainy}},\ }\bibfield  {title} {\enquote {\bibinfo {title} {Exceptional
  points enhance wireless readout},}\ }\href
  {https://doi.org/10.1038/s41928-019-0293-3} {\bibfield  {journal} {\bibinfo
  {journal} {Nat. Electron.}\ }\textbf {\bibinfo {volume} {2}},\ \bibinfo
  {pages} {323--324} (\bibinfo {year} {2019})}\BibitemShut {NoStop}%
\bibitem [{\citenamefont {Kang}\ \emph {et~al.}(2013)\citenamefont {Kang},
  \citenamefont {Liu},\ and\ \citenamefont {Li}}]{Kang2013}%
  \BibitemOpen
  \bibfield  {author} {\bibinfo {author} {\bibfnamefont {M.}~\bibnamefont
  {Kang}}, \bibinfo {author} {\bibfnamefont {F.}~\bibnamefont {Liu}}, \ and\
  \bibinfo {author} {\bibfnamefont {J.}~\bibnamefont {Li}},\ }\bibfield
  {title} {\enquote {\bibinfo {title} {Effective spontaneous
  $\mathcal{PT}$-symmetry breaking in hybridized metamaterials},}\ }\href
  {https://doi.org/10.1103/PhysRevA.87.053824} {\bibfield  {journal} {\bibinfo
  {journal} {Phys. Rev. A}\ }\textbf {\bibinfo {volume} {87}},\ \bibinfo
  {pages} {053824} (\bibinfo {year} {2013})}\BibitemShut {NoStop}%
\bibitem [{\citenamefont {Xu}\ \emph {et~al.}(2016{\natexlab{a}})\citenamefont
  {Xu}, \citenamefont {Mason}, \citenamefont {Jiang},\ and\ \citenamefont
  {Harris}}]{Harris2016}%
  \BibitemOpen
  \bibfield  {author} {\bibinfo {author} {\bibfnamefont {H.}~\bibnamefont
  {Xu}}, \bibinfo {author} {\bibfnamefont {D.}~\bibnamefont {Mason}}, \bibinfo
  {author} {\bibfnamefont {L.}~\bibnamefont {Jiang}}, \ and\ \bibinfo {author}
  {\bibfnamefont {J.~G.~E.}\ \bibnamefont {Harris}},\ }\bibfield  {title}
  {\enquote {\bibinfo {title} {Topological energy transfer in an optomechanical
  system with exceptional points},}\ }\href
  {https://doi.org/10.1038/nature18604} {\bibfield  {journal} {\bibinfo
  {journal} {Nature}\ }\textbf {\bibinfo {volume} {537}},\ \bibinfo {pages}
  {80} (\bibinfo {year} {2016}{\natexlab{a}})}\BibitemShut {NoStop}%
\bibitem [{\citenamefont {Zhu}\ \emph {et~al.}(2014)\citenamefont {Zhu},
  \citenamefont {Ramezani}, \citenamefont {Shi}, \citenamefont {Zhu},\ and\
  \citenamefont {Zhang}}]{Zhu2014}%
  \BibitemOpen
  \bibfield  {author} {\bibinfo {author} {\bibfnamefont {X.}~\bibnamefont
  {Zhu}}, \bibinfo {author} {\bibfnamefont {H.}~\bibnamefont {Ramezani}},
  \bibinfo {author} {\bibfnamefont {C.}~\bibnamefont {Shi}}, \bibinfo {author}
  {\bibfnamefont {J.}~\bibnamefont {Zhu}}, \ and\ \bibinfo {author}
  {\bibfnamefont {X.}~\bibnamefont {Zhang}},\ }\bibfield  {title} {\enquote
  {\bibinfo {title} {$\mathcal{PT}$-symmetric acoustics},}\ }\href
  {https://doi.org/10.1103/PhysRevX.4.031042} {\bibfield  {journal} {\bibinfo
  {journal} {Phys. Rev. X}\ }\textbf {\bibinfo {volume} {4}},\ \bibinfo {pages}
  {031042} (\bibinfo {year} {2014})}\BibitemShut {NoStop}%
\bibitem [{\citenamefont {Fleury}\ \emph {et~al.}(2015)\citenamefont {Fleury},
  \citenamefont {Sounas},\ and\ \citenamefont {Al\`u}}]{Alu2015}%
  \BibitemOpen
  \bibfield  {author} {\bibinfo {author} {\bibfnamefont {R.}~\bibnamefont
  {Fleury}}, \bibinfo {author} {\bibfnamefont {D.}~\bibnamefont {Sounas}}, \
  and\ \bibinfo {author} {\bibfnamefont {A.}~\bibnamefont {Al\`u}},\ }\bibfield
   {title} {\enquote {\bibinfo {title} {An invisible acoustic sensor based on
  parity-time symmetry},}\ }\href {https://doi.org/10.1038/ncomms6905}
  {\bibfield  {journal} {\bibinfo  {journal} {Nat. Commun.}\ }\textbf {\bibinfo
  {volume} {6}},\ \bibinfo {pages} {5905} (\bibinfo {year} {2015})}\BibitemShut
  {NoStop}%
\bibitem [{\citenamefont {Leykam}\ \emph {et~al.}(2017)\citenamefont {Leykam},
  \citenamefont {Bliokh}, \citenamefont {Huang}, \citenamefont {Chong},\ and\
  \citenamefont {Nori}}]{LeykamPRL17}%
  \BibitemOpen
  \bibfield  {author} {\bibinfo {author} {\bibfnamefont {D.}~\bibnamefont
  {Leykam}}, \bibinfo {author} {\bibfnamefont {K.~Y.}\ \bibnamefont {Bliokh}},
  \bibinfo {author} {\bibfnamefont {C.}~\bibnamefont {Huang}}, \bibinfo
  {author} {\bibfnamefont {Y.~D.}\ \bibnamefont {Chong}}, \ and\ \bibinfo
  {author} {\bibfnamefont {F.}~\bibnamefont {Nori}},\ }\bibfield  {title}
  {\enquote {\bibinfo {title} {Edge modes, degeneracies, and topological
  numbers in {non-Hermitian} systems},}\ }\href {\doibase
  10.1103/PhysRevLett.118.040401} {\bibfield  {journal} {\bibinfo  {journal}
  {Phys. Rev. Lett.}\ }\textbf {\bibinfo {volume} {118}},\ \bibinfo {pages}
  {040401} (\bibinfo {year} {2017})}\BibitemShut {NoStop}%
\bibitem [{\citenamefont {Gonz\'alez}\ and\ \citenamefont
  {Molina}(2017)}]{GonzalesPRB17}%
  \BibitemOpen
  \bibfield  {author} {\bibinfo {author} {\bibfnamefont {J.}~\bibnamefont
  {Gonz\'alez}}\ and\ \bibinfo {author} {\bibfnamefont {R.~A.}\ \bibnamefont
  {Molina}},\ }\bibfield  {title} {\enquote {\bibinfo {title} {Topological
  protection from exceptional points in {Weyl} and nodal-line semimetals},}\
  }\href {https://link.aps.org/doi/10.1103/PhysRevB.96.045437} {\bibfield
  {journal} {\bibinfo  {journal} {Phys. Rev. B}\ }\textbf {\bibinfo {volume}
  {96}},\ \bibinfo {pages} {045437} (\bibinfo {year} {2017})}\BibitemShut
  {NoStop}%
\bibitem [{\citenamefont {Hu}\ \emph {et~al.}(2017)\citenamefont {Hu},
  \citenamefont {Wang}, \citenamefont {Shum},\ and\ \citenamefont
  {Chong}}]{HuPRB17}%
  \BibitemOpen
  \bibfield  {author} {\bibinfo {author} {\bibfnamefont {W.}~\bibnamefont
  {Hu}}, \bibinfo {author} {\bibfnamefont {H.}~\bibnamefont {Wang}}, \bibinfo
  {author} {\bibfnamefont {P.~P.}\ \bibnamefont {Shum}}, \ and\ \bibinfo
  {author} {\bibfnamefont {Y.~D.}\ \bibnamefont {Chong}},\ }\bibfield  {title}
  {\enquote {\bibinfo {title} {Exceptional points in a {non-Hermitian}
  topological pump},}\ }\href
  {https://link.aps.org/doi/10.1103/PhysRevB.95.184306} {\bibfield  {journal}
  {\bibinfo  {journal} {Phys. Rev. B}\ }\textbf {\bibinfo {volume} {95}},\
  \bibinfo {pages} {184306} (\bibinfo {year} {2017})}\BibitemShut {NoStop}%
\bibitem [{\citenamefont {Gao}\ \emph {et~al.}(2018)\citenamefont {Gao},
  \citenamefont {Li}, \citenamefont {Estrecho}, \citenamefont {Liew},
  \citenamefont {Comber-Todd}, \citenamefont {Nalitov}, \citenamefont {Steger},
  \citenamefont {West}, \citenamefont {Pfeiffer}, \citenamefont {Snoke},
  \citenamefont {Kavokin}, \citenamefont {Truscott},\ and\ \citenamefont
  {Ostrovskaya}}]{GaoPRL18}%
  \BibitemOpen
  \bibfield  {author} {\bibinfo {author} {\bibfnamefont {T.}~\bibnamefont
  {Gao}}, \bibinfo {author} {\bibfnamefont {G.}~\bibnamefont {Li}}, \bibinfo
  {author} {\bibfnamefont {E.}~\bibnamefont {Estrecho}}, \bibinfo {author}
  {\bibfnamefont {T.~C.~H.}\ \bibnamefont {Liew}}, \bibinfo {author}
  {\bibfnamefont {D.}~\bibnamefont {Comber-Todd}}, \bibinfo {author}
  {\bibfnamefont {A.}~\bibnamefont {Nalitov}}, \bibinfo {author} {\bibfnamefont
  {M.}~\bibnamefont {Steger}}, \bibinfo {author} {\bibfnamefont
  {K.}~\bibnamefont {West}}, \bibinfo {author} {\bibfnamefont {L.}~\bibnamefont
  {Pfeiffer}}, \bibinfo {author} {\bibfnamefont {D.~W.}\ \bibnamefont {Snoke}},
  \bibinfo {author} {\bibfnamefont {A.~V.}\ \bibnamefont {Kavokin}}, \bibinfo
  {author} {\bibfnamefont {A.~G.}\ \bibnamefont {Truscott}}, \ and\ \bibinfo
  {author} {\bibfnamefont {E.~A.}\ \bibnamefont {Ostrovskaya}},\ }\bibfield
  {title} {\enquote {\bibinfo {title} {Chiral modes at exceptional points in
  exciton-polariton quantum fluids},}\ }\href
  {https://link.aps.org/doi/10.1103/PhysRevLett.120.065301} {\bibfield
  {journal} {\bibinfo  {journal} {Phys. Rev. Lett.}\ }\textbf {\bibinfo
  {volume} {120}},\ \bibinfo {pages} {065301} (\bibinfo {year}
  {2018})}\BibitemShut {NoStop}%
\bibitem [{\citenamefont {Liu}\ \emph {et~al.}(2019)\citenamefont {Liu},
  \citenamefont {Zhang}, \citenamefont {Ai}, \citenamefont {Gong},
  \citenamefont {Kawabata}, \citenamefont {Ueda},\ and\ \citenamefont
  {Nori}}]{LiuPRL19}%
  \BibitemOpen
  \bibfield  {author} {\bibinfo {author} {\bibfnamefont {T.}~\bibnamefont
  {Liu}}, \bibinfo {author} {\bibfnamefont {Y.-R.}\ \bibnamefont {Zhang}},
  \bibinfo {author} {\bibfnamefont {Q.}~\bibnamefont {Ai}}, \bibinfo {author}
  {\bibfnamefont {Z.}~\bibnamefont {Gong}}, \bibinfo {author} {\bibfnamefont
  {K.}~\bibnamefont {Kawabata}}, \bibinfo {author} {\bibfnamefont
  {M.}~\bibnamefont {Ueda}}, \ and\ \bibinfo {author} {\bibfnamefont
  {F.}~\bibnamefont {Nori}},\ }\bibfield  {title} {\enquote {\bibinfo {title}
  {Second-order topological phases in {non-Hermitian} systems},}\ }\href
  {\doibase 10.1103/PhysRevLett.122.076801} {\bibfield  {journal} {\bibinfo
  {journal} {Phys. Rev. Lett.}\ }\textbf {\bibinfo {volume} {122}},\ \bibinfo
  {pages} {076801} (\bibinfo {year} {2019})}\BibitemShut {NoStop}%
\bibitem [{\citenamefont {Bliokh}\ \emph {et~al.}(2019)\citenamefont {Bliokh},
  \citenamefont {Leykam}, \citenamefont {Lein},\ and\ \citenamefont
  {Nori}}]{BliokhNat19}%
  \BibitemOpen
  \bibfield  {author} {\bibinfo {author} {\bibfnamefont {K.~Y.}\ \bibnamefont
  {Bliokh}}, \bibinfo {author} {\bibfnamefont {D.}~\bibnamefont {Leykam}},
  \bibinfo {author} {\bibfnamefont {M.}~\bibnamefont {Lein}}, \ and\ \bibinfo
  {author} {\bibfnamefont {F.}~\bibnamefont {Nori}},\ }\bibfield  {title}
  {\enquote {\bibinfo {title} {Topological {non-Hermitian} origin of surface
  {Maxwell} waves},}\ }\href {\doibase 10.1038/s41467-019-08397-6} {\bibfield
  {journal} {\bibinfo  {journal} {Nat. Commun.}\ }\textbf {\bibinfo {volume}
  {10}},\ \bibinfo {pages} {580} (\bibinfo {year} {2019})}\BibitemShut
  {NoStop}%
\bibitem [{\citenamefont {Caspel}\ \emph {et~al.}(2019)\citenamefont {Caspel},
  \citenamefont {Arze},\ and\ \citenamefont {Castillo}}]{MoosSciPost19}%
  \BibitemOpen
  \bibfield  {author} {\bibinfo {author} {\bibfnamefont {M.}~\bibnamefont
  {Caspel}}, \bibinfo {author} {\bibfnamefont {S.~E.~T.}\ \bibnamefont {Arze}},
  \ and\ \bibinfo {author} {\bibfnamefont {I.~P.}\ \bibnamefont {Castillo}},\
  }\bibfield  {title} {\enquote {\bibinfo {title} {{Dynamical signatures of
  topological order in the driven-dissipative {Kitaev} chain}},}\ }\href
  {https://scipost.org/10.21468/SciPostPhys.6.2.026} {\bibfield  {journal}
  {\bibinfo  {journal} {SciPost Phys.}\ }\textbf {\bibinfo {volume} {6}},\
  \bibinfo {pages} {26} (\bibinfo {year} {2019})}\BibitemShut {NoStop}%
\bibitem [{\citenamefont {Ge}\ \emph {et~al.}(2019)\citenamefont {Ge},
  \citenamefont {Zhang}, \citenamefont {Liu}, \citenamefont {Li}, \citenamefont
  {Fan},\ and\ \citenamefont {Nori}}]{Ge2019}%
  \BibitemOpen
  \bibfield  {author} {\bibinfo {author} {\bibfnamefont {Z.-Y.}\ \bibnamefont
  {Ge}}, \bibinfo {author} {\bibfnamefont {Y.-R.}\ \bibnamefont {Zhang}},
  \bibinfo {author} {\bibfnamefont {T.}~\bibnamefont {Liu}}, \bibinfo {author}
  {\bibfnamefont {S.-W.}\ \bibnamefont {Li}}, \bibinfo {author} {\bibfnamefont
  {H.}~\bibnamefont {Fan}}, \ and\ \bibinfo {author} {\bibfnamefont
  {F.}~\bibnamefont {Nori}},\ }\bibfield  {title} {\enquote {\bibinfo {title}
  {Topological band theory for {non-Hermitian} systems from the {D}irac
  equation},}\ }\href {https://doi.org/10.1103/physrevb.100.054105} {\bibfield
  {journal} {\bibinfo  {journal} {Phys. Rev. B}\ }\textbf {\bibinfo {volume}
  {100}},\ \bibinfo {pages} {054105} (\bibinfo {year} {2019})}\BibitemShut
  {NoStop}%
\bibitem [{\citenamefont {Xu}\ \emph {et~al.}(2016{\natexlab{b}})\citenamefont
  {Xu}, \citenamefont {Mason}, \citenamefont {Jiang},\ and\ \citenamefont
  {Harris}}]{Xu2016}%
  \BibitemOpen
  \bibfield  {author} {\bibinfo {author} {\bibfnamefont {H.}~\bibnamefont
  {Xu}}, \bibinfo {author} {\bibfnamefont {D.}~\bibnamefont {Mason}}, \bibinfo
  {author} {\bibfnamefont {Luyao}\ \bibnamefont {Jiang}}, \ and\ \bibinfo
  {author} {\bibfnamefont {J.~G.~E.}\ \bibnamefont {Harris}},\ }\bibfield
  {title} {\enquote {\bibinfo {title} {Topological energy transfer in an
  optomechanical system with exceptional points},}\ }\href
  {https://doi.org/10.1038/nature18604} {\bibfield  {journal} {\bibinfo
  {journal} {Nature}\ }\textbf {\bibinfo {volume} {537}},\ \bibinfo {pages}
  {80--83} (\bibinfo {year} {2016}{\natexlab{b}})}\BibitemShut {NoStop}%
\bibitem [{\citenamefont {Morales}\ \emph {et~al.}(2016)\citenamefont
  {Morales}, \citenamefont {Guerrero}, \citenamefont {{L\'opez-Aguayo}},\ and\
  \citenamefont {{Rodr\'iguez-Lara}}}]{Morales2016}%
  \BibitemOpen
  \bibfield  {author} {\bibinfo {author} {\bibfnamefont {J.~D.~H.}\
  \bibnamefont {Morales}}, \bibinfo {author} {\bibfnamefont {J.}~\bibnamefont
  {Guerrero}}, \bibinfo {author} {\bibfnamefont {S.}~\bibnamefont
  {{L\'opez-Aguayo}}}, \ and\ \bibinfo {author} {\bibfnamefont {B.~M.}\
  \bibnamefont {{Rodr\'iguez-Lara}}},\ }\bibfield  {title} {\enquote {\bibinfo
  {title} {Revisiting the optical $\mathcal{PT}$-symmetric dimer},}\ }\href
  {https://doi.org/10.3390/sym8090083} {\bibfield  {journal} {\bibinfo
  {journal} {Symmetry}\ }\textbf {\bibinfo {volume} {8}},\ \bibinfo {pages}
  {83} (\bibinfo {year} {2016})}\BibitemShut {NoStop}%
\bibitem [{\citenamefont {Boyd}(2003)}]{Boyd2003}%
  \BibitemOpen
  \bibfield  {author} {\bibinfo {author} {\bibfnamefont {R.~W.}\ \bibnamefont
  {Boyd}},\ }\href@noop {} {\emph {\bibinfo {title} {Nonlinear Optics}}}\
  (\bibinfo  {publisher} {Academic Press, New York},\ \bibinfo {year}
  {2003})\BibitemShut {NoStop}%
\bibitem [{\citenamefont {Mandel}\ and\ \citenamefont
  {Wolf}(1995)}]{Mandel1995}%
  \BibitemOpen
  \bibfield  {author} {\bibinfo {author} {\bibfnamefont {L.}~\bibnamefont
  {Mandel}}\ and\ \bibinfo {author} {\bibfnamefont {E.}~\bibnamefont {Wolf}},\
  }\href@noop {} {\emph {\bibinfo {title} {Optical Coherence and Quantum
  Optics}}}\ (\bibinfo  {publisher} {Cambridge Univ. Press, Cambridge},\
  \bibinfo {year} {1995})\BibitemShut {NoStop}%
\bibitem [{\citenamefont {{Pe\v{r}ina~Jr.}}\ and\ \citenamefont
  {Pe\v{r}ina}(2000)}]{PerinaJr2000}%
  \BibitemOpen
  \bibfield  {author} {\bibinfo {author} {\bibfnamefont {J.}~\bibnamefont
  {{Pe\v{r}ina~Jr.}}}\ and\ \bibinfo {author} {\bibfnamefont {J.}~\bibnamefont
  {Pe\v{r}ina}},\ }\bibfield  {title} {\enquote {\bibinfo {title} {Quantum
  statistics of nonlinear optical couplers},}\ }in\ \href
  {https://doi.org/10.1016/S0079-6638(00)80020-7} {\emph {\bibinfo {booktitle}
  {Progress in Optics, Vol. 41}}},\ \bibinfo {editor} {edited by\ \bibinfo
  {editor} {\bibfnamefont {E.}~\bibnamefont {Wolf}}}\ (\bibinfo  {publisher}
  {Elsevier, Amsterdam},\ \bibinfo {year} {2000})\ pp.\ \bibinfo {pages}
  {361---419}\BibitemShut {NoStop}%
\bibitem [{\citenamefont {El-Orany}\ \emph {et~al.}(2004)\citenamefont
  {El-Orany}, \citenamefont {Abdalla},\ and\ \citenamefont
  {Pe{\v{r}}ina}}]{ElOrany2004}%
  \BibitemOpen
  \bibfield  {author} {\bibinfo {author} {\bibfnamefont {F.~A.~A.}\
  \bibnamefont {El-Orany}}, \bibinfo {author} {\bibfnamefont {M.~S.}\
  \bibnamefont {Abdalla}}, \ and\ \bibinfo {author} {\bibfnamefont
  {J.}~\bibnamefont {Pe{\v{r}}ina}},\ }\bibfield  {title} {\enquote {\bibinfo
  {title} {Quantum properties of the codirectional three-mode {K}err nonlinear
  coupler},}\ }\href {\doibase 10.1140/epjd/e2005-00048-2} {\bibfield
  {journal} {\bibinfo  {journal} {Eur. Phys. J. D}\ }\textbf {\bibinfo {volume}
  {33}},\ \bibinfo {pages} {453--463} (\bibinfo {year} {2004})}\BibitemShut
  {NoStop}%
\bibitem [{\citenamefont {Thapliyal}\ \emph {et~al.}({2014})\citenamefont
  {Thapliyal}, \citenamefont {Pathak}, \citenamefont {Sen},\ and\ \citenamefont
  {Pe\v{r}ina}}]{Thapliyal2014}%
  \BibitemOpen
  \bibfield  {author} {\bibinfo {author} {\bibfnamefont {K.}~\bibnamefont
  {Thapliyal}}, \bibinfo {author} {\bibfnamefont {A.}~\bibnamefont {Pathak}},
  \bibinfo {author} {\bibfnamefont {B.}~\bibnamefont {Sen}}, \ and\ \bibinfo
  {author} {\bibfnamefont {J.}~\bibnamefont {Pe\v{r}ina}},\ }\bibfield  {title}
  {\enquote {\bibinfo {title} {{Higher-order nonclassicalities in a
  codirectional nonlinear optical coupler: Quantum entanglement, squeezing, and
  antibunching}},}\ }\href {https://doi.org/10.1103/PhysRevA.90.013808}
  {\bibfield  {journal} {\bibinfo  {journal} {{Phys. Rev. A}}\ }\textbf
  {\bibinfo {volume} {{90}}},\ \bibinfo {pages} {{013808}} (\bibinfo {year}
  {{2014}})}\BibitemShut {NoStop}%
\bibitem [{\citenamefont {Milburn}\ and\ \citenamefont
  {Holmes}(1986)}]{Milburn1986}%
  \BibitemOpen
  \bibfield  {author} {\bibinfo {author} {\bibfnamefont {G.~J.}\ \bibnamefont
  {Milburn}}\ and\ \bibinfo {author} {\bibfnamefont {C.~A.}\ \bibnamefont
  {Holmes}},\ }\bibfield  {title} {\enquote {\bibinfo {title} {Dissipative
  quantum and classical {Liouville} mechanics of the anharmonic oscillator},}\
  }\href {https://doi.org/10.1103/PhysRevLett.56.2237} {\bibfield  {journal}
  {\bibinfo  {journal} {Phys. Rev. Lett.}\ }\textbf {\bibinfo {volume} {56}},\
  \bibinfo {pages} {2237--2240} (\bibinfo {year} {1986})}\BibitemShut {NoStop}%
\bibitem [{\citenamefont {Milburn}\ and\ \citenamefont
  {Holmes}(1991)}]{Milburn1991}%
  \BibitemOpen
  \bibfield  {author} {\bibinfo {author} {\bibfnamefont {G.~J.}\ \bibnamefont
  {Milburn}}\ and\ \bibinfo {author} {\bibfnamefont {C.~A.}\ \bibnamefont
  {Holmes}},\ }\bibfield  {title} {\enquote {\bibinfo {title} {Quantum
  coherence and classical chaos in a pulsed parametric oscillator with a {K}err
  nonlinearity},}\ }\href {https://doi.org/10.1103/physreva.44.4704} {\bibfield
   {journal} {\bibinfo  {journal} {Phys. Rev. A}\ }\textbf {\bibinfo {volume}
  {44}},\ \bibinfo {pages} {4704--4711} (\bibinfo {year} {1991})}\BibitemShut
  {NoStop}%
\bibitem [{\citenamefont {{Kowalewska-Kud{\l}aszyk}}\ \emph
  {et~al.}(2008)\citenamefont {{Kowalewska-Kud{\l}aszyk}}, \citenamefont
  {Kalaga},\ and\ \citenamefont {Leo{\'n}ski}}]{Kowalewska2008}%
  \BibitemOpen
  \bibfield  {author} {\bibinfo {author} {\bibfnamefont {A.}~\bibnamefont
  {{Kowalewska-Kud{\l}aszyk}}}, \bibinfo {author} {\bibfnamefont {J.~K.}\
  \bibnamefont {Kalaga}}, \ and\ \bibinfo {author} {\bibfnamefont
  {W.}~\bibnamefont {Leo{\'n}ski}},\ }\bibfield  {title} {\enquote {\bibinfo
  {title} {Wigner-function nonclassicality as indicator of quantum chaos},}\
  }\href {https://doi.org/10.1103/PhysRevE.78.066219} {\bibfield  {journal}
  {\bibinfo  {journal} {Phys. Rev. E}\ }\textbf {\bibinfo {volume} {78}},\
  \bibinfo {pages} {066219} (\bibinfo {year} {2008})}\BibitemShut {NoStop}%
\bibitem [{\citenamefont {{Kowalewska-Kud{\l}aszyk}}\ \emph
  {et~al.}(2009)\citenamefont {{Kowalewska-Kud{\l}aszyk}}, \citenamefont
  {Kalaga},\ and\ \citenamefont {Leo{\'n}ski}}]{Kowalewska2009}%
  \BibitemOpen
  \bibfield  {author} {\bibinfo {author} {\bibfnamefont {A.}~\bibnamefont
  {{Kowalewska-Kud{\l}aszyk}}}, \bibinfo {author} {\bibfnamefont {J.~K.}\
  \bibnamefont {Kalaga}}, \ and\ \bibinfo {author} {\bibfnamefont
  {W.}~\bibnamefont {Leo{\'n}ski}},\ }\bibfield  {title} {\enquote {\bibinfo
  {title} {Long-time fidelity and chaos for a kicked nonlinear oscillator
  system},}\ }\href {https://doi.org/10.1016/j.physleta.2009.02.022} {\bibfield
   {journal} {\bibinfo  {journal} {Phys. Lett. A}\ }\textbf {\bibinfo {volume}
  {373}},\ \bibinfo {pages} {1334--1340} (\bibinfo {year} {2009})}\BibitemShut
  {NoStop}%
\bibitem [{\citenamefont {Shahinyan}\ \emph {et~al.}(2013)\citenamefont
  {Shahinyan}, \citenamefont {Chew},\ and\ \citenamefont
  {Kryuchkyan}}]{Shahinyan2013}%
  \BibitemOpen
  \bibfield  {author} {\bibinfo {author} {\bibfnamefont {A.~R.}\ \bibnamefont
  {Shahinyan}}, \bibinfo {author} {\bibfnamefont {L.~Y.}\ \bibnamefont {Chew}},
  \ and\ \bibinfo {author} {\bibfnamefont {G.~Y.}\ \bibnamefont {Kryuchkyan}},\
  }\bibfield  {title} {\enquote {\bibinfo {title} {Probing quantum dissipative
  chaos using purity},}\ }\href
  {https://doi.org/10.1016/j.physleta.2013.08.029} {\bibfield  {journal}
  {\bibinfo  {journal} {Phys. Lett. A}\ }\textbf {\bibinfo {volume} {377}},\
  \bibinfo {pages} {2743--2748} (\bibinfo {year} {2013})}\BibitemShut {NoStop}%
\bibitem [{\citenamefont {Leo{\'{n}}ski}\ and\ \citenamefont
  {Tana{\'{s}}}(1994)}]{Leonski1994}%
  \BibitemOpen
  \bibfield  {author} {\bibinfo {author} {\bibfnamefont {W.}~\bibnamefont
  {Leo{\'{n}}ski}}\ and\ \bibinfo {author} {\bibfnamefont {R.}~\bibnamefont
  {Tana{\'{s}}}},\ }\bibfield  {title} {\enquote {\bibinfo {title} {Possibility
  of producing the one-photon state in a kicked cavity with a nonlinear kerr
  medium},}\ }\href {https://doi.org/10.1103/physreva.49.r20} {\bibfield
  {journal} {\bibinfo  {journal} {Phys. Rev. A}\ }\textbf {\bibinfo {volume}
  {49}},\ \bibinfo {pages} {R20--R23} (\bibinfo {year} {1994})}\BibitemShut
  {NoStop}%
\bibitem [{\citenamefont {Miranowicz}\ and\ \citenamefont
  {Leo{\'{n}}ski}(2006)}]{Miranowicz2006}%
  \BibitemOpen
  \bibfield  {author} {\bibinfo {author} {\bibfnamefont {A.}~\bibnamefont
  {Miranowicz}}\ and\ \bibinfo {author} {\bibfnamefont {W.}~\bibnamefont
  {Leo{\'{n}}ski}},\ }\bibfield  {title} {\enquote {\bibinfo {title} {Two-mode
  optical state truncation and generation of maximally entangled states in
  pumped nonlinear couplers},}\ }\href
  {http://stacks.iop.org/0953-4075/39/i=7/a=011} {\bibfield  {journal}
  {\bibinfo  {journal} {{J. Phys. B}}\ }\textbf {\bibinfo {volume} {39}},\
  \bibinfo {pages} {1683--1700} (\bibinfo {year} {2006})}\BibitemShut {NoStop}%
\bibitem [{\citenamefont {Hovsepyan}\ \emph {et~al.}({2014})\citenamefont
  {Hovsepyan}, \citenamefont {Shahinyan},\ and\ \citenamefont
  {Kryuchkyan}}]{Hovsepyan2014}%
  \BibitemOpen
  \bibfield  {author} {\bibinfo {author} {\bibfnamefont {G.~H.}\ \bibnamefont
  {Hovsepyan}}, \bibinfo {author} {\bibfnamefont {A.~R.}\ \bibnamefont
  {Shahinyan}}, \ and\ \bibinfo {author} {\bibfnamefont {G.~Y.}\ \bibnamefont
  {Kryuchkyan}},\ }\bibfield  {title} {\enquote {\bibinfo {title} {{Multiphoton
  blockades in pulsed regimes beyond stationary limits}},}\ }\href
  {https://link.aps.org/doi/10.1103/PhysRevA.90.013839} {\bibfield  {journal}
  {\bibinfo  {journal} {Phys. Rev. A}\ }\textbf {\bibinfo {volume} {{90}}},\
  \bibinfo {pages} {{013839}} (\bibinfo {year} {{2014}})}\BibitemShut {NoStop}%
\bibitem [{\citenamefont {Miranowicz}\ \emph {et~al.}(2014)\citenamefont
  {Miranowicz}, \citenamefont {Bajer}, \citenamefont {Paprzycka}, \citenamefont
  {Liu}, \citenamefont {Zagoskin},\ and\ \citenamefont
  {Nori}}]{Miranowicz2014}%
  \BibitemOpen
  \bibfield  {author} {\bibinfo {author} {\bibfnamefont {A.}~\bibnamefont
  {Miranowicz}}, \bibinfo {author} {\bibfnamefont {J.}~\bibnamefont {Bajer}},
  \bibinfo {author} {\bibfnamefont {M.}~\bibnamefont {Paprzycka}}, \bibinfo
  {author} {\bibfnamefont {{Y.-X.}}\ \bibnamefont {Liu}}, \bibinfo {author}
  {\bibfnamefont {A.~M.}\ \bibnamefont {Zagoskin}}, \ and\ \bibinfo {author}
  {\bibfnamefont {F.}~\bibnamefont {Nori}},\ }\bibfield  {title} {\enquote
  {\bibinfo {title} {State-dependent photon blockade via quantum-reservoir
  engineering},}\ }\href {https://doi.org/10.1103/physreva.90.033831}
  {\bibfield  {journal} {\bibinfo  {journal} {Phys. Rev. A}\ }\textbf {\bibinfo
  {volume} {90}},\ \bibinfo {pages} {033831} (\bibinfo {year}
  {2014})}\BibitemShut {NoStop}%
\bibitem [{\citenamefont {Kowalewska-Kud{\l}aszyk}\ \emph
  {et~al.}(2012)\citenamefont {Kowalewska-Kud{\l}aszyk}, \citenamefont
  {{Leo{\'n}ski}},\ and\ \citenamefont {{Pe\v{r}ina~Jr.}}}]{Kowalewska2012}%
  \BibitemOpen
  \bibfield  {author} {\bibinfo {author} {\bibfnamefont {A.}~\bibnamefont
  {Kowalewska-Kud{\l}aszyk}}, \bibinfo {author} {\bibfnamefont
  {W.}~\bibnamefont {{Leo{\'n}ski}}}, \ and\ \bibinfo {author} {\bibfnamefont
  {J.}~\bibnamefont {{Pe\v{r}ina~Jr.}}},\ }\bibfield  {title} {\enquote
  {\bibinfo {title} {Generalized {B}ell states generation in a parametrically
  excited nonlinear coupler},}\ }\href
  {https://doi.org/10.1088/0031-8949/2012/T147/014016} {\bibfield  {journal}
  {\bibinfo  {journal} {Phys. Scr.}\ }\textbf {\bibinfo {volume} {T147}},\
  \bibinfo {pages} {014016} (\bibinfo {year} {2012})}\BibitemShut {NoStop}%
\bibitem [{\citenamefont {Olsen}(2015{\natexlab{a}})}]{Olsen2015a}%
  \BibitemOpen
  \bibfield  {author} {\bibinfo {author} {\bibfnamefont {M.~K.}\ \bibnamefont
  {Olsen}},\ }\bibfield  {title} {\enquote {\bibinfo {title} {Asymmetric
  steering in coherent transport of atomic population with a three-well
  {Bose-Hubbard} model},}\ }\href {https://doi.org/10.1364/JOSAB.32.000A15}
  {\bibfield  {journal} {\bibinfo  {journal} {J. Opt. Soc. Am. B}\ }\textbf
  {\bibinfo {volume} {32}},\ \bibinfo {pages} {A15--A19} (\bibinfo {year}
  {2015}{\natexlab{a}})}\BibitemShut {NoStop}%
\bibitem [{\citenamefont {Olsen}(2015{\natexlab{b}})}]{Olsen2015b}%
  \BibitemOpen
  \bibfield  {author} {\bibinfo {author} {\bibfnamefont {M.~K.}\ \bibnamefont
  {Olsen}},\ }\bibfield  {title} {\enquote {\bibinfo {title} {Spreading of
  entanglement and steering along small {Bose-Hubbard} chains},}\ }\href
  {https://doi.org/10.1103/PhysRevA.92.033627} {\bibfield  {journal} {\bibinfo
  {journal} {Phys. Rev. A}\ }\textbf {\bibinfo {volume} {92}},\ \bibinfo
  {pages} {033627} (\bibinfo {year} {2015}{\natexlab{b}})}\BibitemShut
  {NoStop}%
\bibitem [{\citenamefont {Kalaga}\ \emph {et~al.}({2016})\citenamefont
  {Kalaga}, \citenamefont {{Kowalewska-Kud{\l}aszyk}}, \citenamefont {{Leo{\'
  n}ski}},\ and\ \citenamefont {Barasi{\' n}ski}}]{Kalaga2016}%
  \BibitemOpen
  \bibfield  {author} {\bibinfo {author} {\bibfnamefont {J.~K.}\ \bibnamefont
  {Kalaga}}, \bibinfo {author} {\bibfnamefont {A.}~\bibnamefont
  {{Kowalewska-Kud{\l}aszyk}}}, \bibinfo {author} {\bibfnamefont
  {W.}~\bibnamefont {{Leo{\' n}ski}}}, \ and\ \bibinfo {author} {\bibfnamefont
  {A.}~\bibnamefont {Barasi{\' n}ski}},\ }\bibfield  {title} {\enquote
  {\bibinfo {title} {{Quantum correlations and entanglement in a model
  comprised of a short chain of nonlinear oscillators}},}\ }\href
  {https://doi.org/10.1103/PhysRevA.94.032304} {\bibfield  {journal} {\bibinfo
  {journal} {{Phys. Rev. A}}\ }\textbf {\bibinfo {volume} {{94}}},\ \bibinfo
  {pages} {{032304}} (\bibinfo {year} {{2016}})}\BibitemShut {NoStop}%
\bibitem [{\citenamefont {Kalaga}\ \emph {et~al.}({2017})\citenamefont
  {Kalaga}, \citenamefont {{Leo{\' n}ski}},\ and\ \citenamefont {Szcz{\c{e}\'
  s}niak}}]{Kalaga2017}%
  \BibitemOpen
  \bibfield  {author} {\bibinfo {author} {\bibfnamefont {J.~K.}\ \bibnamefont
  {Kalaga}}, \bibinfo {author} {\bibfnamefont {W.}~\bibnamefont {{Leo{\'
  n}ski}}}, \ and\ \bibinfo {author} {\bibfnamefont {R.}~\bibnamefont
  {Szcz{\c{e}\' s}niak}},\ }\bibfield  {title} {\enquote {\bibinfo {title}
  {Quantum steering and entanglement in three-mode triangle {Bose-Hubbard}
  system},}\ }\href {https://doi.org/10.1007/s11128-017-1717-5} {\bibfield
  {journal} {\bibinfo  {journal} {Quant. Inf. Proc.}\ }\textbf {\bibinfo
  {volume} {{16}}},\ \bibinfo {pages} {{UNSP 265}} (\bibinfo {year}
  {{2017}})}\BibitemShut {NoStop}%
\bibitem [{\citenamefont {{Pe\v{r}ina~Jr.}}(2016)}]{PerinaJr2016b}%
  \BibitemOpen
  \bibfield  {author} {\bibinfo {author} {\bibfnamefont {J.}~\bibnamefont
  {{Pe\v{r}ina~Jr.}}},\ }\bibfield  {title} {\enquote {\bibinfo {title}
  {Coherent light in intense spatio-spectral twin beams},}\ }\href
  {https://doi.org/10.1103/PhysRevA.93.063857} {\bibfield  {journal} {\bibinfo
  {journal} {Phys. Rev. A}\ }\textbf {\bibinfo {volume} {93}},\ \bibinfo
  {pages} {063857} (\bibinfo {year} {2016})}\BibitemShut {NoStop}%
\bibitem [{\citenamefont {Sukhorukov}\ \emph {et~al.}(2010)\citenamefont
  {Sukhorukov}, \citenamefont {Xu},\ and\ \citenamefont
  {Kivshar}}]{Sukhorukov2010}%
  \BibitemOpen
  \bibfield  {author} {\bibinfo {author} {\bibfnamefont {A.~A.}\ \bibnamefont
  {Sukhorukov}}, \bibinfo {author} {\bibfnamefont {Z.}~\bibnamefont {Xu}}, \
  and\ \bibinfo {author} {\bibfnamefont {Y.~S.}\ \bibnamefont {Kivshar}},\
  }\bibfield  {title} {\enquote {\bibinfo {title} {Nonlinear suppression of
  time reversals in {$\mathcal{PT}$-symmetric} optical couplers},}\ }\href
  {https://doi.org/10.1103/PhysRevA.82.043818} {\bibfield  {journal} {\bibinfo
  {journal} {Phys. Rev. A}\ }\textbf {\bibinfo {volume} {82}},\ \bibinfo
  {pages} {043818} (\bibinfo {year} {2010})}\BibitemShut {NoStop}%
\bibitem [{\citenamefont {Pe\v{r}ina}(1991)}]{Perina1991}%
  \BibitemOpen
  \bibfield  {author} {\bibinfo {author} {\bibfnamefont {J.}~\bibnamefont
  {Pe\v{r}ina}},\ }\href {http://www.doi.org/10.1007/978-94-011-2400-3} {\emph
  {\bibinfo {title} {Quantum Statistics of Linear and Nonlinear Optical
  Phenomena}}}\ (\bibinfo  {publisher} {Kluwer, Dordrecht},\ \bibinfo {year}
  {1991})\BibitemShut {NoStop}%
\bibitem [{\citenamefont {Sargent}\ \emph {et~al.}(1974)\citenamefont
  {Sargent}, \citenamefont {Scully},\ and\ \citenamefont {Lamb}}]{Lamb1974}%
  \BibitemOpen
  \bibfield  {author} {\bibinfo {author} {\bibfnamefont {M.}~\bibnamefont
  {Sargent}}, \bibinfo {author} {\bibfnamefont {M.~O.}\ \bibnamefont {Scully}},
  \ and\ \bibinfo {author} {\bibfnamefont {W.~E.}\ \bibnamefont {Lamb}},\
  }\href@noop {} {\emph {\bibinfo {title} {Laser Physics}}}\ (\bibinfo
  {publisher} {Addison-Wesley},\ \bibinfo {year} {1974})\BibitemShut {NoStop}%
\bibitem [{\citenamefont {Meystre}\ and\ \citenamefont
  {{Sargent~III}}(2007)}]{Meystre2007}%
  \BibitemOpen
  \bibfield  {author} {\bibinfo {author} {\bibfnamefont {P.}~\bibnamefont
  {Meystre}}\ and\ \bibinfo {author} {\bibfnamefont {M.}~\bibnamefont
  {{Sargent~III}}},\ }\href@noop {} {\emph {\bibinfo {title} {Elements of
  Quantum Optics, 4nd edition}}}\ (\bibinfo  {publisher} {Springer, Berlin},\
  \bibinfo {year} {2007})\BibitemShut {NoStop}%
\bibitem [{\citenamefont {Adesso}\ and\ \citenamefont
  {Illuminati}(2007)}]{Adesso2007}%
  \BibitemOpen
  \bibfield  {author} {\bibinfo {author} {\bibfnamefont {G.}~\bibnamefont
  {Adesso}}\ and\ \bibinfo {author} {\bibfnamefont {F.}~\bibnamefont
  {Illuminati}},\ }\bibfield  {title} {\enquote {\bibinfo {title} {Entanglement
  in continuous variable systems: Recent advances and current perspectives},}\
  }\href {https://doi.org/10.1088/1751-8113/40/28/S01} {\bibfield  {journal}
  {\bibinfo  {journal} {J. Phys. A: Math. Theor.}\ }\textbf {\bibinfo {volume}
  {40}},\ \bibinfo {pages} {7821--7880} (\bibinfo {year} {2007})}\BibitemShut
  {NoStop}%
\bibitem [{\citenamefont {Luk\v{s}}\ \emph {et~al.}(1988)\citenamefont
  {Luk\v{s}}, \citenamefont {Pe\v{r}inov\'{a}},\ and\ \citenamefont
  {Pe\v{r}ina}}]{Luks1988}%
  \BibitemOpen
  \bibfield  {author} {\bibinfo {author} {\bibfnamefont {A.}~\bibnamefont
  {Luk\v{s}}}, \bibinfo {author} {\bibfnamefont {V.}~\bibnamefont
  {Pe\v{r}inov\'{a}}}, \ and\ \bibinfo {author} {\bibfnamefont
  {J.}~\bibnamefont {Pe\v{r}ina}},\ }\bibfield  {title} {\enquote {\bibinfo
  {title} {Principal squeezing of vacuum fluctuations},}\ }\href
  {https://doi.org/10.1016/0030-4018(88)90322-7} {\bibfield  {journal}
  {\bibinfo  {journal} {Opt. Commun.}\ }\textbf {\bibinfo {volume} {67}},\
  \bibinfo {pages} {149---151} (\bibinfo {year} {1988})}\BibitemShut {NoStop}%
\bibitem [{\citenamefont {{Pe\v{r}ina~Jr.}}\ and\ \citenamefont
  {Luk\v{s}}(2019)}]{PerinaJr2019x}%
  \BibitemOpen
  \bibfield  {author} {\bibinfo {author} {\bibfnamefont {J.}~\bibnamefont
  {{Pe\v{r}ina~Jr.}}}\ and\ \bibinfo {author} {\bibfnamefont {A.}~\bibnamefont
  {Luk\v{s}}},\ }\bibfield  {title} {\enquote {\bibinfo {title} {Quantum
  behavior of a $\mathcal{PT}$-symmetric two-mode system with {cross-Kerr}
  nonlinearity},}\ }\href@noop {} {\bibfield  {journal} {\bibinfo  {journal}
  {Symmetry}\ }\textbf {\bibinfo {volume} {11}},\ \bibinfo {pages} {1020}
  (\bibinfo {year} {2019})}\BibitemShut {NoStop}%
\bibitem [{\citenamefont {Hill}\ and\ \citenamefont
  {Wootters}(1997)}]{Hill1997}%
  \BibitemOpen
  \bibfield  {author} {\bibinfo {author} {\bibfnamefont {S.}~\bibnamefont
  {Hill}}\ and\ \bibinfo {author} {\bibfnamefont {W.~K.}\ \bibnamefont
  {Wootters}},\ }\bibfield  {title} {\enquote {\bibinfo {title} {Entanglement
  of a pair of quantum bits},}\ }\href
  {https://doi.org/10.1103/PhysRevLett.78.5022} {\bibfield  {journal} {\bibinfo
   {journal} {Phys. Rev. Lett.}\ }\textbf {\bibinfo {volume} {78}},\ \bibinfo
  {pages} {5022} (\bibinfo {year} {1997})}\BibitemShut {NoStop}%
\end{thebibliography}

%

\end{document}